\documentclass[acmsmall,screen,nonacm]{acmart} %

\usepackage{hyperref}
\usepackage[capitalize]{cleveref}
\usepackage{xspace}
\usepackage{subcaption}
\usepackage{multirow}
\usepackage{mathtools}
\usepackage{bbm}
\usepackage{tabularx}
\usepackage{pgfplots}
\pgfplotsset{compat=newest}

\usepgfplotslibrary{groupplots}
\usetikzlibrary{pgfplots.polar}
\usetikzlibrary{pgfplots.statistics}
\usetikzlibrary{arrows.meta}
\usetikzlibrary{overlay-beamer-styles}

\definecolor{veryLightGrey}{HTML}{F2F2F2}
\definecolor{lightGrey}{HTML}{DDDDDD}
\definecolor{colorSimdRecSplit}{HTML}{444444}
\definecolor{colorChd}{HTML}{377EB8}
\definecolor{colorRustFmph}{HTML}{A65628}
\definecolor{colorRustFmphGo}{HTML}{A65628}
\definecolor{colorSicHash}{HTML}{4DAF4A}
\definecolor{colorPthash}{HTML}{984EA3}
\definecolor{colorDensePtHash}{HTML}{377EB8}
\definecolor{colorRecSplit}{HTML}{FF7F00}
\definecolor{colorBbhash}{HTML}{F781BF}
\definecolor{colorShockHash}{HTML}{F8BA01}
\definecolor{colorBipartiteShockHash}{HTML}{F781BF}
\definecolor{colorBipartiteShockHashFlat}{HTML}{984EA3}
\definecolor{colorBmz}{HTML}{000000}
\definecolor{colorBdz}{HTML}{444444}
\definecolor{colorFch}{HTML}{444444}
\definecolor{colorChm}{HTML}{A65628}
\definecolor{colorFiPS}{HTML}{FF7F00}
\definecolor{colorConsensus}{HTML}{4DAF4A}
\definecolor{colorPtrHash}{HTML}{F8BA01}
\definecolor{colorMorphisHash}{HTML}{A65628}
\definecolor{colorMorphisHashFlat}{HTML}{E41A1C}
\definecolor{colorPhast}{HTML}{E41A1C}
\definecolor{colorPhastPlus}{HTML}{34B3BF}

\colorlet{colorBruteForce}{colorSicHash}
\colorlet{colorRotationFitting}{colorChd}
\definecolor{colorCuckoo}{HTML}{E41A1C}
\definecolor{acmBlue}{HTML}{155196}

\pgfdeclareimage[interpolate,height=2.0mm,width=2.0mm]{shockhashMark}{fig/shockhash}
\pgfdeclareplotmark{shockhash}{\pgftext[at=\pgfpointorigin]{\pgfuseimage{shockhashMark}}}
\pgfdeclareimage[interpolate,height=1.85mm,width=1.85mm]{shockhashInverseMark}{fig/shockhashInverse}
\pgfdeclareplotmark{shockhashInverse}{\pgftext[at=\pgfpointorigin]{\pgfuseimage{shockhashInverseMark}}}
\pgfdeclareimage[interpolate,height=1.85mm,width=1.85mm]{phobicMark}{fig/phobic}
\pgfdeclareplotmark{phobic}{\pgftext[at=\pgfpointorigin]{\pgfuseimage{phobicMark}}}
\pgfdeclareimage[interpolate,height=1.85mm,width=1.85mm]{phobicInverseMark}{fig/phobicInverse}
\pgfdeclareplotmark{phobicInverse}{\pgftext[at=\pgfpointorigin]{\pgfuseimage{phobicInverseMark}}}
\pgfdeclareimage[interpolate,height=1.85mm,width=1.85mm]{fingerprintMark}{fig/hand}
\pgfdeclareplotmark{fingerprint}{\pgftext[at=\pgfpointorigin]{\pgfuseimage{fingerprintMark}}}
\pgfdeclareimage[interpolate,height=1.85mm,width=1.85mm]{fingerprintInverseMark}{fig/handInverse}
\pgfdeclareplotmark{fingerprintInverse}{\pgftext[at=\pgfpointorigin]{\pgfuseimage{fingerprintInverseMark}}}
\pgfdeclareimage[interpolate,height=2.0mm,width=2.0mm]{consensusMark}{fig/consensus}
\pgfdeclareplotmark{consensus}{\pgftext[at=\pgfpointorigin]{\pgfuseimage{consensusMark}}}
\pgfdeclareimage[interpolate,height=2.0mm,width=2.0mm]{consensusInverseMark}{fig/consensusInverse}
\pgfdeclareplotmark{consensusInverse}{\pgftext[at=\pgfpointorigin]{\pgfuseimage{consensusInverseMark}}}

\pgfplotscreateplotcyclelist{myColorList}{%
  colorShockHash,mark=triangle\\
  colorBdz,mark=pentagon\\%
  colorChd,mark=square\\%
  colorSicHash,mark=o\\%
  colorRecSplit,mark=diamond\\%
  colorPthash,mark=x\\%
  colorBbhash,mark=flippedTriangle\\%
}
\pgfdeclareplotmark{flippedTriangle}{%
  \pgfpathmoveto{\pgfqpointpolar{-90}{1.2\pgfplotmarksize}}%
  \pgfpathlineto{\pgfqpointpolar{30}{1.2\pgfplotmarksize}}%
  \pgfpathlineto{\pgfqpointpolar{150}{1.2\pgfplotmarksize}}%
  \pgfpathclose%
  \pgfusepath{stroke}
}
\pgfdeclareplotmark{rightTriangle}{%
  \pgfpathmoveto{\pgfqpointpolar{0}{1.2\pgfplotmarksize}}%
  \pgfpathlineto{\pgfqpointpolar{120}{1.2\pgfplotmarksize}}%
  \pgfpathlineto{\pgfqpointpolar{240}{1.2\pgfplotmarksize}}%
  \pgfpathclose%
  \pgfusepath{stroke}%
}
\pgfdeclareplotmark{leftTriangle}{%
  \pgfpathmoveto{\pgfqpointpolar{-180}{1.2\pgfplotmarksize}}%
  \pgfpathlineto{\pgfqpointpolar{-60}{1.2\pgfplotmarksize}}%
  \pgfpathlineto{\pgfqpointpolar{60}{1.2\pgfplotmarksize}}%
  \pgfpathclose%
  \pgfusepath{stroke}%
}

\pgfplotsset{
  mark repeat*/.style={
    scatter,
    scatter src=x,
    scatter/@pre marker code/.code={
      \pgfmathtruncatemacro\usemark{
        or(mod(\coordindex,#1)==0, (\coordindex==(\numcoords-1))
      }
      \ifnum\usemark=0
        \pgfplotsset{mark=none}
      \fi
    },
    scatter/@post marker code/.code={}
  },
  log x ticks with fixed point/.style={
      xticklabel={
        \pgfkeys{/pgf/fpu=true}
        \pgfmathparse{exp(\tick)}%
        \pgfmathprintnumber[fixed relative, precision=3]{\pgfmathresult}
        \pgfkeys{/pgf/fpu=false}
      }
  },
  log y ticks with fixed point/.style={
      yticklabel={
        \pgfkeys{/pgf/fpu=true}
        \pgfmathparse{exp(\tick)}%
        \pgfmathprintnumber[fixed relative, precision=3]{\pgfmathresult}
        \pgfkeys{/pgf/fpu=false}
      }
  },
  every axis/.style={scale only axis},
  major grid style={thin,dotted},
  minor grid style={thin,dotted},
  ymajorgrids,
  yminorgrids,
  every axis/.append style={
    line width=0.9pt,
    tick style={
      line cap=round,
      thin,
      major tick length=4pt,
      minor tick length=2pt,
    },
    mark options={solid},
  },
  legend cell align=left,
  legend style={
    line width=0.7pt,
    /tikz/every even column/.append style={column sep=2mm,black},
    /tikz/every odd column/.append style={black},
    mark options={solid},
    font=\footnotesize,
  },
  title style={yshift=-2pt},
  enlarge x limits=0.04,
  scale only axis,
  /pgf/number format/1000 sep={},
  axis lines*=left,
  xlabel near ticks,
  ylabel near ticks,
  axis lines*=left,
  label style={font=\footnotesize},
  every axis y label/.append style={yshift=-1pt,inner sep=0,outer sep=0},
  tick label style={font=\footnotesize},
  tick align=outside,
  cycle list name=myColorList,
  plotEvalPareto/.style={
    width=57mm,
    height=31mm,
    xmax=5.5,
  },
  plotEvalScaling/.style={
    width=57mm,
    height=28mm,
    ymin=0,
  },
  plotEvalScalingN/.style={
    width=57mm,
    height=28mm,
  },
}

\newcommand{\Oh}[1]{\ensuremath{\mathcal{O}\!\left(#1\right)}\xspace}
\newcommand{\tOh}[1]{\ensuremath{\mathcal{O}(#1)}\xspace}
\newcommand{\etal}{\mbox{et al.}\xspace} %
\newcommand{\rot}[2]{\multirow{#1}{*}{\rotatebox[origin=c]{90}{#2}}}
\newcommand{\mr}[1]{\multirow{2}{*}{#1}}
\newcommand{\mrf}[1]{\multirow{2}{*}{#1}}
\def\consensus{\texorpdfstring{C\scalebox{0.8}{ONSENSUS}}{CONSENSUS}\xspace}
\newcommand{\myparagraph}[1]{\paragraph{#1}}

\newcommand{\psfrage}[1]{{\color{orange}{\sf[PS: #1]}}} %
\newcommand{\hpfrage}[1]{{\color{violet}\sf[HP: #1]}} %
\newcommand{\gpfrage}[1]{{\color{teal}\sf[GP: #1]}} %
\newcommand{\swfrage}[1]{{\color{red}\sf[SW: #1]}} %
\newcommand{\svfrage}[1]{{\color{brown}\sf[SV: #1]}} %
\newcommand{\tmfrage}[1]{{\color{purple}\sf[TM: #1]}} %
\renewcommand{\psfrage}[1]{} \renewcommand{\hpfrage}[1]{} \renewcommand{\gpfrage}[1]{} \renewcommand{\swfrage}[1]{} \renewcommand{\svfrage}[1]{} \renewcommand{\tmfrage}[1]{}

\makeatletter
\renewcommand{\fnum@figure}{Figure \thefigure}
\makeatother
\crefname{figure}{Figure}{Figures}
\Crefname{figure}{Figure}{Figures}
\newif\ifAppendix
\Appendixtrue

\let\oldcite\cite
\renewcommand\cite{\unskip~\oldcite}

\AtBeginDocument{%
  }

\setcopyright{acmlicensed}
\copyrightyear{2018}
\acmYear{2018}
\acmDOI{XXXXXXX.XXXXXXX}

\acmJournal{CSUR}
\acmISBN{978-1-4503-XXXX-X/18/06}

\begin{document}

\title{Modern Minimal Perfect Hashing: A Survey}

\author{Hans-Peter Lehmann}
\email{hans-peter.lehmann@kit.edu}
\orcid{0000-0002-0474-1805}
\affiliation{%
  \institution{Karlsruhe Institute of Technology}
  \country{Germany}
}

\author{Thomas Mueller}
\orcid{0000-0002-6614-3296}
\affiliation{%
  \institution{Independent researcher}
  \country{Switzerland}
}

\author{Rasmus Pagh}
\email{pagh@di.ku.dk}
\orcid{0000-0002-1516-9306}
\affiliation{%
  \institution{BARC, University of Copenhagen}
  \country{Denmark}
}

\author{Giulio Ermanno Pibiri}
\email{giulioermanno.pibiri@unive.it}
\orcid{0000-0003-0724-7092}
\affiliation{%
  \institution{Ca' Foscari University of Venice}
  \country{Italy}
}

\author{Peter Sanders}
\email{sanders@kit.edu}
\orcid{0000-0003-3330-9349}
\affiliation{%
  \institution{Karlsruhe Institute of Technology}
  \country{Germany}
}

\author{Sebastiano Vigna}
\email{sebastiano.vigna@unimi.it}
\orcid{0000-0002-3257-651X}
\affiliation{%
  \institution{Universit\`a degli Studi di Milano}
  \country{Italy}
}

\author{Stefan Walzer}
\email{stefan.walzer@kit.edu}
\orcid{0000-0002-6477-0106}
\affiliation{%
  \institution{Karlsruhe Institute of Technology}
  \country{Germany}
}

\renewcommand{\shortauthors}{Lehmann, Mueller, Pagh, Pibiri, Sanders, Vigna, Walzer}

\begin{abstract}
Given a set $S$ of $n$ keys, a perfect hash function for $S$ maps the keys in $S$ to the first $m \geq n$ integers without collisions.
It may return an arbitrary result for any key not in $S$ and is called \emph{minimal} if $m=n$.
The most important parameters are its space consumption, construction time, and query time.
Years of research now enable modern perfect hash functions to be extremely fast to query, very space-efficient, and scale to billions of keys.
Different approaches give different trade-offs between these aspects.
For example, the smallest constructions get within 0.1\% of the space lower bound of $\log_2 e$ bits per key.
Others are particularly fast to query, requiring only one memory access.
Perfect hashing has many applications, for example to avoid collision resolution in static hash tables, and is used in databases, bioinformatics, and stringology.

Since the last comprehensive survey in 1997, significant progress has been made.
This survey covers the latest developments and provides a starting point for getting familiar with the topic.
Additionally, our extensive experimental evaluation can serve as a guide to select a perfect hash function for use in applications.
\end{abstract}


\maketitle

\section{Introduction}\label{sec:introduction}
Given a set $S \subseteq U$ of $n$ keys drawn from a universe set $U$ of size $u$, a perfect hash function (PHF) for $S$ maps the keys in $S$ to the first $m \geq n$ integers $[m]=\{0, \ldots, m-1\}$ without collisions.
For any key not in $S$, it may return an arbitrary result.
Therefore, it does not need to be able to distinguish between keys in $S$ and keys that it was not constructed with.
As we will see later, this enables a constant amount of space per key regardless of the nature of keys.
Most practical implementations of such functions need, in fact, less than 2 bits/key, depending on the load factor $n/m$.
In contrast, a data structure storing a set of $n$ keys
must use at least $\log_2(eu/n)-o(1)$ bits/key, which could be much larger than 2 when $u \gg n$.
A perfect hash function is called \emph{minimal} (MPHF) if $m=n$, in which case it represents a bijection between the input set and the first $n$ integers.
Intuitively, the smaller output range reduces the flexibility of how keys can be placed, thus reducing the number of functions that are collision-free.
This increases the space consumption compared to the non-minimal case (details in a later paragraph).
For an MPHF, the space lower bound is about $\log_2 e \approx 1.443$ bits per key, and practical applications can already achieve $1.444$ bits per key \cite{lehmann2025consensus}.
In contrast to cryptographic hashing, properties of the inverse function are irrelevant.
We follow the practice in the literature and focus on \emph{minimal} perfect hash functions only.
Note that most of the techniques we describe have configurations to construct non-minimal perfect hash functions as well.
In fact, some techniques construct a PHF first and later convert it to an MPHF (see \cref{s:commonTools}).
Perfect hashing has many applications in various fields of computing,
from bioinformatics to stringology.
Over the past years, there has been significant progress in this research area.
Many different approaches have been proposed, showing a wide variety of ideas on how to solve the perfect hashing problem.

\myparagraph{This Survey in Context\@.}
Lewis and Cook \cite{lewis1988hashing} review early letter-based approaches \cite{sprugnoli1977perfect,cormack1985practical,fredman1984storing,du1983study,jaeschke1981reciprocal,chang1984study,cichelli1980minimal,cook1982letter,haggard1986finding,sager1985polynomial}.
The last survey about minimal perfect hashing by Czech \etal \cite{czech1997perfect} gives a comprehensive review of approaches until the year 1997 \cite{melhorn1984data,schmidt1990spatial,chang1988ordered,fox1991order,mairson1983program,jacobs1986two,winters1990minimal,fox1992practical,anderson1979comments,chang1984ordered,chang1991design,slot1984tape,winters1990minimalLargeSets,yang1985backtracking,lewis1988hashing,ramakrishna1989file,fox1989nlogn,sager1984new,czech1993linear,jaeschke1980cichelli,trono1992undergraduate,gori1989algebraic,tarjan1979storing,brain1990perfect,chang1991letter,brain1994using,dietzfelbinger1992polynomial,dietzfelbinger1990new,havas1994graphs,havas1993graph,pearson1990fast}.
We do not go into detail about these here and instead focus on the \emph{modern} approaches
that have outperformed the former ones on modern hardware.
In particular, our focus is on practical constructions, i.e., on those that have been implemented and shown to perform well in practice.
As such, we also give an extensive evaluation that shows their performance in practice and helps to choose a fitting PHF for a given application.
We nonetheless point out theoretical constructions when necessary.

\myparagraph{Space Lower Bounds\@.}\label{ss:spacebounds}
The space lower bound for representing an MPHF is $n\log_2 e - \Oh{\log n} \approx 1.443n$ bits for large $n$ and $u \to \infty$ \cite{mehlhorn1982program,mairson1983program,fredman1984bounds}.
This bound is quite simple to explain. Take a function $f : S \to [n]$ that is minimal perfect on some set $S$.
Because $f$ can be evaluated with any input key from the universe and outputs only values from $[n]$, there can be additional input sets for which $f$ is minimal perfect.
More precisely, $f$ is minimal perfect for all sets where exactly one input key is in each preimage $f^{-1}(1), f^{-1}(2), \ldots, f^{-1}(n)$.
The number of input sets on which $f$ is minimal perfect is therefore $|f^{-1}(1)| \cdot |f^{-1}(2)| \cdot \ldots \cdot |f^{-1}(n)|$.
This expression is maximized if all preimages have the same size $u/n$.
Therefore, a single function can be minimal perfect for at most $(u/n)^n$ different input sets.
There are ${u \choose n}$ different possible input sets.
Therefore, we need to be able to differentiate between at least ${u \choose n}/(u/n)^n$ different behaviors of our MPHF to be able to cover every possible input set.
A minimal perfect hash function then needs to store which of these behaviors has to be used on the respective input set.
Therefore, the number of bits needed is
$\log_2\left({u \choose n} / \left(\tfrac{u}{n}\right)^n\right)$
which, by using Stirling's approximation of the factorial,
is $\approx \log_2\left(\left(\tfrac{ue}{n}\right)^n / \left(\tfrac{u}{n}\right)^n\right) - \Oh{\log n}
  = n \log_2 e - \Oh{\log n}
  \approx 1.443n$ for large $n$ and $u \to \infty$.
Using a similar combinatorial argument, one can show that the space lower bound to represent non-minimal PHFs with load factor $\alpha=n/m$ is $n\left(\log_2 e + \left(\frac{1}{\alpha}-1\right)\log_2(1-\alpha)\right) - \Oh{\log_2 n}$ \cite{melhorn1984data,belazzougui2009hash,fredman1984bounds,lehmann2024fast}.
For example, a PHF with a load factor of $\alpha=0.5$ already has a lower bound of $\approx 0.443n$ bits.

These lower bounds also hold for the \emph{expected} space requirement of randomized constructions of MPHFs and even if the input set is chosen uniformly at random.
For randomized MPHFs the lower bounds are tight as we will see for $n = m$ in \cref{s:bruteforce}: a randomized brute force construction \cite{melhorn1984data} requires $n \log_2 e - \Omega(\log n)$ bits in expectation, even for infinite universes.

A completely independent lower bound of $\Omega(\log \log u)$ bits in total \cite{fredman1984bounds} holds for \emph{deterministic} constructions of MPHFs in the \emph{worst case} even for $n = 2$ because $\log_2 u$ functions $f : U \rightarrow [2]$ are required for each pair of elements from the universe is separated by at least one such function. Since this lower bound does not apply to randomized constructions and exceeds the more general lower bound only if $n = \Oh{\log \log u}$ we leave it aside for the rest of the paper.

\myparagraph{Variants\@.}\label{s:variants}
There are several variants of perfect hashing.
Because these use independent techniques, we only mention them briefly but do not go into detail.
\emph{$k$-perfect hash functions} are a generalization that allow for up to $k$ collisions on each output value.
\emph{Order-preserving} MPHFs retain any arbitrary order of the keys, which means that they need at least $\log n$ bits per key due to storing one of the $n!$ permutations.
\emph{Monotone} MPHFs, in contrast, use the \emph{intrinsic} order of the keys as implied by an order of the universe.
Surprisingly, MMPHFs achieve a much smaller space consumption of just $\Oh{\log \log \log u}$ bits per key \cite{belazzougui2009monotone,assadi2023tight,kosolobov2024simplified}.
We refer to \ifAppendix \cref{appendix:s:variants} \else the extended version \cite{lehmann2025modern} \fi for details.

\myparagraph{Overview\@.}
We begin our journey with an overview of different applications of perfect hashing in \cref{s:applications}.
In \cref{s:preliminaries}, we give an introduction to common techniques.
We then categorize the modern approaches into three categories based on their main working principle in \cref{s:categorization}.
In \cref{s:retrievalBased,s:bruteforce,s:fingerprinting} we then discuss the approaches in those categories:
Perfect hashing through retrieval in \cref{s:retrievalBased}, brute-force in \cref{s:bruteforce}, and fingerprinting in \cref{s:fingerprinting}.
In \cref{s:theory}, we give theoretical details about the space consumption and construction time of different approaches.
After this overview of the literature, we give a detailed practical evaluation in \cref{ch:evaluation}.
There, \cref{fig:pareto} can serve as a guide to select a perfect hash function in practice, given specific space and time budgets.
We conclude this paper in \cref{s:conclusion}.

\section{Applications}\label{s:applications}
In the following section, we outline some of the wide range of applications of perfect hashing.

\myparagraph{Hash Tables and Retrieval\@.}
Hash tables are one of the most fundamental data structures used today.
When the key set is static, a perfect hash function can be used to directly index the hash table without the need for collision resolution.
This guarantees very fast access times \cite{fredman1984storing} with as few cache faults as well-engineered hash tables
(like, for example, Abseil's ``Swiss'' hash tables \cite{abseil,abseilswisstables}),
and fewer external memory access operations \cite{kurpicz2023pachash}.
If we do not need the ability to query whether some key is stored in the table, we can store only the values and not the keys in the table cells.
We thus obtain a space-efficient dictionary with a static key set and updateable values known as a \emph{value-dynamic} retrieval data structure. As in the static version (see \cref{s:retrieval}), its behavior is undefined for any key not in the original set.
The best known lower and upper bounds for updateable retrieval \cite{updateableRetrieval} suggest that constructions based on perfect hashing are optimal.
We remark that there is also work on retrieval data structures with an incremental or dynamic key set not based on perfect hashing \cite{demaine2006dictionariis,IncAndDynRetrieval,mortensen2005dynamic}. %

\myparagraph{AMQ Data Structures\@.}
An Approximate Membership Query (AMQ) data structure (also called
\emph{filter}) can answer membership queries, ``Is $x \in S$?'',
allowing false-positive answers.
The space lower bound for an AMQ data structure for $S$ with false positive rate $\varepsilon$ is $n\log_2(1/\varepsilon)$ bits \cite{carter1978exact}.
The most widely known example is the Bloom filter \cite{bloom1970space}.
With the optimal number of hash functions, it needs at least $n (\log_2(e) \cdot \log_2(1/\varepsilon))$ bits to be represented \cite{broder2004network},
thus approximately $44.3\%$ more space than the space lower bound for an AMQ
data structure.
Assuming a PHF $f$ taking $c$ bits per key, we can easily achieve
an AMQ data structure with $m(c+\lceil\log_2(1/\varepsilon)\rceil)$ bits by storing $\lceil\log_2(1/\varepsilon)\rceil$-bit fingerprints of the input keys in a packed array with $m$ cells indexed by $f$ \cite{fan2014cuckoo,bender2018bloom,broder2004network,graf2020xor,MARCHET202092}.
For example, with an MPHF we can get close to $n(\log_2(e)+\log_2(1/\varepsilon))$ bits, which is $\frac{\log_2(e)}{\log_2(1/\varepsilon)}\cdot 100\%$ extra space over the lower bound. %
Since MPHFs can have very fast construction and query time, this can be practically valuable because even more space efficient AMQ constructions based on retrieval \cite{dillinger2022burr} have slower queries.

\myparagraph{Databases\@.}
Perfect hashing also finds use in static databases (or \emph{stores})
to map variable-length keys (e.g., strings) from a known universe to smaller identifiers
that can be stored with less space \cite{muller2014retrieval,botelho2007perfect}.
For example, this is the case for Indeed Engineering \cite{indeed-eng}.
The COPR database \cite{reichinger2024copr} uses perfect hashing for efficient search in log files.
Perfect hashing can also be used to speed up finding association rules \cite{agrawal1993mining} in databases \cite{chang2005perfect,hwang2002minimal}.
SNIPS \cite{nygaard2023snips} uses perfect hashing to synchronize the content of a distributed database without communicating the full list of files that each peer stores.
Perfect hashing can also be used to enable faster joins and aggregates in OLAP databases \cite{gaffney2024perfect}.
Finally, \emph{monotone} minimal perfect hash functions can be useful as part of an algorithm for range queries in databases \cite{kurpicz2023pachash,lim2011silt}.

\myparagraph{Bioinformatics\@.}
The (node-centric) De~Bruijn graph is a directed graph where
nodes correspond to strings of length $k$, called $k$-mers, and
edges model $(k-1)$-length overlaps between the nodes \cite{de1946combinatorial,good1946normal}.
Note that there is a one-to-one correspondence between a De~Bruijn graph
and a set of $k$-mers.
This combinatorial object is vital to many applications in bioinformatics,
such as \emph{de novo} genome assembly \cite{pevzner2001eulerian,liao2019current},
transcriptome assembly from RNA \cite{grabherr2011full},
and variant calling \cite{iqbal2012novo}.
Representing such graphs in compact space is therefore very important.
Perfect hashing can be used to enable fast navigation in such graphs \cite{chapman2011meraculous}
and to support membership queries for $k$-mers \cite{almodaresi2018space,marchet2021blight,pibiri2023weighted,pibiri2022sparse,schmidt2023matchtigs}.
Pibiri \etal \cite{pibiri2023locality} also introduce the first MPHF specifically designed for $k$-mer sets.
They exploit the fact that two consecutive $k$-mers overlap by $k-1$ symbols
to beat the space lower bound that applies to general-purpose minimal perfect hashing,
and to even achieve locality of the hash values for overlapping $k$-mers.
Perfect hashing is also used to build a De~Bruijn graph in external memory;
for example to associate some metadata to each $k$-mer to aid construction \cite{khan2022scalable},
and to efficiently implement
union-find data structures for $k$-mers \cite{chikhi2016compacting}.

\myparagraph{Text Indexing\@.}
Perfect hashing can be used to index children of a node of a suffix tree \cite{weiner1973linear} in constant time \cite{belazzougui2014alphabet}.
In general, perfect hashing can be used to store trees, for example in prefix search \cite{belazzougui2010fast}.
In a large external memory lexicon, perfect hashing can be used to ensure that the number of memory accesses is low when accessing specific words \cite{Witten:1999}.
Monotone minimal perfect hashing can be used for constant-time rank queries on an input set.
As a special case, one can build a monotone MPHF on the positions of 1-bits of a bit vector \cite{Navarro:2016book}, possibly undercutting the space consumption of storing the bit vector itself.
The data structure then only supports rank queries for positions that we know have a 1-bit, which can be a common use-case.
Running rank queries on an alphabet is useful to decode the Burrows-Wheeler Transform (BWT) \cite{burrows1994block} using a perfect hash function \cite{belazzougui2014alphabet}.
For reporting the number of occurrences of each character in a substring, Belazzougui \etal \cite{BelazzouguiNV13} store the rank of each character occurrence in an array indexed by a monotone MPHF.
From a list of documents, this then enables efficiently finding the $k$ documents with the most occurrences of a pattern (top-$k$ retrieval) \cite{BelazzouguiNV13,navarro2014spaces}.
Finally, in pattern matching \cite{belazzougui2020linear,gagie2020fully,grossi2010optimal}, monotone MPHFs are applied mostly to integer sequences representing the occurrences of certain characters in a text.

\myparagraph{Natural Language Processing\@.}
$N$-gram language models are a crucial ingredient of natural language processing, machine learning, and spell checking \cite{manning1999foundations}.
Pibiri and Venturini \cite{pibiri2017efficient,pibiri2019handling} introduce a compressed trie representation of $N$-gram language models and use perfect hashing to achieve faster lookups.
For real-time speech recognition at Amazon, Strimel \etal \cite{strimel2020rescore} introduce the compressed $N$-gram data structure DashHashLM, which is particularly focused on fast lookups.
It computes IDs of the $N$-gram context using a minimal perfect hash function.
In contrast to storing the IDs explicitly, this avoids the need for a more complex lookup table.

\myparagraph{Further Applications\@.}
In flow lookup tables in routers \cite{lu2006perfect}, partially dynamic PHFs get rebuilt only if they change too much (see also \ifAppendix \cref{appendix:s:dynamicPerfectHashing}\else the extended version of this paper \cite{lehmann2025modern}\fi).
Perfect hashing helps to perform depth-first search in large implicitly defined graphs
using a value-dynamic retrieval data structure that stores one bit per node while all other data structures can reside in external memory \cite{edelkamp2008semi}.
For example, this can be used for formal verification of software via model-checking.
Finally, monotone MPHFs can be used for efficient queries in encrypted data \cite{boldyreva2011order}.

\section{Preliminaries}\label{s:preliminaries}
In this section, we explain general tools and techniques that we need later in the paper.
We also highlight basic building blocks that are used in almost every MPHF construction.

\subsection{Common Perfect Hashing Tools}\label{s:commonTools}
\vspace{1mm}

\myparagraph{Fully Random Hash Functions\@.}\label{sec:hash-functions}
In the course of this paper, we regularly need access to a fully random hash function $h(x)$, $x \in U$.
In fact, we often need a number of different hash functions, which we here model by adding a second parameter to $h$: we assume that $h(x, s)$ is a fully random function for all values $s \in \mathbbm{N}_0 = \{0,1,2,\ldots\}$ and for any $x \in U$.
We also assume that $h(x,s)$ can be evaluated in constant time
for any $x \in U$ and $s \in \mathbbm{N}_0$.
This is known as the ``simple uniform hashing assumption'' which is common in the literature about hashing \cite{dietzfelbinger1990new,pagh2007linear,pagh2008uniform,lehmann2023shockhash,kurpicz2023pachash,lehmann2023sichash}.
The assumption is an adequate model for practical hash functions and can be justified in theory \cite{dietzfelbinger2009applications,thorup2017fast,bercea2023locally,aamand2020fast}.

\myparagraph{Master Hashes\@.}
Most perfect hash functions first generate an \emph{initial, ``master'', hash}
using a high quality random hash function with large output range
(e.g., with range $[2^{64}]$ or $[2^{128}]$).%
\footnote{Examples include MurmurHash, xxHash, CityHash, and GxHash.
The SMHasher test suite \cite{appleby2021smhasher} offers a comprehensive comparison.}
This makes the construction mostly independent of the input distribution.
Another important advantage is that the construction works with fixed-size
keys (i.e., the hashes themselves) instead of the original, possibly variable-length, keys. Clearly, handling fixed-length keys aids implementation
and improves performance.
The idea is introduced in one of the first perfect hashing papers by Sprugnoli \cite{sprugnoli1977perfect}
(although initially just with a modulo operation).

\myparagraph{Partitioning\@.}
A second fundamental idea introduced by Sprugnoli \cite{sprugnoli1977perfect} is to apply \emph{partitioning} to perfect hashing.
In order to determine a perfect hash function for a larger key set, it is possible to first divide the input keys into smaller subsets of approximately the same size, for example using a random hash function.
After determining perfect hash functions on all subsets independently, adding a prefix sum over the subset sizes to the hash values gives a perfect hash function of the overall set.
While this causes some space and time overhead due to the need to store and
lookup the prefix-sum values, it can help with cache locality and therefore construction performance. Furthermore, each partition can be built independently in parallel using multiple threads.

Another way to partition PHFs that avoids storing prefix sums is to construct a $k$-perfect hash function.
PtrHash \cite{grootkoerkamp2025ptrhash} takes a different approach, to avoid prefix sums, by constructing a non-minimal PHF.
In particular, each partition is allocated the same output range $r$.
If the number of partitions is $P$, then the function is made minimal by \emph{remapping},
i.e., by ignoring the extra $rP - n$ output values as explained below.

\myparagraph{Remapping\@.}
Any perfect hash function with output range $m > n$ can be converted to an MPHF.
Folklore methods include marking the actual outputs in a vector of $m$ bits and then performing
a rank query using the output of the perfect hash function.
An alternative is to store the
unused output values in
a predecessor data structure and then discount the output of the perfect hash function
by the number of unused output values that precede it. Sometimes, in the first case
one does not even need an additional bit vector because a special encoding
of the output can be used to distinguish between used and unused entries~\cite{botelho2013practical}.
Pibiri and Trani~\cite{pibiri2021pthash} remap the values greater than $n$ to the unused output values in
$[n]$ instead.
Therefore, only a small fraction of the keys need to query the remapping data structure.
Constructing a non-minimal PHF first might be faster but might also increase space consumption and query time.

\subsection{Retrieval Data Structures}\label{s:retrieval}
A (\emph{static function}~\cite{genuzio2016fast}) \emph{retrieval data structure} represents a function $f: S \rightarrow \{0, 1\}^r$ that maps each key in $S \subseteq U$ to an $r$-bit integer.
Similarly to an MPHF,
the behavior of a retrieval data structure is undefined for any key not in $S$ and, because it does not need to differentiate between keys
in $S$ and keys not in $S$, it can represent $f$ in close to $rn$ bits.

Demaine \etal \cite{demaine2006dictionariis} point out that given an MPHF $p$ for $S$ it is trivial to construct a retrieval data structure: simply index a packed array of $r$-bit values using $p$.
This results in a space consumption of at least $n \cdot (r + \log_2 e)$ bits.
For small $r$ (such as $r = 1$), the space overhead of at least $\log_2 e$ bits/key is large percentagewise, while for large $r$ (such as $r = 16$), it may be acceptable.
However, the packed array approach issues as least two dependent memory accesses for each query whereas many kinds of retrieval data structures issue independent memory accesses~\cite{dietzfelbinger2008succinct}, which are parallelized efficiently by modern processors.
Classically, three accesses for a good balance between query efficiency and space effectiveness as discussed below.
The converse can be efficient: in \cref{s:retrievalBased} we use retrieval data structures to build perfect hash functions.

One of the main approaches to the construction of retrieval data structures 
is due to Czech~\etal~\cite{czech1992optimal,majewski1996family}, and has been
the source of most development in the area. The idea is that given the function 
$f$, one fixes an integer $k>0$ and chooses $k$ hash
functions $h: U\times[k] \to [m]$, $m\geq n$, and uses them to write a system of $n$ equations in $k$ variables $\{w_i\}$
\begin{equation}
  \label{eq:retrieval}
  w_{h(x, 0)} + w_{h(x, 1)} +  \cdots + w_{h(x, k-1)} = f(x) \qquad\text{for all $x \in S$}.
\end{equation}
Note that `+` denotes the sum operator in an Abelian group. In the original
proposal, it was modular arithmetic, but
is now more commonly a
bitwise XOR operation between bit vectors.
Clearly, storing a solution $w_{0}, \ldots, w_{m-1}$ for this system is sufficient to compute $f$ on all
keys in $S$, but elements in $U\setminus S$ will be mapped to arbitrary values.
The criterion originally used to make the system solvable was to make the ratio
$n/m$ small enough that the system can be almost always put in echelon form in
linear time, instead of the typical cubic
time of Gaussian elimination. For example, for $k=2$ and any $\varepsilon > 0$ one
can obtain a structure with a space of $\approx (2+\varepsilon)rn$,
i.e., more than twice the optimum, whereas
for $k=3$ the overhead decreases to $\approx 23$\%. These overheads come
from \emph{thresholds} for the \emph{peelability} problem discussed
in \cref{s:peelability}.
Arbitrarily lower overheads can be obtained by
increasing $k$ and solving the linear system~\cite{genuzio2016fast} --- for example, for $k=3$ ($10\%$ overhead) or $k=4$ ($3\%$ overhead), as the threshold for solvability is
lower~\cite{pittel2015satisfiability}, and tends to 1 as $k$ grows,
as we discuss in the \cref{s:peelability}.
Recent retrieval data structures achieve very low space consumption: for example, Bumped Ribbon Retrieval (BuRR) \cite{dillinger2022burr} is also based on solving a system of linear equations
where a key's equation uses a random subset of a block of $w$ consecutive variables, e.g. with $w = 64$.
BuRR can achieve a space consumption of less than $1.01rn$ and supports queries in time $O(r)$.
We do not go into detail about its working principle here but still want to mention it because several approaches in \cref{s:multipleChoiceHashing} use BuRR as a black box retrieval data structure.
As noted by Dietzfelbinger and Pagh \cite{dietzfelbinger2008succinct}, retrieval data structures can
also be used to implement AMQ data structures by storing a fingerprint of each key in the set (as discussed in \cref{s:applications}).

\subsection{Graph Properties}\label{s:graphProperties}

Several perfect hash function constructions that we describe later are based on \emph{peeling} and \emph{orienting} graphs.
In this section, we give an introduction to these terms.

\myparagraph{Peelability\@.}\label{s:peelability}
Peeling is the process of iteratively taking away nodes of degree
$1$ from a graph, together with their adjacent edge
\cite{czech1992optimal,majewski1996family,molloy2005cores,janson2007simple,botelho2013practical,walzer2021peeling,walzer2025peelingJournal}.
We call a graph \emph{peelable} if it can be peeled to an empty graph. Clearly
peelability in ordinary graphs is equivalent to acyclicity.
However, once we move to $k$-uniform
hypergraphs, $k\geq 3$, where an edge is a subset of vertices of cardinality
$k$, the two concepts are no longer equivalent.\footnote{Note that early papers sometimes
used the term “acyclic” for “peelable”. In fact, there are several definitions
of acyclicity in a hypergraph~\cite{brault2016acyclicity}.} Peelability is important for retrieval data structures
because we can associate a $k$-uniform hypergraph with the
system in Eq.~(\ref{eq:retrieval}): vertices are elements of $[m]$, and each equation
is a hyperedge containing the vertices $h(x, i)$, $0 \leq i < k$.
It is easy to see that
peelability implies that the system is in echelon form up to row swaps and column swaps, as each peeled edge is an
equation containing a variable not appearing elsewhere. Peeling can be performed
in linear time.
As each edge is peeled due to a degree 1 vertex, the associated equation and variable index are put on a stack.
After peeling, the equations and variables are popped from the stack and processed in reverse order.
Each equation is solved by setting the variable to the required value, exploiting that the variable does not appear in previously solved equations.
At the end of the process,
equations are popped from the stack and solved backwards.

Czech~\etal~\cite{czech1992optimal,majewski1996family} showed 
that a graph with $m$ vertices, $n$ uniformly random edges, and $n/m \leq 1/(2+\varepsilon)$, $\varepsilon > 0$ is peelable with a probability $>0$.
In practice, they suggest to use $n/m
= 0.467$. For $k$-uniform hypergraphs, $k\geq3$, there is a
sharp \emph{peelability threshold} $\mathrm{peel}_k$: a hypergraph with $m$ vertices and $n$ uniformly random hyperedges of size $k$ is peelable with probability $1-o(1)$ if $n/m < \mathrm{peel}_k-\varepsilon$, and peelable with probability $o(1)$ if $n/m > \mathrm{peel}_k+\varepsilon$. For $k=3$ the
threshold is $\mathrm{peel}_k \approx 0.8185$, and for $k>3$ the threshold decreases, so $k=3$ is the most interesting case for practical applications,
providing a $23$\% space overhead for retrieval. Molloy~\cite{molloy2005cores} proved ten
years later that the thresholds conjectured in~\cite{majewski1996family} were
quite accurate.
We remark that a slightly different distribution for the hyperedges yields
larger peelability thresholds \cite{walzer2021peeling,walzer2025peelingJournal},
e.g., $\approx 0.9179$ for $k = 3$, equal to the orientability thresholds we
discuss next, which moreover improve for larger $k$. This has, to our
knowledge, not yet been exploited in MPHF constructions.

\myparagraph{Orientability\@.}
Orientability is a weaker concept than peelability: we call a hypergraph
\emph{orientable} if we can \emph{orient} each edge, that is, select a vertex
from each edge so that no two edges share the same selected vertex. Clearly, a
peelable graph is orientable, as we can select for each edge the vertex from
which it was peeled.\footnote{In general, a hypergraph is $(k,\ell)$-orientable
if it possible to select $k$ vertices in each edge so that no vertex is selected
by more than $\ell$ edges. While this notion might be useful for $k$-perfect
hashing, we do not consider it further in this paper.} Orientability is a well
known property that is analyzed, among others, in the context of cuckoo
hashing~\cite{pagh2004cuckoo}. Similarly to the peelability threshold, the
\emph{orientability threshold} describes how large $m$ has to be with respect to
$n$ to ensure that the hypergraph can be almost always oriented as $n$ goes to
infinity. For $k\geq 3$, this threshold is (surprisingly) equivalent 
to the threshold for the
$k$-XORSAT problem~\cite{pittel2015satisfiability,dietzfelbinger2010tight}: when
considering a random system on $\mathbf F_2$ with $m$ variables, $n$ equations,
and $k$ variables per equation, the system is almost always solvable as $n$ goes
to infinity as long as $n/m$ is is below the $k$-XORSAT solvability threshold, which
is the same as the orientability threshold for $k$-uniform hypergraphs.
For example, $3$-uniform hypergraphs have threshold $\approx 0.9179$,
$4$-uniform hypergraphs have threshold $\approx 0.9768$, and when $k$ increases further the threshold
approaches $1$~\cite{pittel2015satisfiability}.
The threshold has also been extended to larger fields~\cite{goerdt2012satisfiability}.
There is no threshold for $2$-XORSAT, but ordinary graphs have an orientability
threshold of $1/2$, and below the same ratio they are peelable with
positive probability.

\section{Overview}\label{s:categorization}
We give an overview of the origins of perfect hashing in \ifAppendix \cref{appendix:s:birth}\else the extended version of this paper \cite{lehmann2025modern}\fi.
In this survey we focus on modern perfect hash function constructions.
For this, we divide the approaches into three categories, which we explain in the following.

\emph{Perfect hashing through retrieval} (\cref{s:retrievalBased}) builds on retrieval data structures (see \cref{s:retrieval}).
Early approaches explicitly store the desired hash value for each key, leading to a space consumption of $\tOh{\log n}$ bits per key.
Later approaches achieve constant space per key by storing the index of one of several candidate output values determined by different hash functions on the key.

Simple brute-force search for a PHF achieves the optimal space consumption but has exponential running time, making it impractical.
In \emph{perfect hashing through brute-force} (\cref{s:bruteforce}) we describe techniques that still use brute-force at their core, but are practical because they first reduce the size of the sets that need to be handled with brute-force.

\emph{Perfect hashing through fingerprinting} (\cref{s:fingerprinting}) describes another technique that is focused more on construction and query performance than on space consumption.
There are several implementations of the same basic technique, mostly with minor algorithmic changes.

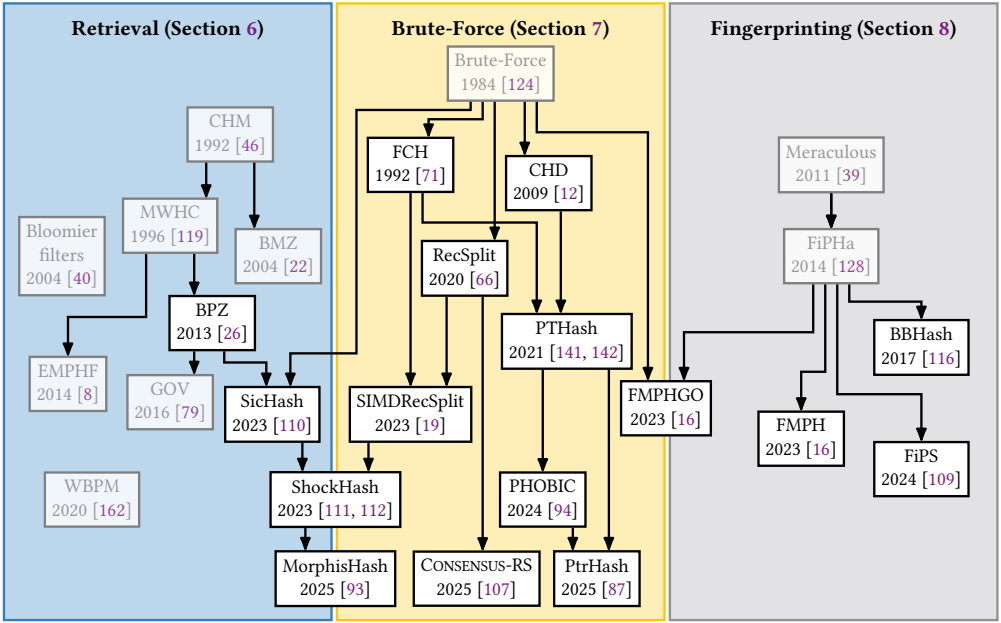
\begin{figure}[t]
  \centering
  \definecolor[named]{historyBlue}{HTML}{377EB8}
\definecolor[named]{historyLightBlue}{HTML}{BCD6EB}
\definecolor[named]{historyYellow}{HTML}{FCC712}
\definecolor[named]{historyLightYellow}{HTML}{FEEEB8}
\definecolor[named]{historyGray}{HTML}{919192}
\definecolor[named]{historyLightGray}{HTML}{E2E2E4}
\scalebox{0.8}{
  \begin{tikzpicture}[every node/.style={font={\small}}]
    \tikzset{algo/.style={draw,fill=white,align=center,line width=1.1pt}}
    \tikzset{algI/.style={algo,draw=gray,text=gray,fill opacity=0.8}}
    \tikzset{connection/.style={{Latex[round]-},line width=1.1pt}}
    \tikzset{largebox/.style={rectangle,minimum width=5.45cm,text depth=9.9cm,text height=4mm,anchor=south,line width=1.1pt,font=\bfseries\normalsize,align=center,text=black}}

    \node[style=largebox,fill=historyLightBlue,draw=historyBlue]     (e) at (-0.54,-1.4) {Retrieval (\cref{s:retrievalBased})};
    \node[style=largebox,fill=historyLightYellow,draw=historyYellow] (e) at (5.0,  -1.4) {Brute-Force (\cref{s:bruteforce})};
    \node[style=largebox,fill=historyLightGray,draw=historyGray]     (e) at (10.54,-1.4) {Fingerprinting (\cref{s:fingerprinting})};

    \node[style=algI] at (0.5,7) (chm) {CHM\\1992 \cite{czech1992optimal}};
    \node[style=algI] at (-0.5,5.5) (mwhc) {MWHC\\1996 \cite{majewski1996family}};
    \node[style=algI] at (-2.3,5) (bloomier) {Bloomier\\filters\\2004 \cite{chazelle2004bloomier}};
    \node[style=algI] at (1.3,5) (bmz) {BMZ\\2004 \cite{botelho2004new}};
    \node[style=algo] at (0.2,3.9) (bdz) {BPZ\\2013 \cite{botelho2013practical}};
    \node[style=algI] at (-0.5,2.6) (gov) {GOV\\2016 \cite{genuzio2016fast}};
    \node[style=algo] at (1.2,2.4) (sichash) {SicHash\\2023 \cite{lehmann2023sichash}};
    \node[style=algI] at (-1.8,1) (wbpm) {WBPM\\2020 \cite{weaver2020constructing}};
    \node[style=algI] at (-2.2,2.9) (emphf) {EMPHF\\2014 \cite{belazzougui2014cache}};

    \node[style=algo] at (2.25,1.0) (shockhash) {ShockHash\\2023 \cite{lehmann2023shockhash,lehmann2023bipartite}};
    \node[style=algo] at (2.25,-0.3) (morphishash) {MorphisHash\\2025 \cite{hermann2025morphishash}};

    \node[style=algI] at (5,8.0) (bruteforce) {Brute-Force\\1984 \cite{melhorn1984data}};
    \node[style=algo] at (3.5,6.5) (fch) {FCH\\1992 \cite{fox1992faster}};
    \node[style=algo] at (5.8,6.2) (chd) {CHD\\2009 \cite{belazzougui2009hash}};
    \node[style=algo] at (6.1,3.6) (pthash) {PTHash\\2021 \cite{pibiri2021pthash,pibiri2021parallel}};
    \node[style=algo] at (4.4,4.8) (recsplit) {RecSplit\\2020 \cite{esposito2020recsplit}};
    \node[style=algo] at (3.5,2.4) (simdrecsplit) {SIMDRecSplit\\2023 \cite{bez2023high}};
    \node[style=algo] at (5.7,1.9) (phobic) {PHOBIC\\2024 \cite{hermann2024phobic}};
    \node[style=algo] at (6.6,0.6) (ptrhash) {PtrHash\\2025 \cite{grootkoerkamp2025ptrhash}};
    \node[style=algo] at (6.6,-0.7) (phast) {PHast\\2025 \cite{beling2025phast}};
    \node[style=algo] at (4.6,-0.3) (consensusrs) {\consensus-RS\\2025 \cite{lehmann2025consensus}};

    \node[style=algo] at (7.75,2.5) (fmphgo) {FMPHGO\\2023 \cite{beling2023fingerprinting}};

    \node[style=algI] at (10.5,6.5) (meraculous) {Meraculous\\2011 \cite{chapman2011meraculous}};
    \node[style=algI] at (10.5,5) (fipha) {FiPHa\\2014 \cite{muller2014retrieval}};
    \node[style=algo] at (12,3.5) (bbhash) {BBHash\\2017 \cite{limasset2017fast}};
    \node[style=algo] at (10,2) (fmph) {FMPH\\2023 \cite{beling2023fingerprinting}};
    \node[style=algo] at (12,1.5) (fips) {FiPS\\2024 \cite{lehmann2024fast}};

    \draw[style=connection] ($(shockhash.north) - (0.55,0)$) -- ($(sichash.south) + (0.5, 0)$);
    \draw[style=connection] ($(shockhash.north) + (0.55,0)$) -- ($(simdrecsplit.south) - (0.7, 0)$);
    \draw[style=connection] ($(simdrecsplit.north) + (0.6,0)$) -- ($(recsplit.south) - (0.3, 0)$);
    \draw[style=connection] ($(simdrecsplit.north)$) -- ($(fch.south)$);
    \draw[style=connection] ($(pthash.north) - (0.5,0)$) |- ($(pthash.north)!0.5!(fch.south) + (0,0.5)$) -| ($(fch.south) + (0.2,0)$);
    \draw[style=connection] ($(pthash.north) - (0.1,0)$) -- ($(chd.south) + (0.2,0)$);
    \draw[style=connection] (phobic.north) -- ($(pthash.south) - (0.4, 0)$);
    \draw[style=connection] (fipha.north) -- (meraculous.south);
    \draw[style=connection] ($(ptrhash.north) - (0.4, 0)$) -- ($(phobic.south) + (0.5,0)$);
    \draw[style=connection] ($(ptrhash.north) + (0.2, 0)$) -- ($(pthash.south) + (0.7, 0)$);
    \draw[style=connection] ($(consensusrs.north) + (0.1,0)$) -- ($(recsplit.south) + (0.3,0)$);
    \draw[style=connection] ($(bbhash.north)$) -- ($(bbhash.north)!0.5!(fipha.south) + (0.75,0)$) -| ($(fipha.south) + (0.3,0)$);
    \draw[style=connection] (fipha.north) -- (meraculous.south);
    \draw[style=connection] (fmph.north) -- ($(fmph.north)!0.5!(fipha.south) - (0.25,0.4)$) -| ($(fipha.south) - (0.1,0)$);
    \draw[style=connection] (fips.north) -- ($(bbhash.south) - (0,0.3)$) -| ($(fipha.south) + (0.1,0)$);
    \draw[style=connection] ($(sichash.north) - (0.1,0)$) |- ($(sichash.north)!0.5!(bdz.south) + (0,0.1)$) -| ($(bdz.south) + (0.2,0)$);
    \draw[style=connection] ($(gov.north) + (0.4,0)$) -- ($(bdz.south)-(0.3,0)$);
    \draw[style=connection] ($(mwhc.north) + (0.6,0)$) -- ($(chm.south) - (0.4,0)$);
    \draw[style=connection] ($(bmz.north) - (0.4,0)$) -- ($(chm.south) + (0.4,0)$);
    \draw[style=connection] ($(bdz.north) - (0.3,0)$) -- ($(mwhc.south) + (0.4,0)$);
    \draw[style=connection] ($(fmphgo.north) + (0.3,0)$) |- ($(fmphgo.north)!0.5!(fipha.south) - (0.5,0)$) -| ($(fipha.south) - (0.3,0)$);
    \draw[style=connection] ($(fmphgo.north) - (0.3,0)$) |- ($(bruteforce.south) + (0.8,-0.5)$) -| ($(bruteforce.south) + (0.6,0)$);
    \draw[style=connection] ($(fch.north) + (0.3,0)$) |- ($(fch.north)!0.5!(bruteforce.south)$) -| ($(bruteforce.south) - (0.3,0)$);
    \draw[style=connection] ($(chd.north) - (0.4,0)$) -- ($(bruteforce.south) + (0.4,0)$);
    \draw[style=connection] ($(sichash.north) + (0.3,0)$) |- ($(sichash.north)!0.5!(bruteforce.south)-(0.5,1.8)$) -- ($(sichash.north)!0.5!(bruteforce.south) - (0.5,-2.2)$) -| ($(bruteforce.south) - (0.5,0)$);
    \draw[style=connection] ($(recsplit.north) + (0.5,0)$) -- ($(bruteforce.south) - (0.1,0)$);
    \draw[style=connection] ($(emphf.north)$) |- ($(emphf.north)!0.5!(mwhc.south) - (0.5,0.2)$) -| ($(mwhc.south) - (0.4,0)$);
    \draw[style=connection] ($(morphishash.north) - (0.5,0)$) -- ($(shockhash.south) - (0.5,0)$);
    \draw[style=connection] ($(phast.north) - (0,0)$) -- ($(ptrhash.south) - (0,0)$);
  \end{tikzpicture}
  }
  \caption{%
      Perfect hashing approaches and how they influence each other.
      While we describe them in this paper, we do not evaluate the performance of approaches given in gray color because they are clearly larger than current ones or because they do not have a publicly available implementation.
  }
  \label{fig:history}
\end{figure}

In addition to the techniques themselves, we give a comparison of the asymptotic performance in \cref{s:theory}.
\Cref{fig:history} gives an overview of all approaches that we consider in this survey, and shows how the ideas influence each other.
We can see that the three main categories each have their own strands of research.
Recent approaches start using ideas from multiple categories, as we show in the figure by placing them in between the categories.
Interestingly, the number of publications in the last few years increased massively, after several years of mostly inactivity in the early 2000s.
With so much new development in the field, this survey gives structure and an overview over the state of the art.

\section{Perfect Hashing through Retrieval}\label{s:retrievalBased}
We recall from \cref{s:retrieval} that a retrieval data structure represents a function $f: S \rightarrow \{0, 1\}^r$ that maps each key in $S \subseteq U$ to an $r$-bit integer.
In the following, we explain a range of perfect hash function constructions that use retrieval data structures (see \cref{s:retrieval}).
In \cref{s:retrievalPhfs}, we explain perfect hash functions that store the final output hash value in the retrieval data structure.
This needs a space consumption of $\tOh{\log n}$ bits per key.
Then, in \cref{s:multipleChoiceHashing}, we explain approaches where the retrieval data structures only store the index of one of multiple choices, enabling constant space per key.
Some of the approaches we present here do not have an explicit name in their original paper.
We therefore follow the convention of previous papers and name the approaches after the first letters of the authors' last names.

\begin{table}[t]
  \caption{%
      Overview of perfect hashing through retrieval.
      We use the hypergraph interpretation here. %
      The number $k$ of nodes per edge corresponds to the number of choices for each
      key.
      ${}^*$ The original paper \cite{chazelle2004bloomier} discusses only PHFs, not MPHFs, and with weaker bounds.
      We report here the actual bounds for the hypergraph technique~\cite{majewski1996family}.
      ${}^{**}$ The retrieval data structure used cannot represent all orientable graphs, so some have to be skipped.
  }
  \label{tab:retrievalMethods}
  \centering
  \scalebox{0.9}{
  \begin{tabular}[t]{lllllll}
    \toprule
    Stored & Approach                       & Year & $k$            & Graph Property & $|V|$                 & Space Usage \\ \midrule
    \rot{3}{\parbox{1.2cm}{\centering Output\\value}}
    & CHM \cite{czech1992optimal}           & 1992 & 2              & Acyclic        & $2.09n$               & $2.09\cdot n\log_2 n$ \\
    & MWHC \cite{majewski1996family}        & 1996 & 3              & Peelable       & $1.23n$               & $1.23\cdot n\log_2 n$ \\
    & BMZ \cite{botelho2004new}             & 2004 & 2              & |Critical| $< n/2$ & $1.15n$  & $1.15\cdot n\log_2 n$ \\ \midrule
    \rot{7}{\parbox{2cm}{\centering Hash function\\index}}
    & Bloomier filters \cite{chazelle2004bloomier} & 2004 & 3       & Peelable
    & $1.23n$               & $2.46n^*$ \\
    & BPZ \cite{botelho2007external}        & 2007 & 3              & Peelable       & $1.23n$               & $2.62n$ \\
    & GOV \cite{genuzio2016fast}            & 2016 & 3              & Orientable     & $1.10n$               & $2.24n$ \\
    & WBPM \cite{weaver2020constructing}    & 2020 & $\log_2 n$     & Bipartite      & $n$                   & $1.83n$ \\
    & SicHash \cite{lehmann2023sichash}     & 2023 & Mix 2, 4, 8    & Orientable     & $(1+\varepsilon)n$    & $2n$ \\
    & ShockHash \cite{lehmann2023shockhash} & 2023 & 2              & Orientable     & $n$                   & $1.443n$ \\
    & MorphisHash \cite{hermann2025morphishash} & 2025 & 2    & Orientable${}^{**}$  & $n$                   & $1.443n$ \\
    \bottomrule
  \end{tabular}
  }
\end{table}

\subsection{Storing Perfect Hash Values}\label{s:retrievalPhfs}
Retrieval data structures can be used to explicitly store a desired perfect hash value for each key using about $\log_2 n$ bits per key.
Several early MPHF constructions actually implement a retrieval data structure and store the desired hash values \cite{czech1992optimal,majewski1996family}, even if the authors do not explicitly explain their approach as a retrieval data structure.
In this section, we nevertheless look at those to illustrate the progress on perfect hashing made over the past years.
We give an overview in \cref{tab:retrievalMethods}.
Remember from \cref{s:retrieval} that retrieval data structures can be constructed from the following system of equations:
$w_{h(x, 0)} + w_{h(x, 1)} +  \cdots + w_{h(x, k-1)} = f(x)$ for all $x \in S$.
All the perfect hash function constructions in this section solve a system of that form.
Remember also from \cref{s:graphProperties} the concepts of \emph{peelability} (repeatedly taking away nodes of degree 1) and \emph{orientability} (directing edges such that each node has indegree 1) of graphs.

\myparagraph{CHM\@.}
Given a list of keys, Czech~\etal~\cite{czech1992optimal} use peelability on a
random graph to solve the system in Eq.~(\ref{eq:retrieval}), where $f$ maps each key to its
rank in the list. If the graph is not peelable, which rarely happens below
the peelability threshold, CHM retries with another set of hash functions. The
construction is a general retrieval data structure, but the authors only use it
to construct an MPHF. The space consumption is $2.09n\log_2 n$
bits. As we noted in \cref{s:graphProperties}, this paper introduces the idea of peelability, that
is used in many later constructions.

\myparagraph{MWHC\@.}\label{s:mwhc}
The authors of CHM then joined forces with Wormald~\cite{majewski1996family}
to extend the idea of CHM to 3-uniform hypergraphs. Once again,
the construction is a general retrieval data structure, but the authors use it
to construct an MPHF, lowering the space consumption to $1.23n\log_2 n$ bits.
Belazzougui~\etal~\cite{belazzougui2014cache} improve the peeling step of MWHC
through better cache locality.

\myparagraph{BMZ\@.}
Botelho \etal \cite{botelho2004new} enhance the idea of CHM into another direction than MWHC.
In contrast, the graph does not need to be peelable --- only orientable.
Like CHM, the hash value is given by the sum of the values stored at the two ends of each edge.
When assigning values to nodes, BMZ first uses the normal peeling process that is known from the previous approaches.
When there are no more nodes of degree~$1$, we have arrived at the 2-core.
If that 2-core consists of only nodes with degree $2$, the remaining edges form loops.
BMZ calls them \emph{critical} and assigns them first using breadth-first-search.
This ensures that different keys get different values, but BMZ no longer supports an arbitrary value for each key.
As such, in contrast to CHM and MWHC, it is no longer a general-purpose retrieval data structure.
Instead, it is a step towards smaller perfect hash functions.
After the critical edges, BMZ then assigns the remaining nodes similar to the algorithms above.
At all times, it keeps a list of output values that are not assigned yet to find collision-free mapping.
The fact that BMZ does not require peelability has the advantage that $|V|$ can be much closer to $n$.
This reduces the amount of space needed to only $1.15n\log_2 n$ bits.

\subsection{Multiple Choice Hashing}\label{s:multipleChoiceHashing}
\begin{figure}[t]
    \begin{minipage}[c]{0.67\textwidth}
        \includegraphics[scale=0.9]{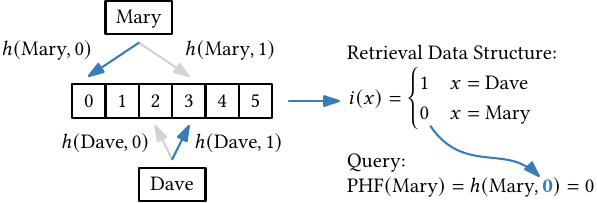}
    \end{minipage}\hfill
    \begin{minipage}[c]{0.3\textwidth}
        \caption{%
          Multiple Choice Hashing.
          Each key has the choice between several locations, in this case two.
          A retrieval data structure stores which of the choices to take.
        }
        \label{fig:multipleChoiceHashing}
    \end{minipage}
\end{figure}

In the following, we present several ideas that we categorize as \emph{multiple choice hashing}.
In multiple choice hashing, each input key has a set of candidate output values.
Here, these candidate values are determined by evaluating different hash functions $h(\cdot, i)$ on the input key.
Let $i(x)$ be a function assigning a number to each key, indicating which of its candidate values should be selected in order to give a collision-free mapping.
Then $h(x, i(x))$ is unique for each key $x$, and only the hash function index $i(x)$ needs to be stored in a retrieval data structure.
In contrast to \cref{s:retrievalPhfs}, the approaches can therefore break the space consumption of $\Omega(\log n)$ bits per key.
We illustrate this idea in \cref{fig:multipleChoiceHashing}.
Multiple choice hashing can also be interpreted as a hypergraph, where each key corresponds to a hyperedge, connecting its candidate output values.
Then an orientation of this graph gives a mapping from keys to candidate values.
In this interpretation, the similarities to the older approaches in the previous section become clear.
In the following, we explain approaches from the literature in more detail.
\Cref{tab:retrievalMethods} gives an overview of the differences between the approaches.
We want to mention the similarity of multiple choice hashing to \emph{cuckoo hashing} \cite{pagh2004cuckoo}, which is a collision resolution strategy in hash tables.
Cuckoo hashing uses the \emph{power of multiple choices} \cite{mitzenmacher2001power} to reduce the load of hash table cells.

\myparagraph{Bloomier filters\@.} Chazelle~\etal~\cite{chazelle2004bloomier}
introduce the concept of \emph{Bloomier filters}, which are thought of as a
generalization of Bloom filter. The setting is slightly different, as they
assume that each key is representable with $r$ bits, and they create a
retrieval data structure that maps each key to itself. The purpose of the
mapping is to create an approximate set membership data structure, but
as in the previous case the construction can be used for general
retrieval. In fact, they
rediscover the idea of $k$-uniform hypergraph peeling introduced by 
Majewski~\etal~\cite{majewski1996family}, but they use a completely
different formalization, thus missing the connection, and
not citing the previous results.
Moreover, the peelability thresholds they provide are much weaker than the ones
conjectured in~\cite{majewski1996family}. However, they make an important
observation that provides the first example of multiple-choice hashing:
since each edge (i.e., key) is associated uniquely with the vertex it was peeled from, we can associate a unique vertex with
each key by arbitrarily numbering the vertices in each edge, and storing
the function $i$ mapping each edge $x$ to the index of the vertex $i(x)$ it was peeled
from. This mapping just requires $\log k$ bits and is a perfect hash 
function.

\myparagraph{BPZ\@.}
Botelho~\etal~\cite{botelho2007simple} rediscover the perfect hash functions
described in the Bloomier filter paper~\cite{chazelle2004bloomier}, but they do not
cite it, likely because the different formalization.
However, since their goal is minimal perfect hashing, they consider the problem
of how to convert the Bloomier filter~\cite{chazelle2004bloomier} idea to an MPHF.
As we mentioned in the introduction, a standard technique involves a ranked bit
vector: in this case, the vector would track the vertices that have been used to
peel an edge, and thus are output by the perfect hash function for some input
key. The authors note however that in the optimal case of $3$-uniform
hypergraphs one uses two bits to store three values. They suggest to store $0$
as such for vertices that are not output by the perfect hash function, and to
store $1$, $2$, or $3$ for those which are. Then, after invoking the perfect hash function, one
has just to count how many nonzero pairs of bits appear before the output, which
can be done in constant time and $o(n)$ space adapting standard ranking
techniques~\cite{jacobson1989space}. In this way, they obtain a minimal perfect
hash function using just $2.62n$ bits \cite{botelho2013practical}.\footnote{Recall that the perfect hash function
can be stored in $2.46n$ bits but additional space is required for ranking.}

\myparagraph{GOV\@.}
Genuzio~\etal~\cite{genuzio2016fast}, reduce the space overhead of
retrieval data structures and MPHFs.
They exploit the fact that, as we already mentioned, the following thresholds are the same \cite{dietzfelbinger2011cuckoo,goerdt2012satisfiability}:
(1) the threshold for orientability of
$k$-uniform hypergraphs, (2) for solvability of random systems on $\mathbf F_2$ with
$k$ variables per equations, and (3) even for the same kind of random systems on
larger fields.
First they orient
a random $3$-uniform hypergraph, thus assigning a distinct vertex to each edge.
Then they solve the system of equations on $\mathbf F_3$ representing the function that
assigns to each edge the vertex selected by the orientation. To tame the cubic
time of Gaussian elimination, GOV introduces a range of engineering tricks:
first of all, keys are hashed into small partitions (see \cref{s:commonTools}).
The systems
associated with each partition are first peeled as much as possible to reduce the number
of equations. Then operations on equations use~\emph{broadword
programming}~\cite{knuth2011art4A} by packing multiple values into a single 64-bit word
and running calculations on all of them at once. Finally, the paper introduces
\emph{lazy Gaussian elimination}, which is a heuristic for fast solution of
random linear systems. Using three hash functions, and therefore two bits per
key, and exploiting the same trick of Botelho~\etal~\cite{botelho2007simple} to perform
ranking on the output, they reduce the space consumption of minimal perfect
hash functions to $2.24n$ bits.\footnote{As in the case of BPZ, the additional
$0.04n$ bits are used for the ranking structure.}

\myparagraph{WBPM\@.}
Weaver and Heule \cite{weaver2020constructing} use a larger number of $\Oh{\log n}$ candidate output values.
When selecting which candidate to use for each key, they prefer choices with smaller indices.
This is modeled as a weighted graph connecting each key to its candidates, where the edge of hash function $i$ has weight $i$.
WBPM then determines a minimum weight bipartite perfect matching (WBPM), which can be shown to reach a weight of $1.83n$.
The selected edge indicates which hash function to use for each key.
WBPM stores the hash function indices bit-by-bit in unary coding using a 1-bit retrieval data structure.
The matching weight of $1.83n$ therefore also equals the space consumption of the final data structure, except for overheads like prefix sums due to partitioning.

\myparagraph{SicHash\@.}
Most earlier approaches couple the task of retrieval and perfect hashing, which means that the approaches have to store the same number of bits for every input key.
SicHash \cite{lehmann2023sichash}, in contrast, separates the two tasks of key placement and retrieval.
This makes it possible to efficiently use more than one retrieval data structure and store values of different widths for different keys.
Using the same width for every key, even with optimal retrieval, we can achieve no better than about 2.2 bits per key using 4 choices and remapping values to be minimal perfect.
SicHash's main innovation is using a hash function to determine how many choices each key has.
Some of the keys have 2 choices, some have 4 choices, and some keys have 8 choices.
As such, it stores a 1, 2, and 3-bit retrieval data structure, each handling a portion of the input keys.
Using the space-efficient BuRR \cite{dillinger2022burr} retrieval data structure, this makes it possible to reduce the space consumption compared to previous approaches.
SicHash achieves a favorable space-performance trade-off when being allowed 2--3 bits of space per key.
It also achieves a rather limited gain in space efficiency by \emph{overloading} the corresponding graph beyond the orientability threshold.
Using small partitions, SicHash exploits the variance in the number of keys that can fit.
Also, it uses the fact that the load factor at which construction likely fails converges to the load threshold from \emph{above} as $n$ grows.

\myparagraph{ShockHash\@.}\label{s:shockhash}
ShockHash \cite{lehmann2023shockhash} can be seen as an extreme version of SicHash that directly tries to orient a graph with $n$ nodes and $n$ edges.
Each key corresponds to one edge, connecting the two candidate output values of the key.
The graph can be oriented if and only if it is a \emph{pseudoforest} --- a graph where each component contains as many edges as nodes.
Because ShockHash works far above the orientability threshold, it needs to retry with many different graphs before it samples one that is orientable.
ShockHash then stores the choice between the two candidate values of each key in a 1-bit retrieval data structure taking close to $n$ bits.
Additionally, it stores the hash function seed, which can be shown to need about $0.44n$ bits.
This means that the majority of the data originates from a simple linear time orientation of the graph.
Only $0.44n$ bits need to be determined by exponential time brute-force.
Compared to a simple brute-force technique (see \cref{s:bruteforce}), which needs $e^n$ tries in expectation, ShockHash needs only $(e/2)^n$ tries, while still reaching the space lower bound.
We refer to \cref{s:theory} for more details on asymptotic performance.
A key idea for making ShockHash practical is the introduction of a simple bit-parallel filter that checks whether all output values are hit by some key.
If one output is not hit, the graph cannot be orientable.
Because the construction is still exponential time, ShockHash first partitions the keys to smaller subsets using RecSplit (see \cref{s:recursiveSplitting}).
This makes the overall construction time linear in expectation.
ShockHash has configurations that achieve 1.52 bits per key using a construction time similar to RecSplit with 1.6 bits per key.

\myparagraph{Bipartite ShockHash\@.}
\emph{Bipartite} ShockHash \cite{lehmann2023bipartite} enhances the ShockHash idea.
The approach stores two independent hash function seeds, one for each end of the edges.
During construction, it builds a pool of hash function seeds and tests all combinations of seed pairs.
By hashing each end of an edge to disjoint output ranges, the hash function pool can be filtered before building the pairs, which enables additional exponential speedups.
Bipartite ShockHash-RS achieves a space consumption of $1.489$ bits per key.
In addition, bipartite ShockHash introduces a variant called ShockHash-Flat.
The approach uses a $k$-perfect hash function to partition the keys to smaller subsets instead of RecSplit.
This leads to higher space consumption but much faster queries due to not having to traverse the splitting tree (see \cref{s:recursiveSplitting}).

\myparagraph{MorphisHash\@.}
ShockHash orients pseudoforests, so each component of the graph is a cycle with trees branching from it.
The cycle can be oriented in two ways, which leads to about 1 bit of redundancy for every component.
MorphisHash \cite{hermann2025morphishash} constructs a special retrieval data structure that avoids this redundancy.
The idea is to formulate orienting the graph as a system of linear equations indicating that each node needs an indegree of 1 (mod 2).
The solution to the equation system is then a retrieval data structure that can return the orientation of each edge.
By reducing the number of columns of the equation system, MorphisHash gives a trade-off between space consumption and success probability.
Because it stores many small retrieval data structures next to the hash function seeds, MorphisHash has better cache locality during queries than ShockHash.

\section{Perfect Hashing through Brute-Force}\label{s:bruteforce}
The fact that brute-force can be useful for constructing perfect hash functions is mentioned early after the discovery of perfect hashing \cite{cichelli1980minimal}.
Today, most of the approaches achieving the best space efficiency are based on brute-force techniques.
In the following, we first describe a very simple brute-force construction and an approach that uses SAT solving for more structured brute-force search.
We then illustrate the practical approaches:
perfect hashing through bucket placement in \cref{s:bucketPlacement}, and recursive splitting in \cref{s:recursiveSplitting}.

\myparagraph{A Simple Brute-Force Construction\@.}\label{s:simpleBruteForce}
The idea of the simple brute-force construction is to try random hash functions (identified by different seeds) until one happens to be minimal perfect \cite{melhorn1984data}.
The data structure simply needs to store the found seed in binary coding.
We can expect to try $n^n/n! \approx e^n$ different hash function seeds because there are $n^n$ functions, from which $n!$ are minimal perfect \cite{melhorn1984data,esposito2020recsplit}.
This means that it reaches the space lower bound of minimal perfect hashing, but because of its exponential running time, it is not practical for large $n$.
Still, brute-force is used as a building block on smaller sets in a range of practical constructions, which we explain below.

\myparagraph{SAT Solving\@.}
Weaver and Heule \cite{weaver2020constructing} describes a solution based on SAT solving.
The data structure consists of a simple sequence of bits, and a hash function on each key determines which of these bits (possibly flipped) should be concatenated to give the final hash function value.
To determine the values of the bits in the sequence, the algorithm searches for an assignment using SAT solving.
The SAT encoding is based on an \emph{all-different} constraint and is feasible for $n \leq 40$.
The SAT solver can detect correlations between the bits in the data structure to prune the search space more efficiently than brute-force.
Still, due to its large search space of all possible functions, we categorize it as brute-force.
While the construction using SAT encoding achieves space consumption very close to the lower bound, it is much slower in practice than other approaches.

\subsection{Bucket Placement}\label{s:bucketPlacement}

Perfect hashing through bucket placement is achieved via a three-step algorithm:
\emph{mapping}, \emph{ordering}, and \emph{searching} (MOS) \cite{fox1992faster,fox1991order,fox1992practical,fox1989nlogn,belazzougui2009hash,pagh1999hashdisplace,cichelli1980minimal,sager1985polynomial}.
The general idea is as follows.
(1) \emph{Mapping.}
Each key $x$ is mapped to a small bucket $B_i = \{ x | b(x)=i \}$,
using a \emph{bucket assignment function} $b : U \to [t]$ and $t$ buckets.
(2) \emph{Ordering.}
The buckets $\{ B_i \}_{i=0}^{t-1}$ are then sorted; usually by falling size.
(3) \emph{Searching.}
For each bucket $B_i$ in the order, we determine a value $v_i$, using brute-force,
that the final function $f$ uses to place all the keys in the bucket to the output domain $[n]$
\emph{without collisions} with keys of previously processed buckets and with one another.
Once the value $v_i$ has been determined, we say that the bucket $B_i$ has been ``placed''.
(4) \emph{Encoding.}
The collection of values $\{v_i\}_{i=0}^{t-1}$ is then stored in an array $V[i]=v_i$
and compressed with a mechanism that retains the ability to quickly retrieve
$V[i]$ for any random $i \in [t]$.
The array $V$ is what the MPHF data structure actually stores.

\begin{figure}[t]
    \begin{minipage}[t]{.54\textwidth}
        \includegraphics[scale=0.9]{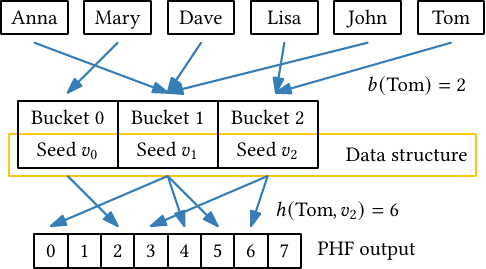}
        \caption{Illustration of perfect hashing through bucket placement,
            where the $n=6$ keys are first mapped to buckets and then, for each bucket $i$, a seed value
            $v_i$ is computed to place the keys in the bucket without collisions
            to the output domain $[m=8]$.
            Here, $h$ has range $[m]$.
        }
        \label{fig:bucketplacementIllustration}
    \end{minipage}
    \hfill
    \begin{minipage}[t]{.42\textwidth}
        \input{fig/bucketfunction.tex}
        \caption{
            Relative sizes of different buckets when using various bucket mapping functions.
            The legend gives the original paper even if later techniques use the same functions.
            Adapted from \cite{hermann2024phobic}.
        }
        \label{fig:bucketplacementFunction}
    \end{minipage}
\end{figure}

At the beginning of the search, almost all values in the output
domain $[m]$ are available, hence the first buckets are significantly easier to place.
Buckets with fewer keys are easier to place as well.
This is intuitively why, to accelerate the search, it is useful to sort the buckets
by falling size and process them in this order.
A query for a key $x$ then only needs to retrieve $V[i]$, where $i=b(x)$,
and use it to compute the resulting hash value.
\Cref{fig:bucketplacementIllustration} presents an illustration of bucket placement and \Cref{tab:bucketPlacementMethods} outlines the approaches discussed in this section.

\begin{table}[t]
  \caption{%
      Overview of perfect hashing approaches using bucket placement.
  }
  \label{tab:bucketPlacementMethods}
  \centering
  \scalebox{0.9}{
  \begin{tabular}[t]{llllll}
    \toprule
      Approach                                & Year & Bucket Assignment     & Seed Compression   & Search Type        & Partitions \\ \midrule
      FCH \cite{fox1992faster}                & 1992 & Skewed                & Fixed width        & Additive           & None \\
      CHD \cite{belazzougui2009hash}          & 2009 & Uniform               & SDC                & Double Additive    & None \\
      PTHash \cite{pibiri2021pthash}          & 2021 & Skewed                & Multiple options   & XOR                & Large \cite{pibiri2021parallel} \\
      PHOBIC \cite{hermann2024phobic}         & 2024 & Optimized             & Interleaved        & Additive+Retry     & Small \\
      PtrHash \cite{grootkoerkamp2025ptrhash} & 2025 & Approximate optimized & Byte array         & Cuckoo             & Medium \\
      PHast \cite{beling2025phast}            & 2025 & Uniform               & Fixed width        & Windowed           & None \\
    \bottomrule
  \end{tabular}}
\end{table}

\myparagraph{FCH\@.}
FCH \cite{fox1992faster} is the first perfect hash function construction that uses the bucket placement idea.
It uses $t = \lceil cn/(1+\log_2 n) \rceil$ buckets, where $c > \log_2 e$ is a
parameter that trades space usage for construction speed.
To further amplify the effect that the first buckets are easier to place,
FCH uses a bucket assignment function that produces a \emph{skewed} distribution
of bucket sizes: it hashes 60\% of the keys to 30\% of the buckets.
The information $v_i$ used for placing a bucket consists of two parts, $d_i$ and $s_i$.
For each bucket $B_i$,
a displacement value $d_i \in [n]$ is determined so that all the keys
$x$ in $B_i$ are placed and assigned hash values $(h(x,s_i)+d_i) \bmod n$.
In the following (and in \Cref{tab:bucketPlacementMethods}),
we refer to this form of hash function as \emph{additive displacement}.
The seed $s_i \in \{0,1\}$ gives more flexibility during the search
for bucket $B_i$:
if all of the displacements cause collisions for $s_i=0$, FCH repeats the search for $s_i=1$.
If both seeds do not work, the construction fails.

To accelerate the identification of $d_i$, an auxiliary data structure formed
by two integer arrays, taking $\Theta(n \log n)$ bits overall, is used to maintain
the output values available in $[n]$ during the search.
Given a random available output value $p \in [n]$, $d_i$ is computed by
``aligning'' an arbitrary key of $B_i$, say $x_i$, to $p$:
let $q = h(x_i,s_i)\bmod n$, then $d_i = (p-q) \bmod n$.
This random alignment is rather critical for the search
as it guarantees that all output values have the same probability to be occupied.
On the other hand, this auxiliary structure has to be updated as well during
the search, involving some extra cache misses.

Since each displacement $d_i$ can be coded in $\lceil\log_2 n \rceil$ bits,
all the pairs $\{(s_i,d_i)\}$ are stored in an array of $t(1+\lceil\log_2 n\rceil)$ bits.
It follows that FCH uses about $c$ bits/key.
By decreasing the number of buckets, it is possible to lower the space usage
at the cost of a larger construction time.
However, in practice it is unfeasible
to go below $2.5$ bits/key for large values of $n$.
On the other hand, FCH shines for its very fast query time because
the pair $(s_i,d_i)$ is coded in fixed-width.
That is, FCH just spends one memory access per query (plus some inexpensive
bit manipulations), which is optimal.

\myparagraph{CHD\@.}\label{s:chd}
CHD \cite{belazzougui2009hash} maps the keys uniformly to $t=\lceil n/\lambda \rceil$ buckets,
where $\lambda > 0$ is a chosen parameter representing the average bucket size.
For each bucket $B_i$, a displacement \emph{pair} $(d_{i,0},d_{i,1})$ is computed
to ensure that all keys $x$ within the bucket can be placed collision-free
to output values
$(h(x,0) + d_{i,0}h(x,1) + d_{i,1}) \bmod m$.
CHD chooses for the search an output range larger than $n$, hence the values
are in $[m]$, where $m=\lceil n/\alpha \rceil$ for a suitable load factor
$0 < \alpha < 1$.
The domain of the function is then restricted back to $[n]$ using a rank query
as we discussed in \cref{s:commonTools}.
Differently from FCH, CHD maintains a bitmap of $m$ bits to mark values in $[m]$
that are still available.

Instead of explicitly storing a displacement pair $(d_{i,0},d_{i,1})$ for each bucket,
CHD encodes the position of the pair in the predefined sequence
$(0,0), \dots, (0, m-1), \dots, (m-1, 0), \dots, (m-1, m-1)$.
Very importantly, CHD tries positions in increasing order starting from the smallest position, i.e., following the order $[0, 1, 2, \ldots]$, and not at random as FCH. This intuitively helps compression because smaller numbers are tried first.
The sequence of pairs' positions is then compressed within entropy bounds
using the Simple Dense Coding (SDC) mechanism by Fredriksson and Nikitin \cite{fredriksson2007simple},
while permitting to retrieve an index in constant time.
With $\alpha=0.99$ and $\lambda=6$, CHD achieves $2$ bits/key and it is much faster to build
compared to FCH (albeit significantly slower to query).

\myparagraph{PTHash\@.}\label{s:pthash}
Under proper configuration, PTHash \cite{pibiri2021pthash} combines the query time of FCH and
the space effectiveness of CHD, with fast construction time.
It uses the same imbalance trick of FCH by hashing 60\% of the keys to 30\% of $t=\lceil cn/\log_2 n \rceil$ buckets.
Like CHD, it stores the values $\{ v_i \}$ in compressed form but,
rather than coding them using SDC, PTHash generalizes the encoding
step to use a wide range of mechanisms (e.g., fixed-width, dictionary-based, Elias-Fano \cite{elias74efficient,fano71number}, and others).
Using an appropriate coding scheme, only a single memory access is required to find the hash value
for a key and the remaining operations use simple arithmetic.

Since the first processed buckets are much easier to place, the values $\{v_i\}$ for the first 30\% of the buckets have a much smaller entropy than those for the other buckets.
The \emph{front} of the array $V$, of length $\approx 0.3t$, can thus be compressed better than the remaining \emph{back} part.
This allows PTHash to tune two
different compressors for the front and the back parts respectively:
a light-weight compressors for the front and a more-succinct (but slower to query) compressor for the back part.

Like CHD, it also first generates a non-minimal function
with range $m = \lceil n/\alpha \rceil$ for some load factor $0 < \alpha < 1$,
and uses a bitmap of $m$ bits to mark taken output values.
It then scales the output range back to $[n]$ by remapping values
larger than $n$ using an explicit list of $m-n$ free output values
that can be effectively compressed using Elias-Fano.

To accelerate the search for the values $\{v_i\}$, PTHash uses the so-called
\emph{XOR displacement}:
the keys of a bucket are mapped to output values given by
$(h(x,s) \oplus h(v_i,s)) \bmod m$, where $(A \oplus B)$ denotes the bit-wise XOR between the
integers $A$ and $B$, and $s$ is a randomly chosen seed.
As also noted for CHD, the values $\{v_i\}$ are tried in linear order to improve compression, while still achieving random displacement
of keys via the XOR operator.
Compared to FCH, the table of random displacements is avoided altogether. Compared to CHD, instead,
long clusters of already occupied slots are skipped.
Also note that the seed $s$ does not change during construction, hence
all the keys can be hashed only once instead of for every seed (which is particularly important to save time
when hashing long, variable-length, keys).
On the other hand, this XOR-based approach, in combination with $\bmod$, only works for large output domains \cite{hermann2024phobic}.

PTHash-HEM \cite{pibiri2021parallel} is a parallel construction of PTHash that
first partitions the input and then builds each partition (roughly holding a few million keys) independently in parallel.

\myparagraph{PHOBIC\@.}\label{s:phobic}
PHOBIC \cite{hermann2024phobic} is based on PTHash.
It introduces an optimal bucket assignment function $b(\cdot)$ that,
instead of generating two expected bucket sizes as in FCH, gives a different expected size to \emph{every} bucket. This allows each bucket to have
the same probability to be placed successfully, which
minimizes the construction time (in asymptotic terms) and makes the expected value of each $v_i$ equal.
The optimal bucket assignment function practically accelerates the construction compared to PTHash for large values of $\lambda$.
\Cref{fig:bucketplacementFunction} plots the relative expected bucket sizes using different bucket assignment functions. As apparent from the plot,
PHOBIC makes the first (and easiest to place) buckets much larger than those generated by FCH.

PHOBIC uses partitioning as a key design feature.
In particular, it uses small partitions (of about 2000--3000 keys),
differently from PTHash-HEM.
Small partitions lead to better cache locality since
the bitmap used by the search is small and accesses are more localized in memory.
Small partitions also enable efficient GPU acceleration.
Since small partitions are created, it might be impossible to place keys without collisions using the XOR-based search.
PHOBIC therefore relies on an additive displacement function and computes
$v_i = s_i \cdot P + d_i$ where $s_i$ is a seed for the bucket $B_i$, $P$ is the size of the partition, and $d_i \in [P]$ is the displacement value.
The hash value of a key $x$, relative to the partition size $P$, is then
$(h(x,s_i) + d_i) \bmod P$. This is similar to the function computed by FCH, but
the seed $s_i$ is not necessarily restricted in $\{0,1\}$:
if no displacement $d_i \in [P]$ works for the bucket $B_i$, the seed $s_i$ is incremented (and all the keys of the bucket hashed again), and the search for a new $d_i$ continues.
The pair $(s_i,d_i)$ is computed from $v_i$ as follows:
$s_i = \lfloor v_i/P \rfloor$ because $d_i < P$, and clearly $d_i = v_i \bmod P$.
Another contribution of PHOBIC is \emph{interleaved coding} which exploits that the values of the $j$-th buckets of all partitions have approximately the same distribution. This is a natural generalization of the \emph{front-back} compression strategy of PTHash and gives PHOBIC a slight space advantage over PTHash without significantly compromising query time.

\myparagraph{PtrHash\@.}\label{s:ptrhash}
PtrHash \cite{grootkoerkamp2025ptrhash} builds on both PTHash and PHOBIC
with the intent of simplifying the design of the data structure to accelerate queries.
The two main simplifications are: (1) a value $v_i$ is coded using exactly 1 byte,
hence PtrHash stores a plain byte array at its core;
(2) the allocation of the same number of output slots for each partition.
The values stored in PtrHash are therefore less than $256$ by design.
If none of the values $\{0,\ldots,255\}$ is able to
place a bucket, PtrHash bumps out already placed buckets
in a similar way to cuckoo hashing \cite{pagh2004cuckoo}.
Like PHOBIC, PtrHash partitions the input into small partitions
and uses not only the same number of buckets per partition
but also the same number of output slots.
This has the net advantage that, during a query for key $x$, the size of the partition
to which $x$ belongs to does not have to be retrieved from an encoded sequence.
PtrHash further uses Taylor approximations of the optimal bucket assignment function
introduced in PHOBIC, which are more efficient to compute.
Lastly, it supports explicit prefetching of batches of keys to increase query throughput.

\myparagraph{PHast\@.}
Perfect hashing made fast (PHast) \cite{beling2025phast} simplifies several aspects of the bucket placement idea for fast queries.
It uses a uniform bucket assignment function, fixed width coding of seeds, and no partitioning.
This is possible due to a new search strategy for seeds.
Each bucket only has a small window of possible output values, which overlaps with the windows of other buckets.
Instead of taking the first collision-free seed, PHast evaluates all seeds of a bucket and takes the one giving the smallest hash values among the collision-free seeds.
The order in which to place the buckets is determined by a combination of both bucket index (lowest first) and size (largest first).
Therefore, taking the seed with the smallest hash values leaves more flexibility for later buckets.
The few buckets that cannot be placed with any of their available seeds are bumped to a more expensive fallback data structure.
PHast${}^+$ designates variants that use additive placement functions.
This accelerates queries and enables bit-parallel seed search,
actually making PHast as fast as, or faster than, non-brute-force techniques.

\subsection{Recursive Splitting}\label{s:recursiveSplitting}
The simple brute-force construction explained at the beginning of this section needs $e^n$ trials and therefore is not practical for large $n$.
The recursive splitting idea introduced in RecSplit \cite{esposito2020recsplit} enables space-efficiently reducing the input set into small sets of size $\ell$.
These are then small enough that the brute-force construction becomes feasible.
The following paragraphs introduce RecSplit, as well as improvements in later papers.
\Cref{tab:recursiveSplitting} gives an overview.

\begin{table}[t]
  \caption{%
      Overview of approaches using recursive splitting.
      The base case column gives the technique used after splitting the input set to small subsets of size $\ell$.
      We also include ShockHash \cite{lehmann2023shockhash,lehmann2023bipartite} again, even though it is actually explained in \cref{s:bucketPlacement}, because it uses recursive splitting to partition the keys.
  }
  \label{tab:recursiveSplitting}
  \centering
  \scalebox{0.9}{
  \begin{tabular}[t]{llllll}
    \toprule
    Approach                                      & Year &  $\ell$ & Base case        & Seed encoding & Fanout \\ \midrule
    RecSplit \cite{esposito2020recsplit}          & 2020 &   8--16 & Brute-force      & Golomb-Rice   & Depending on $\ell$ \\
    SIMDRecSplit \cite{bez2023high}               & 2023 &   8--18 & Rotation fitting & Golomb-Rice   & Depending on $\ell$ \\
    ShockHash-RS \cite{lehmann2023shockhash}      & 2023 &  40--60 & ShockHash        & Golomb-Rice   & Depending on $\ell$ \\
    Bip. ShockHash-RS \cite{lehmann2023bipartite} & 2023 & 90--120 & Bip. ShockHash   & Golomb-Rice   & Depending on $\ell$ \\
    \consensus-RS \cite{lehmann2025consensus}     & 2025 &       1 & None, $\ell=1$   & \consensus    & 2 \\
    \bottomrule
  \end{tabular}
  }
\end{table}

\begin{figure}[t]
    \begin{minipage}[c]{0.57\textwidth}
        \centering
        \includegraphics[scale=0.9]{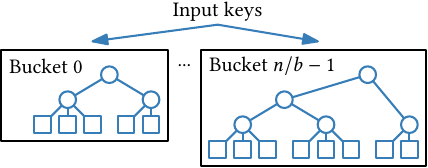}
    \end{minipage}\hfill
    \begin{minipage}[c]{0.4\textwidth}
        \caption{%
            Illustration of the overall RecSplit data structure.
            Within each bucket, it constructs a splitting tree.
            Circular nodes represent splittings, squares represent bijections.
        }
        \label{fig:recsplit}
    \end{minipage}
\end{figure}

\myparagraph{RecSplit\@.}\label{s:recsplit}
RecSplit \cite{esposito2020recsplit} uses brute-force in a novel way to enable large key sets with space close to the lower bound.
It first maps all keys to buckets of expected size $b$, where $b$ is a tuning parameter usually in the range 100--2000.
In each bucket, it then constructs an independent \emph{splitting tree}, as illustrated in \cref{fig:recsplit}.
The splitting tree partitions the keys into smaller and smaller sets until we arrive at the leaves which have a small configurable size $\ell$.
At each inner node, RecSplit tries random hash functions to find one that distributes the keys to the child nodes according to a \emph{splitting strategy}.
The splitting strategy also decides the \emph{fanout} (number of child nodes) of the tree and is optimized for balancing the work between splittings and leaves.
If the number of keys in a node is not a multiple of the child node sizes, the last node might be smaller. RecSplit adjusts the hash function accordingly.
The lowest level of the splitting tree is called \emph{leaf level}.
Each leaf, except for possibly the last, contains exactly $\ell$ keys.
Usual values for the leaf size are about 8--16 keys.
This is small enough that it is feasible to use the simple brute-force construction.
The shape of the splitting tree depends only on the leaf size $\ell$ and the total number of keys in the bucket.
Therefore, RecSplit does not need to store structural information of the tree.
It is enough to store the seed for each node in DFS order, encoded using Golomb-Rice codes \cite{golomb1966run,rice1979some}.
The key theoretical observation of RecSplit is that the probability of finding an
MPHF by brute force is the same as the probability of finding splittings and
MPHFs for the leaves of the tree, so the space needed to represent the
MPHF does not change with splitting (modulo small overheads due to approximations), 
but splitting makes the construction much faster.
Overall, the overhead over the lower bound is constant for each node of the splitting tree.

RecSplit can be queried by traversing the splitting tree from the root to a leaf by applying the splitting hash functions.
During traversal, it accumulates the number of keys stored in children to the left of the one descended into.
The final hash value is then the sum of the value of leaf bijection, the number of keys to the left in the splitting tree, and the total size of previous buckets.
Due to having to traverse the tree, queries are relatively slow compared to bucket placement techniques.
There are configurations that need only $1.56$ bits per key with reasonable construction time.

\myparagraph{SIMDRecSplit and GPURecSplit\@.}
SIMDRecSplit \cite{bez2023high} enhances RecSplit to use parallelism on multiple levels to improve the construction throughput.
The paper proposes a new technique for searching for bijections called \emph{rotation fitting}.
Instead of just applying hash functions on the keys in a leaf directly, rotation fitting splits the keys into two sets using a 1-bit hash function.
It then hashes each of the two sets individually, forming two words where the bits indicate which hash values are occupied.
Then it tries to cyclically rotate the second word, such that the empty output values left by the first set are filled by the values of the second set.
Each rotation essentially gives a new chance for a bijection \cite{bez2023high}, so it is a way to quickly evaluate additional hash function seeds in a bit-parallel way.
In addition to bit parallelism, SIMDRecSplit uses parallelism on the level of words, through SIMD instructions.
For this, it tests different hash function seeds or rotations simultaneously.
Finally, it uses multi-threading to construct different buckets in parallel.

GPURecSplit \cite{bez2023high} is an implementation of SIMDRecSplit on the GPU.
An important observation here is that all buckets of the same size also lead to splitting trees of the same shape.
Therefore, they can be constructed together using the same set of kernel calls.
GPURecSplit then uses CUDA \emph{streams} to construct different \emph{tree shapes} in parallel.

\myparagraph{\consensus-RecSplit\@.}
Perfect hashing through brute-force often stores the seed for each subtask individually using a variable-length code like Golomb-Rice \cite{golomb1966run,rice1979some}.
With \consensus \cite{lehmann2025consensus}, each seed has a fixed number of choices just barely larger than its expected number of choices.
If a subtask runs out of choices without finding a successful seed, \consensus backtracks to a previous subtask and looks for another successful seed.
The key idea making this feasible is that hashing takes into consideration not just the current seed but the \emph{concatenation} of all previous seeds.
With this, backtracking gives a completely new chance for a seed to be successful.
\consensus needs less space than storing the seed of each subtask individually, even when assuming an entropy-optimal coding.
\consensus-RecSplit uses RecSplit with a leaf size $\ell=1$ which would otherwise be inefficient.
However, it stores the seeds efficiently using \consensus.
Both RecSplit and \consensus-RecSplit can achieve a space consumption of $(1 + \varepsilon)\log_2 e$ bits per key, arbitrarily close to optimal.
However, while RecSplit has a running time of $\Oh{n \cdot e^{1/\varepsilon}}$, \consensus-RecSplit is the first approach to achieve this in just $\Oh{n/\varepsilon}$.
We refer to \cref{s:theory} for details.
A practical implementation achieves a space consumption of just $1.444$ bits per key, very close to the lower bound of $\log_2 e \approx 1.443$ bits per key.
However, the queries are about 60\% slower than RecSplit.

\section{Perfect Hashing through Fingerprinting}\label{s:fingerprinting}
The idea of perfect hashing through fingerprinting is to assign a small ($\Theta(n)$) fingerprint to each input key using a hash function.
We then resolve collisions between the fingerprints using recursion on the colliding keys.
An advantage is the very simple and easily parallelizable construction.
It is originally introduced by Chapman \etal \cite{chapman2011meraculous} in the context of bioinformatics, but not described as a perfect hashing data structure of general interest.
Müller \etal \cite{muller2014retrieval} then enhance and describe the idea from a data structure perspective.
\Cref{tab:fingerprintingMethods} gives an overview of the development.

\begin{figure}[t]
    \centering
    \includegraphics[scale=0.9]{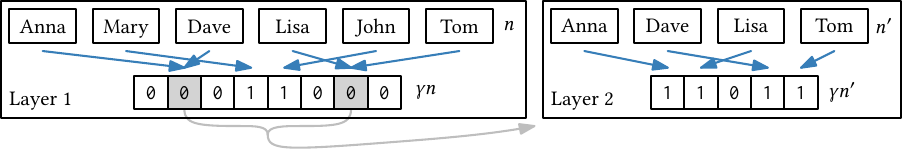}
    \caption{Illustration of PHFs using fingerprinting. Collisions are handled in the next layer.}
    \label{fig:fingerprinting}
\end{figure}

\myparagraph{FiPHa\@.}
Perfect hashing through fingerprinting \cite{muller2014retrieval} hashes the $n$ keys to $\gamma n$ fingerprints using an ordinary hash function, where $\gamma \geq 1$ is a tuning parameter.
A bit vector of length $\gamma n$ indicates fingerprints to which exactly one key was mapped.
At query time, when a key is the only one mapping to its location, a rank operation on the bit vector gives the MPHF value.
The bit vector indicates with a 0-bit that the key was not the only one mapping to that location.
In this case, an additional layer of the same data structure needs to be queried.
\Cref{fig:fingerprinting} illustrates this idea.
The most space-efficient choice $\gamma=1$ leads to a space consumption of $e$ bits per key \cite{muller2014retrieval}, which is quite far from the lower bound of $\log_2 e$ bits per key.
In this case, a query traverses $e$ layers of the data structure in expectation.
FiPHa was developed in cooperation with the German company SAP, and the source code is not publicly available.

\myparagraph{BBHash\@.}
The first publicly available implementation, BBHash \cite{limasset2017fast}, iterates over all keys to count the collisions.
Then it iterates over the keys again to write the fingerprint bit vector and to extract colliding keys.
During construction, this causes up to three random memory accesses per key and level, while we will later see that FiPS \cite{lehmann2024fast} achieves much higher cache locality.
For parallelization, BBHash uses a large number of atomic operations in the arrays.
While FiPHa already introduces the idea to scale the bit vector of fingerprints, BBHash makes this more explicit by introducing the $\gamma$ parameter that we already used in the description of FiPHa.

\begin{table}[t]
  \caption{%
      Overview of perfect hashing through fingerprinting.
  }
  \label{tab:fingerprintingMethods}
  \centering
  \scalebox{0.9}{
  \begin{tabular}[t]{lll}
      \toprule
      Approach                                & Year & New idea \\ \midrule
      Meraculous \cite{chapman2011meraculous} & 2011 & Original idea \\
      FiPHa \cite{muller2014retrieval}        & 2014 & First description as data structure \\
      BBHash \cite{limasset2017fast}          & 2017 & First public implementation \\
      FMPH \cite{beling2023fingerprinting}    & 2023 & Partitioning for faster construction \\
      FMPHGO \cite{beling2023fingerprinting}  & 2023 & Brute-Force trials for fewer levels \\
      FiPS \cite{lehmann2024fast}             & 2024 & Sorting for faster construction, interleaved rank data structure \\
    \bottomrule
  \end{tabular}
  }
\end{table}

\myparagraph{FMPH\@.}
FMPH \cite{beling2023fingerprinting} is a fast implementation of the idea in the Rust programming language.
The approach still uses the construction algorithm with two passes.
For parallelization, FMPH distributes the keys to multiple threads and relies on a large number of atomic operations, just like BBHash.
FMPH achieves impressive speedups compared to BBHash, even though it essentially implements the same algorithm.

\myparagraph{FiPS\@.}
FiPS \cite{lehmann2024fast} ---
Fingerprint Perfect hashing through Sorting
--- is a third implementation of the approach.
In contrast to the approaches above, FiPS focuses on cache locality when filling the bit vector during construction.
It does so by sorting the fingerprints and then determining collisions by a simple scan.
Using integer sorting, the construction takes linear time.
This approach through sorting is already described in the FiPHa paper \cite{muller2014retrieval} but without a publicly available implementation.
Using existing sorting libraries, the construction can be parallelized efficiently.
FiPS interleaves the select data structures and the bit vector.
More precisely, in each cache line, FiPS stores both the bit vector indicating fingerprints that did not collide, and the number of bits set in previous cache lines.
Different cache line sizes provide different trade-offs between space consumption and query performance.

\myparagraph{FMPHGO\@.}
FMPH with group optimization (FMPHGO) \cite{beling2023fingerprinting} is a new spin on the fingerprinting idea, combining it with a few brute-force tries.
The idea is to hash the keys to buckets like in FCH or CHD.
FMPHGO then tries a (small) number of different hash functions for each bucket and invests additional space to store which hash function should be used.
It selects the hash function that causes the least collisions in the fingerprint array.
Overall, this reduces the number of required layers and the required space.
Compared to FMPH, FMPHGO reduces the storage space by up to $0.7$ bits/key, while the query performance stays mostly the same.

\section{Asymptotic Performance}\label{s:theory}
In this section, we discuss the asymptotic space consumption, construction time, and query time of different families of approaches.
This includes how the configuration parameters can be tuned to achieve a space consumption that asymptotically approaches the lower bound.

We say a perfect hash function has space overhead $\varepsilon$ if it uses $(1+\varepsilon) n \log_2 e + o(n)$ bits of space (see \cref{ss:spacebounds}).
It is called \emph{succinct} if $\varepsilon \in o(1)$.
Many of the approaches we describe in this survey admit succinct configurations, such as PTHash \cite{pibiri2021pthash,pibiri2021parallel}, PHOBIC \cite{hermann2024phobic}, RecSplit \cite{esposito2020recsplit,bez2023high}, and ShockHash \cite{lehmann2023shockhash,lehmann2023bipartite}.
Here, we go one step further and discuss the $o(1)$ term in more detail.
We are specifically interested in approaches with tuning parameters that can bring $\varepsilon$ arbitrarily close to 0 without the construction or query time increasing too much.
In contrast, earlier papers mostly model the tuning parameters as constants.
In \cref{tab:asymptotics}, we give an overview of the asymptotic construction time depending on the parameters.
Note that all approaches can trivially get linear time construction through partitioning (see \cref{s:commonTools}).
However, the partitions introduce some space overhead.
If we want to get arbitrarily close to the lower bound, we might have to select larger partition sizes, leading to construction and query times that depend on the parameter.
We only use partitioning where used in the original paper.
We start with brute-force approaches that are only suitable for small input sets of size $\ell$.

\myparagraph{Simple Brute-Force\@.}
The simple brute-force technique (see \cref{s:simpleBruteForce}) takes time $\Oh{e^\ell}$ to construct.
The technique reaches the space lower bound but encoding the result loses a constant number of bits.
Therefore, in order to decrease the space overhead, we have to increase $\ell$.
Queries take constant time assuming that the seed of $\approx \ell \log_2 e$ bits fits into a machine word of size $w$.

\myparagraph{ShockHash\@.}
ShockHash \cite{lehmann2023shockhash} (see \cref{s:shockhash}) is a hybrid between brute-force and multiple choice hashing.
Like brute-force, its construction time is exponential in $\ell$, however with a much smaller base.
Bipartite ShockHash \cite{lehmann2023bipartite} reduces the base even further.
Queries of ShockHash access a seed of $\approx \ell (\log_2(e) - 1)$ bits and a 1-bit retrieval data structure.
Both operations take constant time, assuming $\ell \in \Oh{w}$.
While ShockHash uses the BuRR \cite{dillinger2022burr} retrieval data structure (see \cref{s:retrieval}) in practice, we use a different construction \cite{dietzfelbinger19constant} here because its space overhead shrinks more quickly.
In the following, we now consider techniques for larger input sets.

\myparagraph{Multiple Choice Hashing\@.}
Multiple choice hashing (see \cref{s:multipleChoiceHashing}) needs a constant number of bits/key due to storing the index of the choice.
Using a smart choice of hash function indices we can bring this down to 1.83 bits/key \cite{weaver2020constructing}.
However, no matter how we tune the number of choices for each key, classical multiple choice hashing does not reach the lower bound of $\log_2 e$ bits/key.

\myparagraph{Bucket Placement\@.}
PHOBIC \cite{hermann2024phobic} shows that perfect hashing through bucket placement (see \cref{s:bucketPlacement}) achieves a space consumption of $\log_2(e) + \Oh{\log(\lambda)/ \lambda}$ bits per key.
Therefore, by selecting large average bucket sizes $\lambda$, we can get arbitrarily close to the optimal space consumption.
The construction time stays linear in $n$, but it increases exponentially in $\lambda$.
Depending on the bucket assignment function and the order in which the buckets are placed, it is possible to achieve a construction time of $\Oh{ne^\lambda/\lambda}$.
The query time is constant.
Note that PtrHash \cite{grootkoerkamp2025ptrhash} uses fixed size seeds for improved query time in practice, which makes it not succinct.
For a similar reason, PHast \cite{beling2025phast} likely is not succinct either. However, its space consumption as a function of $\lambda$ is an open problem.

\myparagraph{Recursive Splitting\@.}
Each split in RecSplit's splitting tree loses a constant number of bits \cite{esposito2020recsplit}.
The original implementation uses the simple brute-force technique as its base case in each leaf.
Then, the space overhead in each leaf of size $\ell$ is in $\Oh{1/\ell}$.
For buckets of size $b$, storing the prefix sum of bucket sizes causes an additional overhead of $\Oh{\log(b)/b}$ bits.
Queries traverse the splitting tree, which takes at least time $\Oh{\log(b/\ell)}$, and then evaluate the base case in constant time.
Note that this assumes $\ell,b \in \Oh{w}$.
Constructing the splittings takes polynomial time in the bucket size, in addition to the base case construction.
Note that now $\ell$ is a tuning parameter.
This means that for any constant factor $\alpha$, we get a space overhead of $\Oh{\alpha/\ell} = \Oh{1/\ell}$.
However, the construction time of the base case is $\Oh{e^{\alpha\ell}} = \Oh{(e^\alpha)^\ell}$.
Therefore, any exponential base case (such as ShockHash) only gives a constant factor of space improvement when included in RecSplit.

\begin{table}[t]
    \caption{%
        Asymptotic construction and query time of different approaches.
        The variable $\varepsilon$ denotes the space overhead over the lower bound.
        While almost all approaches have some tuning parameters, not all make it possible to get arbitrarily close to the lower bound.
        With $w$, we denote the size of a machine word.
        Top: Brute-force approaches for small sets of size $\ell$, assuming $\ell \in \Oh{w}$.
        Bottom: Approaches for large sets of size $n$, assuming $\ell, b, \lambda \in \Oh{w}, \varepsilon \in \omega(1/\sqrt{n})$.
    }
    \label{tab:asymptotics}
    \centering
    \setlength{\tabcolsep}{4pt}
    \renewcommand*{\arraystretch}{1.1}
    \scalebox{0.9}{
        \begin{tabularx}{145mm}{ m{60mm} m{25mm} m{30mm} m{30mm} }
            \toprule
            Approach                      & Space overhead                         & Construction time   & Query time \\ \midrule
            Simple Brute-Force \cite{melhorn1984data}
                                          & $\Oh{\frac{1}{\ell}}$                  & $\Oh{e^{\ell}}$     & $\Oh{1}$ \\
            ShockHash \cite{lehmann2023shockhash}
                                          & $\Oh{\frac{1}{\ell}}$                  & $\Oh{1.36^\ell}$    & $\Oh{1}$ \\
            Bip. ShockHash \cite{lehmann2023bipartite}
                                          & $\Oh{\frac{1}{\ell}}$                  & $\Oh{1.17^\ell}$    & $\Oh{1}$ \\
            \midrule
            Multiple Choice \cite{botelho2007simple,genuzio2016fast,weaver2020constructing,lehmann2023sichash}
                                          & $\Oh{1}$                               & $\Oh{n}$                                  & $\Oh{1}$ \\
            Bucket Placement \cite{fox1992faster,belazzougui2009hash,pibiri2021pthash,pibiri2021parallel,hermann2024phobic} %
                                          & $\Oh{\frac{1}{\lambda}}$               & $\Oh{n\frac{e^\lambda}{\lambda}}$         & $\Oh{1}$ \\[-1mm]
            Recursive Splitting \cite{esposito2020recsplit,bez2023high}\newline\phantom{...} + any brute-force base case
                                          & \Oh{\frac{1}{\ell} + \frac{\log b}{b}} & $n(e^{\Oh{\ell}} + \textrm{poly}(b))$     & $\Oh{\log \frac{b}{\ell}}$ \\
            \consensus-RecSplit \cite{lehmann2025consensus}
                                          & \Oh{\varepsilon}                       & $\Oh{\frac{n}{\varepsilon}}$              & $\Oh{\log \frac{1}{\varepsilon}}$ \\
            Fingerprinting \cite{chapman2011meraculous,muller2014retrieval,limasset2017fast,beling2023fingerprinting,lehmann2024fast}
                                          & $\Oh{1}$                               & $\Oh{n}$                                  & $\Oh{1}$ \\
            Hagerup and Tholey \cite{HagTho01}
                                          & $\Oh{\frac{(\log \log n)^2}{\log n}}$  & $\Oh{n}$                                  & $\Oh{1}$ \\
            \bottomrule
        \end{tabularx}
    }
\end{table}

\myparagraph{\consensus-RecSplit\@.}
Bucketed \consensus-RecSplit \cite{lehmann2025consensus} is currently the only known approach where the construction time increases linearly in the inverse of the space overhead.
This is because its space consumption is not influenced by an exponential-time base case.
Through buckets of size $\Oh{1/\varepsilon}$, it achieves a query time of $\Oh{\log(1/\varepsilon)}$.
Consider the approaches using the parameter $\ell$ (or $\lambda$).
Assume the word size is $w$, e.g., $w \in \Theta(\log n)$.
If we wanted to reduce the space overhead to $o(1/w)$ by using $\ell \in \omega(w)$, then not only would the construction time become a problem (after all, it is exponential in $\ell$), but the seed values of $\Theta(\ell)$ bits would also no longer fit the word size.
This means that the query time would start to increase linearly in $\ell$.
\consensus-RecSplit is currently the only approach for which this is not the case.

\myparagraph{Fingerprinting\@.}
The fingerprint based approaches (see \cref{s:fingerprinting}) have a tuning parameter $\gamma$ that influences the space consumption.
However, when selecting the smallest possible value of $\gamma=1$, we get a space consumption of $e$ bits per key, quite far from the lower bound of $\log_2 e$ bits per key.
FMPHGO \cite{beling2023fingerprinting} introduces the idea of group optimization to improve the space consumption of fingerprinting.
In a hypothetical case where we would encode the group seeds with variable length and retry until there are no collisions, we get perfect hashing through bucket placement.
FMPHGO can therefore interpolate between Fingerprinting and bucket placement.

\myparagraph{Hagerup and Tholey\@.}
Finally, we look at a purely theoretical result that is not included in the main part of the survey.
Hagerup and Tholey \cite{HagTho01} first reduce the range of input universe to $[n^3]$ by repeatedly taking the keys modulo a prime number.
Then they hash the input to small sets of size $\Oh{\log n / \log \log n}$.
For each possible small input set, they use brute-force to determine a minimal perfect hash function.
Then they map each actual small set to one of the pre-computed MPHFs using a lookup table.
Overall, Hagerup and Tholey achieve linear construction time, constant query time, and space $1+o(1)$ times the lower bound.
More precisely, the space overhead is $\Oh{(\log \log n)^2 / \log n}$ and therefore only depends on $n$.
This means that in order to reduce the overhead, we have to increase the input size.
In contrast, other approaches have a tuning parameter that can control the space overhead independently of $n$.
Note that even for $n = 2^{256}$ we have $(\log \log n)^2/\log n > 1$, so the overhead is large even for astronomic sizes of $n$.
The approach has never been implemented (and is, in fact, not properly defined for $n < 2^{150}$ \cite{botelho2013practical}).

\section{Evaluation}
\label{ch:evaluation}
In this section, we compare the performance of the different state-of-the-art perfect hash function constructions.
Our detailed measurements can help to pick the most fitting perfect hash function for any given application.
For this comparison, we look at the three most important parameters for minimal perfect hash functions: space consumption, construction time, and query time.

\myparagraph{Experimental Setup\@.}\label{evaluation:s:setup}

\def\totalDatapoints{9290}

\def\totalDays{32}

We run most of our experiments on an Intel i7 11700 processor with 8 cores (16 hardware threads), pinned to its base clock speed of 2.5 GHz, and supporting AVX-512.
In our experiments, pinning the clock speed does not influence the relative performance of the approaches.
The machine runs Rocky Linux 9.5 with Linux 5.14.0.
The sizes of the L1 and L2 data caches are 48 KiB and 512 KiB per core, and the L3 cache has a size of 16 MiB.
We use the GNU C++ compiler version 14.2.0 with optimization flags \texttt{-O3 -march=native} and Rust 1.85.0 with release mode and \texttt{target-cpu=native}.
We also give experiments on an ARM processor in \ifAppendix \cref{appendix:arm}\else the extended version \cite{lehmann2025modern}\fi.
Many of the implementations we evaluate are actively maintained and get updated with performance improvements.
Our measurements refer to their state from October 12, 2025.
As input data, we use strings of uniform random length in the range $[10..50]$ containing random characters except for the zero byte.
This is because essentially all approaches support string inputs.
Note that, as a first step, almost all competitors generate an initial hash of each key using a high quality hash function.
This makes the remaining computation largely independent of the input distribution.
The code and scripts needed to reproduce our experiments are available on GitHub under the General Public License \cite{sourceCodeMphfExperiments}.
In total, we measure \totalDatapoints{} data points with a cumulative duration of over \totalDays{} days.

\myparagraph{Dominance Maps\@.}
Visual plots give a good understanding of the trade-offs between different perfect hash functions, especially if each approach has a wide number of configurations.
In order to print all three main parameters --- space consumption, construction performance, and query performance --- on paper, we use a type of plot that we call \emph{dominance map} \cite{lehmann2024fast,dillinger2022burr}.
We start with a 2D projection and then color each point based on the best competitor along the third dimension.
This therefore provides a ``front view'' of the Pareto space.
Pareto optimal points are the points that are not dominated in all three dimensions simultaneously.
This results in a cleaner plot because it does not include data points for dominated approaches.
When having a specific time and/or space budget, dominance maps show the fastest approach given that restriction.

\myparagraph{Overview\@.}
We start our evaluation with Pareto fronts and dominance maps containing a wide range of configurations for each competitor.
In \cref{s:construction,s:query} we focus on the construction time trade-off and the query time trade-off.
From these configurations, we pick representative ones for each approach, details in \ifAppendix \cref{appendix:s:table}\else the extended version of this paper \cite{lehmann2025modern}\fi.
We then look at how the approaches scale in the input size in \cref{s:scalingN} and in the number of threads in \cref{s:scalingThreads}.

\subsection{Construction Trade-Off}\label{s:construction}

\begin{figure*}[p]
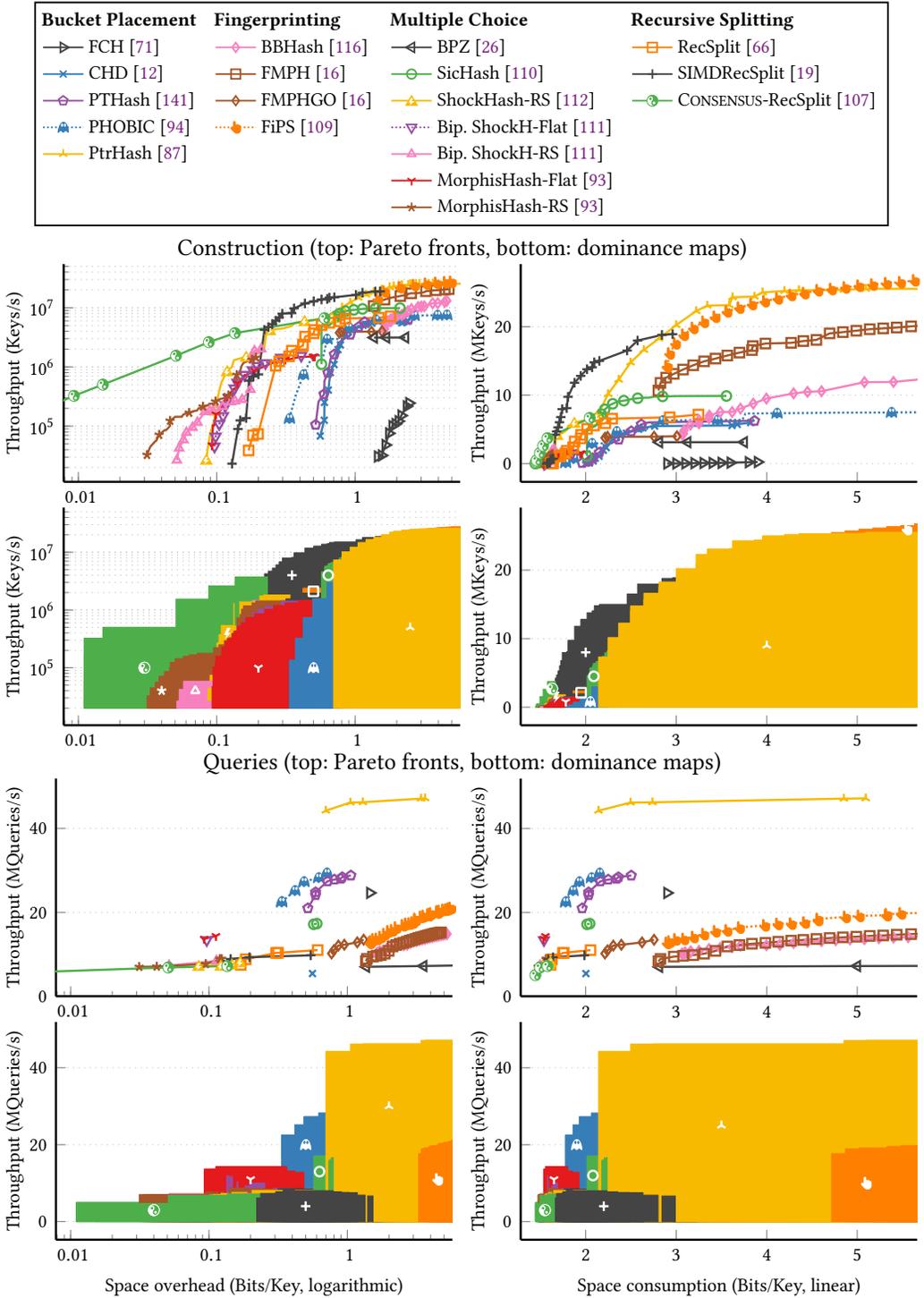

  \vspace{-6mm}
  \center
  \begin{tikzpicture}
    \pgfplotslegendfromname{legendEvalParetoConstruction}
  \end{tikzpicture}

  \textbf{Construction (top: Pareto fronts, bottom: dominance maps)}\\
  \hfill
  \begin{subfigure}[t]{0.49\textwidth}
    \input{fig/paretoConstructionLog}
  \end{subfigure}%
  \hspace{1mm}%
  \begin{subfigure}[t]{0.49\textwidth}
    \input{fig/paretoConstruction}
  \end{subfigure}
  \\[-1mm]
  \hfill
  \begin{subfigure}[t]{0.49\textwidth}
    \input{fig/paretoConstructionHeatmapLog}
  \end{subfigure}%
  \hspace{1mm}%
  \begin{subfigure}[t]{0.49\textwidth}
    \input{fig/paretoConstructionHeatmap}
  \end{subfigure}
  \\[-2mm]
  \textbf{Queries (top: Pareto fronts, bottom: dominance maps)}\\
  \hfill
  \begin{subfigure}[t]{0.49\textwidth}
        \begin{tikzpicture}
        \begin{axis}[
            plotEvalPareto,
            ylabel={Throughput (Queries/s)},
            legend columns=2,
            ymin=0,
            xmode=log,
            ymode=log,
            xmin=0.01,
            xmax=4.5,
            ymin=8e5,
            ymax=1e8,
            log x ticks with fixed point,
            ytick distance=10,
          ]
          \addplot[mark=diamond,color=colorBbhash,solid] coordinates { (1.61538,5070993) (1.71358,5165556) (1.8098,5283457) (1.92019,5503274) (2.04478,5621767) (2.17547,5736247) (2.31363,5880277) (2.75311,5975857) (2.90529,6130080) (3.05946,6154982) (3.21529,6250781) (3.3725,6305965) (3.53089,6362537) (3.69028,6437077) (3.85055,6490556) (4.01156,6538084) (4.17323,6587615) (4.33547,6636580) (4.49823,6655131) (4.66143,6695232) (4.82504,6703311) (4.989,6764069) (5.31788,6797172) };
          \addlegendentry{BBHash \cite{limasset2017fast}}
          \addplot[mark=leftTriangle,color=colorBdz,solid] coordinates { (1.3698,4532475) };
          \addlegendentry{BPZ \cite{botelho2013practical}}
          \addplot[mark=flippedTriangle,color=colorBipartiteShockHashFlat,densely dotted] coordinates { (0.09623,5932957) };
          \addlegendentry{Bip. ShockH-Flat \cite{lehmann2023bipartite}}
          \addplot[mark=triangle,color=colorBipartiteShockHash,solid] coordinates { (0.05756,3069367) (0.06847,3220923) (0.11233,3701373) (0.1131,3724117) };
          \addlegendentry{Bip. ShockH-RS \cite{lehmann2023bipartite}}
          \addplot[mark=x,color=colorChd,solid] coordinates { (0.59854,2642775) (0.6004,2669015) };
          \addlegendentry{CHD \cite{belazzougui2009hash}}
          \addplot[mark=rightTriangle,color=colorFch,solid] coordinates { (1.55731,10896807) };
          \addlegendentry{FCH \cite{fox1992faster}}
          \addplot[mark=square,color=colorRustFmph,solid] coordinates { (1.36037,5971218) (1.38371,6320313) (1.63286,6509145) (1.74496,7018036) (1.99193,7203572) (2.25717,7693491) (2.3944,7936507) (2.53355,8118201) (2.81798,8567511) (3.10664,8929368) (3.2525,9164222) (3.69554,9372949) (4.14284,9632055) (4.29275,9731413) (4.44294,9759906) (4.60365,9828009) (4.75442,9891196) };
          \addlegendentry{FMPH \cite{beling2023fingerprinting}}
          \addplot[mark=diamond,color=colorRustFmphGo,solid] coordinates { (0.7698,6957489) (0.79831,7751937) (0.86231,9062075) (0.95114,9369436) (1.05827,9842519) (1.17981,10353038) (1.31285,10556317) (1.45566,11103708) (2.09309,11373976) (2.26392,11586142) };
          \addlegendentry{FMPHGO \cite{beling2023fingerprinting}}
          \addplot[mark=fingerprint,color=colorFiPS,densely dotted] coordinates { (1.45722,6525285) (1.48693,6633059) (1.55262,6945409) (1.64269,7221259) (1.74892,7490636) (1.86562,7733952) (2.11884,7746533) (2.12169,7926442) (2.25778,8412551) (2.39671,8591803) (2.68353,8625894) (2.82618,9122422) (2.97784,9234463) (3.42821,9297136) (3.57616,9678668) (3.72864,9787608) (4.04063,9902951) (4.19517,10006003) (4.34989,10038144) (4.64575,10108157) (4.66078,10193679) (4.83533,10238558) (4.96755,10328444) (5.21835,10416666) (5.41073,10522992) (5.60358,10595465) (5.79711,10674637) (5.9912,10728462) (6.18556,10779346) (6.38032,10848340) (6.57526,10913456) };
          \addlegendentry{FiPS \cite{lehmann2024fast}}
          \addplot[mark=Mercedes star flipped,color=colorMorphisHashFlat] coordinates { (0.0923,6541933) (0.10159,6552221) (0.11227,6556947) };
          \addlegendentry{MorphisHash-Flat \cite{hermann2025morphishash}}
          \addplot[mark=star,color=colorMorphisHash] coordinates { (0.03603,3108292) (0.03833,3125879) (0.042,3137156) (0.04607,3164757) (0.04955,3198157) (0.05171,3262110) (0.06483,3277506) (0.09462,3539697) (0.10333,3544339) (0.11324,3568752) (0.11943,4250255) (0.12357,4255500) };
          \addlegendentry{MorphisHash-RS \cite{hermann2025morphishash}}
          \addplot[mark=phobic,color=colorDensePtHash,densely dotted] coordinates { (0.48999,7656967) (0.517,7674008) (0.53024,8480325) (0.82597,10689470) (0.95206,11042402) (1.22571,11057054) };
          \addlegendentry{PHOBIC \cite{hermann2024phobic}}
          \addplot[mark=otimes,color=colorPhast,mark repeat*=4] coordinates { (0.36076,23084025) (0.36359,23607176) (0.37737,24148756) (0.41659,24912805) (0.44516,25516713) (0.48084,26171159) (0.50019,26652452) (0.51037,27034333) (0.52306,27344818) (0.55538,27540622) (0.59511,28224668) (0.615,29163021) (0.6533,29515938) (0.6832,30674846) };
          \addlegendentry{PHast \cite{beling2025phast}}
          \addplot[mark=oplus,color=colorPhastPlus,mark repeat*=4] coordinates { (0.37941,22583559) (0.37947,23137436) (0.3805,24740227) (0.38955,26462026) (0.40549,26645350) (0.58014,27909572) (0.58802,28288543) (0.6443,28530670) (0.72488,28555111) (0.77364,28563267) (0.93858,29788501) (0.9614,29976019) (1.00475,30084235) (1.04228,30138637) (1.08506,30165912) (1.13293,30175015) (1.1783,30202355) };
          \addlegendentry{PHast${}^+$ \cite{beling2025phast}}
          \addplot[mark=pentagon,color=colorPthash,solid] coordinates { (0.49421,9078529) (0.57319,10009008) (0.58836,10231225) (0.70563,11757789) (0.71842,12030798) (0.76127,12227928) (0.88456,13140604) };
          \addlegendentry{PTHash \cite{pibiri2021pthash}}
          \addplot[mark=Mercedes star,color=colorPtrHash] coordinates { (0.7262,19308746) (0.93185,19361084) (1.09916,20508613) (1.30481,20622808) };
          \addlegendentry{PtrHash \cite{grootkoerkamp2025ptrhash}}
          \addplot[mark=square,color=colorRecSplit,solid] coordinates { (0.18256,3601786) (0.19869,4009623) (0.25685,4241421) (0.31596,4783773) };
          \addlegendentry{RecSplit \cite{esposito2020recsplit}}
          \addplot[mark=+,color=colorSimdRecSplit,solid] coordinates { (0.14245,3599323) (0.14777,3614022) (0.15889,4156448) (0.21513,4747661) (0.53614,4907734) };
          \addlegendentry{SIMDRecSplit \cite{bez2023high}}
          \addplot[mark=shockhash,color=colorShockHash,solid] coordinates { (0.0894,2974154) (0.0958,2988196) (0.10342,3167363) (0.13383,3172085) (0.14659,3280732) (0.17151,3415300) (0.20405,3469331) };
          \addlegendentry{ShockHash-RS \cite{lehmann2023shockhash}}
          \addplot[mark=o,color=colorSicHash,solid] coordinates { (0.57168,7590708) (0.60215,7631840) (0.66815,7661074) (0.72976,7724393) (0.76027,7756146) (0.78396,7777864) (0.79083,7798487) (0.81421,7811889) };
          \addlegendentry{SicHash \cite{lehmann2023sichash}}
          \addplot[mark=consensus,color=colorConsensus] coordinates { (0.00211,2225882) (0.05069,2700513) (0.13547,2796968) };
          \addlegendentry{\consensus-RecSplit \cite{lehmann2025consensus}}

          \legend{}
        \end{axis}
    \end{tikzpicture}%
  \end{subfigure}%
  \hspace{1mm}%
  \begin{subfigure}[t]{0.49\textwidth}
        \begin{tikzpicture}
        \begin{axis}[
            plotEvalPareto,
            ylabel={Throughput (MQueries/s)},
            legend to name=legendEvalParetoQuery,
            legend columns=3,
            ytick distance=10,
            ymin=-0.2,
          ]
          \addplot[mark=diamond,color=colorBbhash,solid] coordinates { (3.05808,5.07099) (3.15628,5.16556) (3.2525,5.28346) (3.36289,5.50327) (3.48748,5.62177) (3.61817,5.73625) (3.75633,5.88028) (4.19581,5.97586) (4.34799,6.13008) (4.50216,6.15498) (4.65799,6.25078) (4.8152,6.30597) (4.97359,6.36254) (5.13298,6.43708) (5.29325,6.49056) (5.45426,6.53808) (5.61593,6.58762) (5.77817,6.63658) (5.94093,6.65513) (6.10413,6.69523) (6.26774,6.70331) (6.4317,6.76407) (6.76058,6.79717) };
          \addlegendentry{BBHash \cite{limasset2017fast}}
          \addplot[mark=leftTriangle,color=colorBdz,solid] coordinates { (2.8125,4.53248) };
          \addlegendentry{BPZ \cite{botelho2013practical}}
          \addplot[mark=flippedTriangle,color=colorBipartiteShockHashFlat,densely dotted] coordinates { (1.53893,5.93296) };
          \addlegendentry{Bip. ShockH-Flat \cite{lehmann2023bipartite}}
          \addplot[mark=triangle,color=colorBipartiteShockHash,solid] coordinates { (1.50026,3.06937) (1.51117,3.22092) (1.55503,3.70137) (1.5558,3.72412) };
          \addlegendentry{Bip. ShockH-RS \cite{lehmann2023bipartite}}
          \addplot[mark=x,color=colorChd,solid] coordinates { (2.04124,2.64278) (2.0431,2.66902) };
          \addlegendentry{CHD \cite{belazzougui2009hash}}
          \addplot[mark=rightTriangle,color=colorFch,solid] coordinates { (3.00001,10.8968) };
          \addlegendentry{FCH \cite{fox1992faster}}
          \addplot[mark=square,color=colorRustFmph,solid] coordinates { (2.80307,5.97122) (2.82641,6.32031) (3.07556,6.50915) (3.18766,7.01804) (3.43463,7.20357) (3.69987,7.69349) (3.8371,7.93651) (3.97625,8.1182) (4.26068,8.56751) (4.54934,8.92937) (4.6952,9.16422) (5.13824,9.37295) (5.58554,9.63206) (5.73545,9.73141) (5.88564,9.75991) (6.04635,9.82801) (6.19712,9.8912) };
          \addlegendentry{FMPH \cite{beling2023fingerprinting}}
          \addplot[mark=diamond,color=colorRustFmphGo,solid] coordinates { (2.2125,6.95749) (2.24101,7.75194) (2.30501,9.06208) (2.39384,9.36944) (2.50097,9.84252) (2.62251,10.353) (2.75555,10.5563) (2.89836,11.1037) (3.53579,11.374) (3.70662,11.5861) };
          \addlegendentry{FMPHGO \cite{beling2023fingerprinting}}
          \addplot[mark=fingerprint,color=colorFiPS,densely dotted] coordinates { (2.89992,6.52529) (2.92963,6.63306) (2.99532,6.94541) (3.08539,7.22126) (3.19162,7.49064) (3.30832,7.73395) (3.56154,7.74653) (3.56439,7.92644) (3.70048,8.41255) (3.83941,8.5918) (4.12623,8.62589) (4.26888,9.12242) (4.42054,9.23446) (4.87091,9.29714) (5.01886,9.67867) (5.17134,9.78761) (5.48333,9.90295) (5.63787,10.006) (5.79259,10.0381) (6.08845,10.1082) (6.10348,10.1937) (6.27803,10.2386) (6.41025,10.3284) (6.66105,10.4167) (6.85343,10.523) (7.04628,10.5955) (7.23981,10.6746) (7.4339,10.7285) (7.62826,10.7793) (7.82302,10.8483) (8.01796,10.9135) };
          \addlegendentry{FiPS \cite{lehmann2024fast}}
          \addplot[mark=Mercedes star flipped,color=colorMorphisHashFlat] coordinates { (1.535,6.54193) (1.54429,6.55222) (1.55497,6.55695) };
          \addlegendentry{MorphisHash-Flat \cite{hermann2025morphishash}}
          \addplot[mark=star,color=colorMorphisHash] coordinates { (1.47873,3.10829) (1.48103,3.12588) (1.4847,3.13716) (1.48877,3.16476) (1.49225,3.19816) (1.49441,3.26211) (1.50753,3.27751) (1.53732,3.5397) (1.54603,3.54434) (1.55594,3.56875) (1.56213,4.25026) (1.56627,4.2555) };
          \addlegendentry{MorphisHash-RS \cite{hermann2025morphishash}}
          \addplot[mark=phobic,color=colorDensePtHash,densely dotted] coordinates { (1.93269,7.65697) (1.9597,7.67401) (1.97294,8.48033) (2.26867,10.6895) (2.39476,11.0424) (2.66841,11.0571) };
          \addlegendentry{PHOBIC \cite{hermann2024phobic}}
          \addplot[mark=otimes,color=colorPhast,mark repeat*=4] coordinates { (1.80346,23.084) (1.80629,23.6072) (1.82007,24.1488) (1.85929,24.9128) (1.88786,25.5167) (1.92354,26.1712) (1.94289,26.6525) (1.95307,27.0343) (1.96576,27.3448) (1.99808,27.5406) (2.03781,28.2247) (2.0577,29.163) (2.096,29.5159) (2.1259,30.6748) };
          \addlegendentry{PHast \cite{beling2025phast}}
          \addplot[mark=oplus,color=colorPhastPlus,mark repeat*=4] coordinates { (1.82211,22.5836) (1.82217,23.1374) (1.8232,24.7402) (1.83225,26.462) (1.84819,26.6454) (2.02284,27.9096) (2.03072,28.2885) (2.087,28.5307) (2.16758,28.5551) (2.21634,28.5633) (2.38128,29.7885) (2.4041,29.976) (2.44745,30.0842) (2.48498,30.1386) (2.52776,30.1659) (2.57563,30.175) (2.621,30.2024) };
          \addlegendentry{PHast${}^+$ \cite{beling2025phast}}
          \addplot[mark=pentagon,color=colorPthash,solid] coordinates { (1.93691,9.07853) (2.01589,10.009) (2.03106,10.2312) (2.14833,11.7578) (2.16112,12.0308) (2.20397,12.2279) (2.32726,13.1406) };
          \addlegendentry{PTHash \cite{pibiri2021pthash}}
          \addplot[mark=Mercedes star,color=colorPtrHash] coordinates { (2.1689,19.3087) (2.37455,19.3611) (2.54186,20.5086) (2.74751,20.6228) };
          \addlegendentry{PtrHash \cite{grootkoerkamp2025ptrhash}}
          \addplot[mark=square,color=colorRecSplit,solid] coordinates { (1.62526,3.60179) (1.64139,4.00962) (1.69955,4.24142) (1.75866,4.78377) };
          \addlegendentry{RecSplit \cite{esposito2020recsplit}}
          \addplot[mark=+,color=colorSimdRecSplit,solid] coordinates { (1.58515,3.59932) (1.59047,3.61402) (1.60159,4.15645) (1.65783,4.74766) (1.97884,4.90773) };
          \addlegendentry{SIMDRecSplit \cite{bez2023high}}
          \addplot[mark=shockhash,color=colorShockHash,solid] coordinates { (1.5321,2.97415) (1.5385,2.9882) (1.54612,3.16736) (1.57653,3.17209) (1.58929,3.28073) (1.61421,3.4153) (1.64675,3.46933) };
          \addlegendentry{ShockHash-RS \cite{lehmann2023shockhash}}
          \addplot[mark=o,color=colorSicHash,solid] coordinates { (2.01438,7.59071) (2.04485,7.63184) (2.11085,7.66107) (2.17246,7.72439) (2.20297,7.75615) (2.22666,7.77786) (2.23353,7.79849) (2.25691,7.81189) };
          \addlegendentry{SicHash \cite{lehmann2023sichash}}
          \addplot[mark=consensus,color=colorConsensus] coordinates { (1.44481,2.22588) (1.49339,2.70051) (1.57817,2.79697) };
          \addlegendentry{\consensus-RecSplit \cite{lehmann2025consensus}}
        \end{axis}
    \end{tikzpicture}%
  \end{subfigure}
  \\[-1mm]
  \hfill
  \begin{subfigure}[t]{0.49\textwidth}
    \input{fig/paretoQueryHeatmapLog}
  \end{subfigure}%
  \hspace{1mm}%
  \begin{subfigure}[t]{0.49\textwidth}
    \input{fig/paretoQueryHeatmap}
  \end{subfigure}
  \\[-3mm]
  \caption{
      Trade-off construction time, query time, and space consumption.
      Single-threaded, $n=100$~million string keys of length [10..50].
      We do not show all markers to increase readability.
      The $x$-axis on the left is logarithmic to the space lower bound, so an overhead of 0 bits would correspond to~$-\infty$.
  }
  \label{fig:pareto}
\end{figure*}

We start our comparison by looking at the single-threaded construction time.
\Cref{fig:pareto} gives both a Pareto front showing all approaches, and the dominance map described above.
The dominance map plots the space consumption and the construction throughput on the axes, and colors each point with the approach that has the fastest queries.
For example, with a space consumption of 4 bits per key, FiPS \cite{lehmann2024fast} has the fastest construction.
However, by sacrificing just a little bit of construction performance by looking further down, we see that we can get the faster queries of PtrHash \cite{grootkoerkamp2025ptrhash}.
Because some approaches are focused on very small space consumption, we additionally give a plot that uses logarithmic $x$- and $y$-axes.
In the following, we look at the different perfect hash function constructions in more detail.
We focus on the construction time but also briefly mention the query time from the dominance maps.
For details on query time, we refer to \cref{s:query}.

\myparagraph{Multiple Choice Hashing\@.}
SicHash \cite{lehmann2023sichash} is up to two times faster to construct than PTHash \cite{pibiri2021pthash} and PHOBIC \cite{hermann2024phobic}, up to a space consumption of about 2 bits per key.
However, like most other approaches, its construction throughput stays quite far below SIMDRecSplit.
Even though BPZ \cite{botelho2013practical} is a lot older than the other approaches, it still holds up in terms of construction time.
However, the dominance map shows that newer approaches cover it almost completely with a better trade-off.
Looking at even more space-efficient approaches, mainly variants of ShockHash \cite{lehmann2023shockhash} achieve below 1.55 bits per key (0.1 bits per key overhead).
Especially the logarithmic plot shows how ShockHash and bipartite ShockHash improve the space consumption significantly.
Bipartite ShockHash-Flat \cite{lehmann2023bipartite} trades larger space consumption and slower construction for better query speed.
MorphisHash \cite{hermann2025morphishash}, as an enhancement to ShockHash, offers similar construction time but has lower space consumption.
Because its queries are slightly faster due to the integrated retrieval data structure, it mostly covers ShockHash in the dominance maps.

\myparagraph{Bucket Placement\@.}
PTHash \cite{pibiri2021pthash} and CHD \cite{belazzougui2009hash} have a similar trade-off between construction time and space consumption.
PHOBIC \cite{hermann2024phobic} improves the construction time of PTHash significantly through its optimized bucket assignment function.
The construction time of FCH \cite{fox1992faster} is rather far from the other approaches.
PtrHash \cite{grootkoerkamp2025ptrhash}, an extension of PTHash and PHOBIC, achieves significantly faster construction, allthough at a larger space consumption.
Finally, PHast \cite{beling2025phast} is focused mainly on query speed but also achieves very good construction performance and small space consumption.
Because of its fast queries that we will look at in \cref{s:query}, it fills most of the dominance map about space and construction time.
Because PtrHash and PHast cover such a large range of configurations, they are a good initial choice in practice.

\myparagraph{Recursive Splitting\@.}
RecSplit \cite{esposito2020recsplit} is a construction that achieved a significant step towards the space lower bound at the time of its publication.
Today, it is mostly dominated by the SIMD-parallel implementation.
Just like RecSplit, SIMDRecSplit \cite{bez2023high} is originally designed for small space consumption.
We make the surprising observation that SIMDRecSplit not only wins for the most space-efficient configurations, but also has very competitive construction time for less space-efficient cases.
However, being based on the splitting tree of RecSplit, it has quite slow queries, so in the dominance map, it only appears in areas that no other approaches can reach.
\consensus-RS \cite{lehmann2025consensus} needs the smallest amount of space.
As visible in the logarithmic versions of the plots, it significantly outperforms previous approaches in terms of space efficiency.

\myparagraph{Fingerprinting\@.}
The goal of perfect hashing through fingerprinting \cite{muller2014retrieval} is to offer efficient queries and fast construction at the cost of larger space consumption.
BBHash \cite{limasset2017fast} is dominated by more recent implementations.
FiPS \cite{lehmann2024fast} and recent version of FMPH \cite{beling2023fingerprinting} achieve almost twice the construction throughput and lower minimal space consumption.
FMPHGO \cite{beling2023fingerprinting} reduces the space consumption of FMPH but also has a much slower construction.
Its construction performance is similar to SicHash but has slower queries.

\subsection{Query Trade-Off}\label{s:query}
We now discuss the approaches again, this time focusing on the query time.
\Cref{fig:pareto} also includes a Pareto front and a dominance map for the query times.
The dominance map plots the space consumption and the query throughput on the axes, and colors each point with the approach that has the fastest construction.
In \ifAppendix \cref{appendix:s:table}\else \cite{lehmann2025modern}\fi, we briefly consider the cache misses per query.

\myparagraph{Multiple Choice Hashing\@.}
ShockHash-RS \cite{lehmann2023shockhash} and bipartite ShockHash-RS \cite{lehmann2023bipartite} use the RecSplit splitting tree but need an additional access to a retrieval data structure.
However, their query performance is still very close to RecSplit.
The reason is that they have fewer tree layers to traverse.
It shows that the overhead of the retrieval operation is small compared to the work for traversing the heavily compressed tree.
Bipartite ShockHash-Flat \cite{lehmann2023bipartite} is a variant focused on faster queries.
For this, it sacrifices some of the space consumption of bipartite ShockHash-RS.
For the same space consumption, it achieves 30\% faster queries, which brings the query performance of very space-efficient MPHFs much closer to competitors that are not focused on space consumption.
For both ShockHash variants, MorphisHash \cite{hermann2025morphishash} uses a more cache-efficient retrieval data structure but its main contribution is smaller space consumption.
SicHash \cite{lehmann2023sichash} provides a middle ground between construction and query performance.

\myparagraph{Bucket Placement\@.}
PHOBIC \cite{hermann2024phobic} improves the construction time of PTHash \cite{pibiri2021pthash} without sacrificing too much query performance.
However, PHast \cite{beling2025phast} is the clear winner in terms of query performance.
It is almost three times faster than most other competitors, and only PtrHash \cite{grootkoerkamp2025ptrhash} gets close.
Simultaneously, PHast achieves good space consumption.
CHD \cite{belazzougui2009hash} is much slower to query, mostly due to decoding variable-length seeds.
FCH \cite{fox1992faster} has fast queries, but needs more~space.

\myparagraph{Recursive Splitting\@.}
RecSplit \cite{esposito2020recsplit} and SIMDRecSplit \cite{bez2023high} are slow to query because they have to traverse the splitting tree, decoding variable-bitlength data in each step.
However, even though the operations are much more complex, the performance is still solid and not too far away from FiPS and FMPH.
For extremely small space consumption, \consensus-RS is the only competitor, so it fills a large area of the dominance maps.
Its query performance is similar to SIMDRecSplit.

\myparagraph{Fingerprinting\@.}
Perfect hashing through fingerprinting \cite{muller2014retrieval} is originally designed to offer fast queries.
Of the different implementations, FiPS \cite{lehmann2024fast} and FMPH \cite{beling2023fingerprinting} offer the best query throughput, being much faster than BBHash \cite{limasset2017fast}.
However, it is still far away from PtrHash.
This holds even for a rather large space consumption of 3.5 bits per key ($\gamma=2$), where about 73\% of the keys can be handled in the first recursion layer.
This indicates that the rank operation is rather costly, even when the rank data structure is interleaved with the bit vector.
FMPHGO \cite{beling2023fingerprinting} achieves a smaller space consumption than the other fingerprinting approaches through some brute-force retries.

\begin{figure*}[tp]
    \centering
    \begin{tikzpicture}
        \pgfplotslegendfromname{legendEvalParetoConstruction}
    \end{tikzpicture}

    Space consumption $\in (1.44, 1.7]$ bits/key.
        \hfill
    \begin{tikzpicture}
        \begin{axis}[
            plotEvalScalingN,
            xticklabel=\empty,
            ylabel={Throughput (MKeys/s)},
            xmode=log,
            ymin=0,
            ytick distance=1,
          ]
          \addplot[mark=flippedTriangle,color=colorBipartiteShockHashFlat,densely dotted] coordinates { (1000000,0.473934) (2000000,0.467071) (5000000,0.467873) (10000000,0.465549) (20000000,0.459132) (50000000,0.460431) (1e+08,0.452876) (2e+08,0.452043) };
          \addplot[mark=triangle,color=colorBipartiteShockHash,solid] coordinates { (1000000,0.0709157) (2000000,0.0693137) (5000000,0.0706567) (10000000,0.0705) (20000000,0.0702943) (50000000,0.0703101) (1e+08,0.0703786) (2e+08,0.0704367) };
          \addplot[mark=Mercedes star flipped,color=colorMorphisHashFlat] coordinates { (1000000,0.0965321) (2000000,0.0958436) (5000000,0.0948791) (10000000,0.0954326) (20000000,0.0939093) (50000000,0.0951089) (1e+08,0.0950269) (2e+08,0.0947237) };
          \addplot[mark=star,color=colorMorphisHash] coordinates { (1000000,0.0521037) (2000000,0.0512444) (5000000,0.0518665) (10000000,0.051873) (20000000,0.0517341) (50000000,0.051801) (1e+08,0.0518802) (2e+08,0.0518516) };
          \addplot[mark=shockhash,color=colorShockHash,solid] coordinates { (1000000,0.244798) (2000000,0.243063) (5000000,0.24357) (10000000,0.242175) (20000000,0.242296) (50000000,0.241236) (1e+08,0.241369) (2e+08,0.241277) };
          \addplot[mark=consensus,color=colorConsensus] coordinates { (1000000,1.83824) (2000000,1.8315) (5000000,1.82149) (10000000,1.82083) (20000000,1.81315) (50000000,1.79218) (1e+08,1.75645) (2e+08,1.78282) };

        \end{axis}
    \end{tikzpicture}
    \hspace{3mm}
    \begin{tikzpicture}
        \begin{axis}[
            plotEvalScalingN,
            xticklabel=\empty,
            ylabel={Throughput (MQueries/s)},
            xmode=log,
            ymin=0,
          ]
          \addplot[mark=shockhash,color=colorShockHash,solid] coordinates { (1000000,4.29378) (2000000,4.17851) (5000000,3.97046) (10000000,3.78673) (20000000,3.72787) (50000000,3.61716) (1e+08,3.15956) (2e+08,2.54466) };
          \addplot[mark=flippedTriangle,color=colorBipartiteShockHashFlat,densely dotted] coordinates { (1000000,9.03383) (2000000,8.47793) (5000000,7.98467) (10000000,7.79727) (20000000,7.28226) (50000000,6.75767) (1e+08,5.35045) (2e+08,4.91159) };
          \addplot[mark=triangle,color=colorBipartiteShockHash,solid] coordinates { (1000000,3.88259) (2000000,3.76449) (5000000,3.55181) (10000000,3.43289) (20000000,3.39409) (50000000,3.31631) (1e+08,2.76091) (2e+08,2.37971) };
          \addplot[mark=Mercedes star flipped,color=colorMorphisHashFlat] coordinates { (1000000,10.6423) (2000000,10.2987) (5000000,9.43752) (10000000,9.1391) (20000000,8.91742) (50000000,8.05023) (1e+08,6.17665) (2e+08,5.48607) };
          \addplot[mark=star,color=colorMorphisHash] coordinates { (1000000,4.72333) (2000000,4.6262) (5000000,4.36478) (10000000,4.22565) (20000000,4.1176) (50000000,4.12031) (1e+08,3.4303) (2e+08,2.80269) };
          \addplot[mark=consensus,color=colorConsensus] coordinates { (1000000,7.17952) (2000000,6.6726) (5000000,5.36692) (10000000,4.68955) (20000000,4.24538) (50000000,3.71388) (1e+08,2.4233) (2e+08,1.78846) };

        \end{axis}
    \end{tikzpicture}
    \hspace{8mm}
    \\[-2mm]
    Space consumption $\in (1.7, 2.5]$ bits/key.
        \hfill
    \begin{tikzpicture}
        \begin{axis}[
            plotEvalScalingN,
            xticklabel=\empty,
            ylabel={Throughput (MKeys/s)},
            xmode=log,
            ymin=0,
          ]
          \addplot[mark=x,color=colorChd,solid] coordinates { (1000000,1.55642) (2000000,1.43438) (5000000,1.23254) (10000000,1.15307) (20000000,1.04682) (50000000,0.829518) (1e+08,0.658627) (2e+08,0.554796) };
          \addplot[mark=phobic,color=colorDensePtHash,densely dotted] coordinates { (1000000,1.5163) (2000000,1.49925) (5000000,1.55183) (10000000,1.56482) (20000000,1.58272) (50000000,1.59964) (1e+08,1.6165) (2e+08,1.63671) };
          \addplot[mark=otimes,color=colorPhast] coordinates { (1000000,1.40203) (2000000,1.3934) (5000000,1.38479) (10000000,1.38055) (20000000,1.37774) (50000000,1.34221) (1e+08,1.3303) (2e+08,1.33058) };
          \addplot[mark=oplus,color=colorPhastPlus] coordinates { (1000000,2.28964) (2000000,3.03644) (5000000,3.70462) (10000000,3.98883) (20000000,4.13052) (50000000,3.93515) (1e+08,3.89075) (2e+08,3.84823) };
          \addplot[mark=pentagon,color=colorPthash,solid] coordinates { (1000000,1.14548) (2000000,1.11815) (5000000,1.01916) (10000000,0.925883) (20000000,0.906721) (50000000,0.864588) (1e+08,0.879554) (2e+08,0.605895) };
          \addplot[mark=square,color=colorRecSplit,solid] coordinates { (1000000,0.556483) (2000000,0.575705) (5000000,0.604351) (10000000,0.612895) (20000000,0.614845) (50000000,0.610754) (1e+08,0.610046) (2e+08,0.613495) };
          \addplot[mark=+,color=colorSimdRecSplit,solid] coordinates { (1000000,4.47928) (2000000,4.3956) (5000000,4.3054) (10000000,4.158) (20000000,4.24809) (50000000,4.29037) (1e+08,4.32077) (2e+08,4.35114) };
          \addplot[mark=o,color=colorSicHash,solid] coordinates { (1000000,1.11328) (2000000,1.70164) (5000000,2.49626) (10000000,2.95858) (20000000,3.22243) (50000000,3.334) (1e+08,3.40391) (2e+08,3.40038) };

        \end{axis}
    \end{tikzpicture}
    \hspace{3mm}
    \begin{tikzpicture}
        \begin{axis}[
            plotEvalScalingN,
            xticklabel=\empty,
            ylabel={Throughput (MQueries/s)},
            xmode=log,
            ymin=0,
          ]
          \addplot[mark=x,color=colorChd,solid] coordinates { (1000000,5.15026) (2000000,4.92465) (5000000,4.52557) (10000000,4.26876) (20000000,3.99712) (50000000,3.24802) (1e+08,2.51509) (2e+08,2.34434) };
          \addplot[mark=square,color=colorRecSplit,solid] coordinates { (1000000,7.16204) (2000000,6.99399) (5000000,6.60008) (10000000,6.42385) (20000000,6.17856) (50000000,5.39491) (1e+08,4.24016) (2e+08,4.19569) };
          \addplot[mark=pentagon,color=colorPthash,solid] coordinates { (1000000,15.9791) (2000000,15.1745) (5000000,13.5665) (10000000,12.3553) (20000000,11.4125) (50000000,9.70622) (1e+08,8.0429) (2e+08,6.7325) };
          \addplot[mark=+,color=colorSimdRecSplit,solid] coordinates { (1000000,7.11845) (2000000,6.93225) (5000000,6.53937) (10000000,6.35566) (20000000,6.21813) (50000000,5.02109) (1e+08,4.21408) (2e+08,3.76081) };
          \addplot[mark=o,color=colorSicHash,solid] coordinates { (1000000,10.1338) (2000000,9.61251) (5000000,10.1126) (10000000,9.97208) (20000000,9.74058) (50000000,7.66127) (1e+08,6.52061) (2e+08,5.90667) };
          \addplot[mark=phobic,color=colorDensePtHash,densely dotted] coordinates { (1000000,13.7934) (2000000,12.9889) (5000000,11.924) (10000000,10.7848) (20000000,10.1031) (50000000,8.18777) (1e+08,7.00771) (2e+08,5.96445) };
          \addplot[mark=otimes,color=colorPhast] coordinates { (1000000,57.5374) (2000000,51.9391) (5000000,46.4253) (10000000,45.3926) (20000000,44.4642) (50000000,24.9377) (1e+08,21.3767) (2e+08,21.9974) };
          \addplot[mark=oplus,color=colorPhastPlus] coordinates { (1000000,59.0667) (2000000,54.7246) (5000000,47.8469) (10000000,46.8384) (20000000,45.5581) (50000000,27.3224) (1e+08,22.3214) (2e+08,22.9779) };

        \end{axis}
    \end{tikzpicture}
    \hspace{8mm}
    \\[-2mm]
    Space consumption $\in (2.5, \infty]$ bits/key.
        \hfill
    \begin{tikzpicture}
        \begin{axis}[
            plotEvalScalingN,
            xlabel={Input keys},
            ylabel={Throughput (MKeys/s)},
            xmode=log,
            ymin=0,
          ]
          \addplot[mark=diamond,color=colorBbhash,solid] coordinates { (1000000,9.87654) (2000000,10.5634) (5000000,11.5473) (10000000,11.2803) (20000000,9.65251) (50000000,6.64894) (1e+08,5.95167) (2e+08,5.42962) };
          \addplot[mark=leftTriangle,color=colorBdz,solid] coordinates { (1000000,3.26264) (2000000,2.85171) (5000000,2.3116) (10000000,2.28467) (20000000,2.12642) (50000000,1.83493) (1e+08,1.74213) (2e+08,1.41238) };
          \addplot[mark=rightTriangle,color=colorFch,solid] coordinates { (1000000,1.88058) (2000000,1.66574) (5000000,1.47522) (10000000,1.50387) (20000000,1.10056) (50000000,1.05296) (1e+08,0.475357) (2e+08,0.671344) };
          \addplot[mark=square,color=colorRustFmph,solid] coordinates { (1000000,25.974) (2000000,24.4898) (5000000,23.2558) (10000000,22.3214) (20000000,19.1663) (50000000,15.4799) (1e+08,14.5012) (2e+08,7.53523) };
          \addplot[mark=diamond,color=colorRustFmphGo,solid] coordinates { (1000000,2.81294) (2000000,3.77358) (5000000,4.8844) (10000000,4.78813) (20000000,2.62778) (50000000,2.18895) (1e+08,2.05179) (2e+08,0.874852) };
          \addplot[mark=fingerprint,color=colorFiPS,densely dotted] coordinates { (1000000,15.2091) (2000000,14.4928) (5000000,13.587) (10000000,12.4146) (20000000,13.089) (50000000,13.5099) (1e+08,13.5575) (2e+08,13.9841) };
          \addplot[mark=Mercedes star,color=colorPtrHash] coordinates { (1000000,7.40741) (2000000,7.33496) (5000000,7.28863) (10000000,7.19942) (20000000,7.13521) (50000000,6.92905) (1e+08,7.05766) (2e+08,7.02593) };

        \end{axis}
    \end{tikzpicture}
    \hspace{3mm}
    \begin{tikzpicture}
        \begin{axis}[
            plotEvalScalingN,
            xlabel={Input keys},
            ylabel={Throughput (MQueries/s)},
            xmode=log,
            ymin=0,
          ]
          \addplot[mark=rightTriangle,color=colorFch,solid] coordinates { (1000000,18.8501) (2000000,18.7782) (5000000,18.4888) (10000000,14.9948) (20000000,12.4875) (50000000,10.8554) (1e+08,9.17263) (2e+08,10.02) };
          \addplot[mark=leftTriangle,color=colorBdz,solid] coordinates { (1000000,9.49127) (2000000,9.02093) (5000000,8.65601) (10000000,8.1182) (20000000,5.83465) (50000000,4.61595) (1e+08,4.30256) (2e+08,3.55442) };
          \addplot[mark=diamond,color=colorBbhash,solid] coordinates { (1000000,14.573) (2000000,14.2274) (5000000,14.0213) (10000000,13.2083) (20000000,9.79816) (50000000,7.38771) (1e+08,6.71231) (2e+08,6.3012) };
          \addplot[mark=square,color=colorRustFmph,solid] coordinates { (1000000,19.1571) (2000000,18.5231) (5000000,18.0397) (10000000,17.2414) (20000000,12.9416) (50000000,11.8371) (1e+08,11.3662) (2e+08,10.9481) };
          \addplot[mark=diamond,color=colorRustFmphGo,solid] coordinates { (1000000,22.1214) (2000000,20.8943) (5000000,20.1153) (10000000,19.425) (20000000,13.598) (50000000,11.3404) (1e+08,12.2699) (2e+08,11.7702) };
          \addplot[mark=fingerprint,color=colorFiPS,densely dotted] coordinates { (1000000,24.8973) (2000000,24.2092) (5000000,23.4668) (10000000,19.5274) (20000000,15.083) (50000000,13.1027) (1e+08,10.6474) (2e+08,10.1092) };
          \addplot[mark=Mercedes star,color=colorPtrHash] coordinates { (1000000,52.8914) (2000000,45.5466) (5000000,43.6427) (10000000,43.1406) (20000000,33.6172) (50000000,23.0805) (1e+08,22.2222) (2e+08,21.5579) };

        \end{axis}
    \end{tikzpicture}
    \hspace{8mm}

    \caption{
        Comparison of construction and query performance by number of input keys $n$.
        Note that each approach has configuration parameters giving a wide trade-off between space, construction performance, and query performance.
        For the configurations used here, we refer to \ifAppendix \cref{appendix:s:table}\else the extended version \cite{lehmann2025modern}\fi.
    }
    \label{fig:scalingN}
\end{figure*}
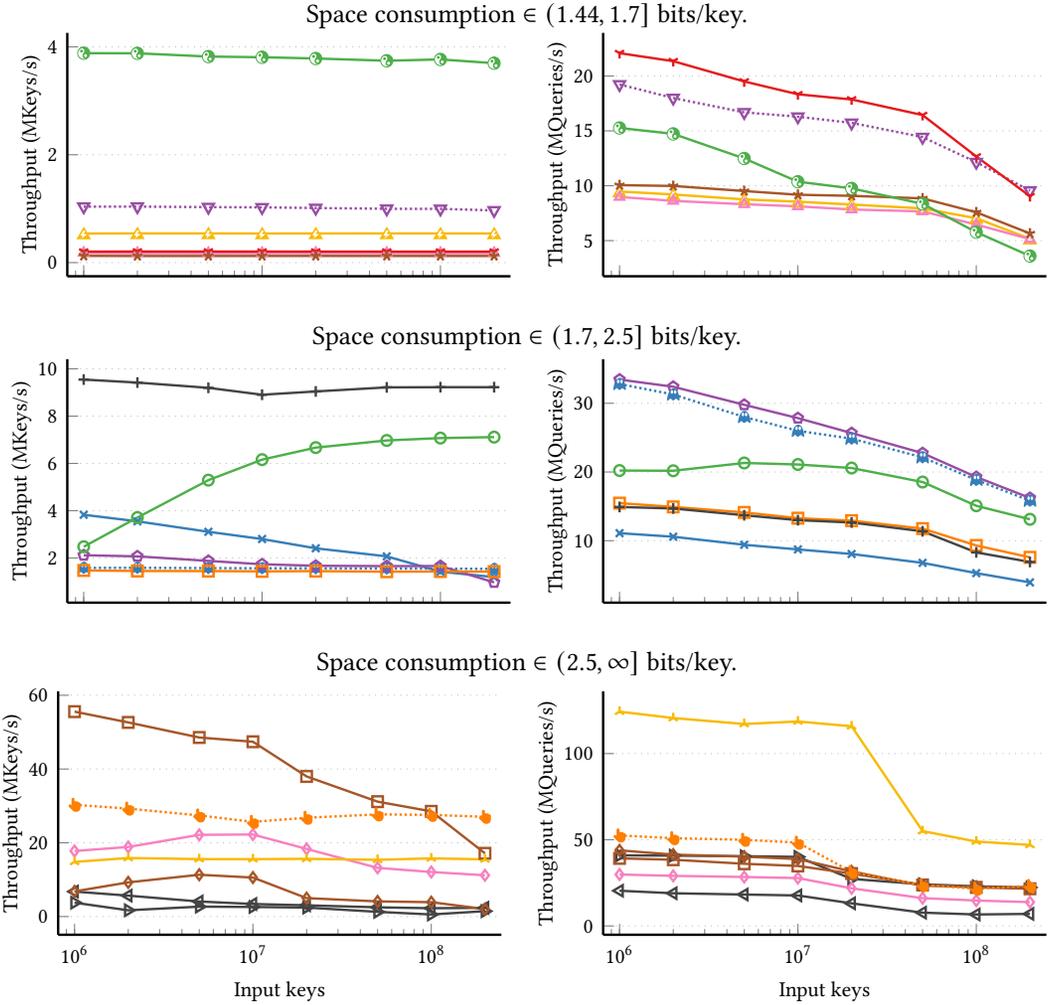

\subsection{Scaling in the Input Size}\label{s:scalingN}
We now look at the construction and query performance of different perfect hash functions, depending on the number of input keys $n$.
The goal is usually to have linear construction time.
To achieve this, it is always possible to partition the input to smaller MPHFs.
However, partitioning comes with a performance penalty, both during construction and queries.
\Cref{fig:scalingN} starts at rather small input sets of 1~million keys and goes all the way up to 200 million.
In case partitions have to be used (e.g., for a parallel implementation), the plot can be used to decide on a partition size.
For each approach, we use a configuration that is typical for it.
Therefore, all competitors have a different space consumption and construction time to avoid using unfair configurations that approaches are not designed for.
Due to the different space consumption, we give different plots for different space consumption.
\ifAppendix
Refer to \cref{appendix:s:table} for the exact configurations we use.
\else
In the extended version \cite{lehmann2025modern} we give the exact configurations.
\fi

\myparagraph{Construction\@.}
\Cref{fig:scalingN} shows how the construction throughput scales in the number of input keys $n$.
The throughput of CHD \cite{belazzougui2009hash}, BBHash \cite{limasset2017fast}, and FMPH \cite{beling2023fingerprinting} decreases most for larger input sizes.
For the approaches based on fingerprinting, the slowdown can be explained by the random accesses to the large bit vector.
FiPS \cite{lehmann2024fast} avoids these cache inefficient access patterns through the use of sorting, making it less sensitive to input size changes.
SicHash and PHast$^+$ have a large startup overhead and become relatively faster to construct with more keys.
The remaining approaches scale almost linearly in $n$.

\myparagraph{Queries\@.}
The query throughput of all approaches in \cref{fig:scalingN} drops when increasing the input size $n$.
This is expected because the data structures get larger and do not fit in cache.
In general, all approaches perform pretty well regarding their query time.
For techniques below 2.5 bits/key, PHast \cite{beling2025phast} by far remains the approach with the fastest queries for the entire range of input sizes.
PtrHash \cite{grootkoerkamp2025ptrhash} comes close for large $n$.
The query performance of PtrHash and PHast drops significantly when passing $10^7$ keys.
This can be explained by the fact that PtrHash queries are close to the RAM throughput.
As the data structure becomes larger than the cache, the throughput drops.
Due to PtrHash needing more space than PHast, it drops earlier.
RecSplit based approaches (RecSplit \cite{esposito2020recsplit}, SIMDRecSplit \cite{bez2023high}, ShockHash-RS \cite{lehmann2023shockhash,lehmann2023bipartite}, MorphisHash \cite{hermann2025morphishash}, \consensus-RS \cite{lehmann2025consensus}) have much slower queries for the entire range of input sizes but offer faster construction.
Over a large range of input sizes, their query performance has a very small slope.
This indicates that they are more limited by the computation than memory access.

\subsection{Multi-Threaded Construction}\label{s:scalingThreads}
Modern processors have many cores available, and testing single-threaded code leaves a lot of processing power unused.
Additionally, most data structures like perfect hash functions are not used in isolation in actual applications.
There are always other processes running on the machines.
Performing multi-threaded measurements can, to a certain extent, account for this.
Note that perfect hashing can be parallelized trivially by partitioning, which many approaches use.
For these approaches, a major factor to their multi-threaded scaling is how efficiently they implement their partitioning step.
This makes it less interesting algorithmically and causes a bias in the measurements.
We still give multi-threaded measurements because they are very relevant in applications.
Additionally, because there are fewer memory channels than threads, the cache locality of approaches is important in parallel measurements.

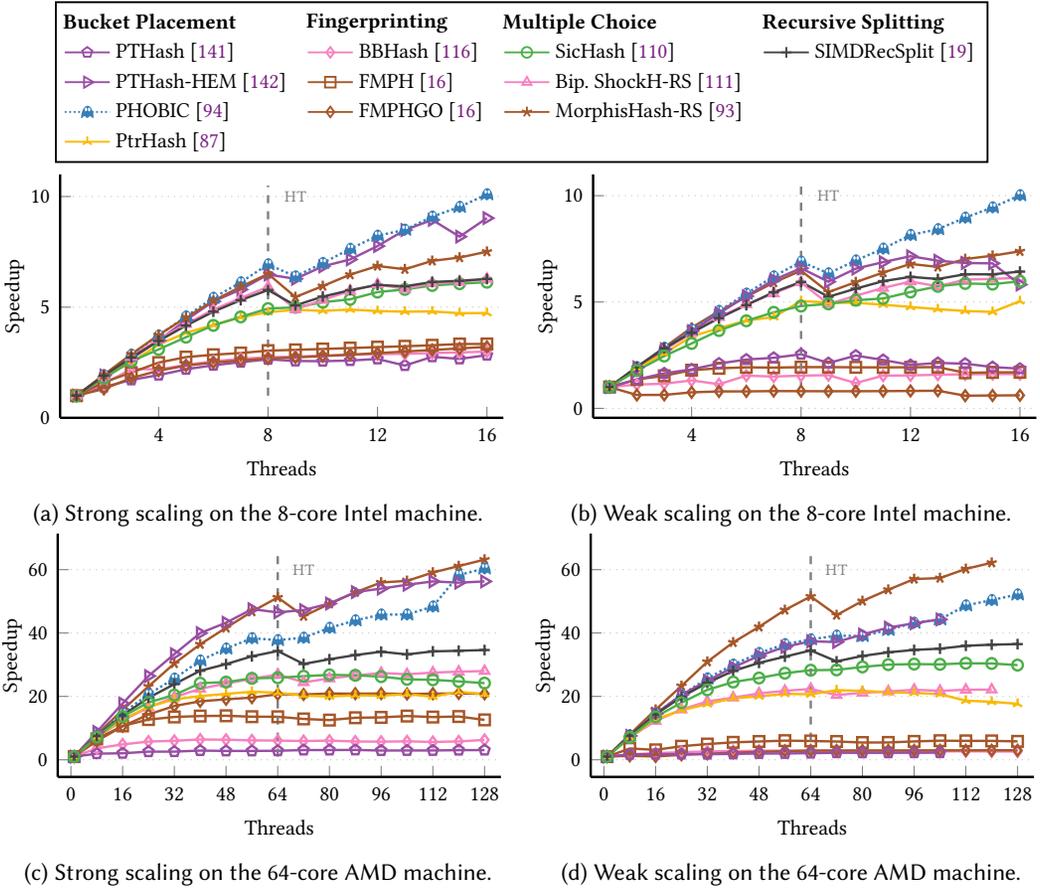
\begin{figure*}[t]
    \centering
    \begin{tikzpicture}
        \pgfplotslegendfromname{legendEvalScaling}
    \end{tikzpicture}
    \vspace{1mm}

    \begin{subfigure}[t]{0.49\textwidth}
        \centering
\begin{tikzpicture}
    \begin{axis}[
        plotEvalScaling,
        legend columns=7,
        transpose legend,
        legend to name=legendEvalScaling,
        ylabel={Speedup},
        xlabel={Threads},
        xtick distance=4,
        ymax=15,
      ]

      \addplot[no marks,draw=none,opacity=0] coordinates { (1,1.0) };
      \addlegendentry{\hspace{-7mm}\textbf{Bucket Placement}}
      \addplot[mark=pentagon,color=colorPthash,solid] coordinates { (1,1.0) (2,1.542) (3,1.96303) (4,2.29593) (5,2.50307) (6,2.75674) (7,3.12261) (8,3.3117) (9,3.34499) (10,3.38488) (11,3.46865) (12,3.32266) (13,3.7359) (14,3.55294) (15,3.87633) (16,3.69739) };
      \addlegendentry{PTHash \cite{pibiri2021pthash}}
      \addplot[mark=rightTriangle,color=colorPthash,solid] coordinates { (1,1.0) (2,2.09736) (3,3.00506) (4,4.13208) (5,5.10634) (6,5.96677) (7,7.05682) (8,7.9606) (9,8.03847) (10,8.98914) (11,9.56566) (12,10.5252) (13,10.9311) (14,11.8275) (15,12.4315) (16,13.0266) };
      \addlegendentry{PTHash-HEM \cite{pibiri2021parallel}}
      \addplot[mark=phobic,color=colorDensePtHash,densely dotted] coordinates { (1,1.0) (2,1.98071) (3,2.90757) (4,3.85575) (5,4.45496) (6,5.66177) (7,6.53111) (8,7.0641) (9,6.69694) (10,8.02588) (11,8.65503) (12,9.41843) (13,9.95147) (14,10.5467) (15,11.1871) (16,11.8281) };
      \addlegendentry{PHOBIC \cite{hermann2024phobic}}
      \addplot[mark=otimes,color=colorPhast] coordinates { (1,1.0) (2,1.91988) (3,2.85786) (4,3.80043) (5,4.67396) (6,5.54895) (7,6.39688) (8,7.22873) (9,6.9267) (10,7.60509) (11,8.23892) (12,8.87621) (13,9.26357) (14,9.85961) (15,10.3757) (16,10.9073) };
      \addlegendentry{PHast \cite{beling2025phast}}
      \addplot[mark=oplus,color=colorPhastPlus] coordinates { (1,1.0) (2,1.80386) (3,2.63721) (4,3.46425) (5,4.27903) (6,5.03817) (7,5.67969) (8,6.34404) (9,6.07728) (10,6.55792) (11,7.01111) (12,7.42566) (13,7.62257) (14,7.98488) (15,8.36978) (16,8.63518) };
      \addlegendentry{PHast${}^+$ \cite{beling2025phast}}
      \addplot[mark=Mercedes star,color=colorPtrHash] coordinates { (1,1.0) (2,1.9797) (3,2.87086) (4,3.69164) (5,4.44343) (6,5.11541) (7,5.70811) (8,6.27007) (9,6.4977) (10,6.74249) (11,6.87026) (12,7.08012) (13,7.26194) (14,7.3679) (15,7.46121) (16,7.77295) };
      \addlegendentry{PtrHash \cite{grootkoerkamp2025ptrhash}}
      \addplot[no marks,draw=none,opacity=0] coordinates { (1,1.0) };
      \addlegendentry{\hspace{-7mm}\textbf{Fingerprinting}}
      \addplot[mark=diamond,color=colorBbhash,solid] coordinates { (1,1.0) (2,1.84823) (3,2.31003) (4,2.83041) (5,3.28973) (6,3.60273) (7,3.97109) (8,4.27217) (9,4.44941) (10,4.66545) (11,4.8099) (12,4.94144) (13,5.12739) (14,5.26544) (15,5.35535) (16,5.48071) };
      \addlegendentry{BBHash \cite{limasset2017fast}}
      \addplot[mark=square,color=colorRustFmph,solid] coordinates { (1,1.0) (2,1.75934) (3,2.46118) (4,3.17443) (5,3.80055) (6,4.34006) (7,4.76715) (8,5.1451) (9,5.3367) (10,5.4988) (11,5.63852) (12,5.76131) (13,5.82966) (14,5.93017) (15,5.99216) (16,6.01311) };
      \addlegendentry{FMPH \cite{beling2023fingerprinting}}
      \addplot[mark=diamond,color=colorRustFmphGo,solid] coordinates { (1,1.0) (2,1.30529) (3,1.79101) (4,2.35506) (5,2.82012) (6,3.32324) (7,3.76882) (8,4.11002) (9,4.26886) (10,4.41392) (11,4.56233) (12,4.6879) (13,4.80821) (14,4.91792) (15,5.02445) (16,5.09007) };
      \addlegendentry{FMPHGO \cite{beling2023fingerprinting}}
      \addplot[no marks,draw=none,opacity=0] coordinates { (1,1.0) };
      \addlegendentry{\phantom{Dummy}}
      \addplot[no marks,draw=none,opacity=0] coordinates { (1,1.0) };
      \addlegendentry{\phantom{Dummy}}
      \addplot[no marks,draw=none,opacity=0] coordinates { (1,1.0) };
      \addlegendentry{\phantom{Dummy}}
      \addplot[no marks,draw=none,opacity=0] coordinates { (1,1.0) };
      \addlegendentry{\hspace{-7mm}\textbf{Multiple Choice}}
      \addplot[mark=o,color=colorSicHash,solid] coordinates { (1,1.0) (2,1.94177) (3,2.89305) (4,3.66964) (5,4.57183) (6,5.23361) (7,6.07612) (8,6.52827) (9,6.99401) (10,7.28166) (11,7.8175) (12,8.38038) (13,8.73624) (14,9.19257) (15,9.36646) (16,9.66711) };
      \addlegendentry{SicHash \cite{lehmann2023sichash}}
      \addplot[mark=triangle,color=colorBipartiteShockHash,solid] coordinates { (1,1.0) (2,1.96207) (3,2.88263) (4,3.78389) (5,4.65286) (6,5.476) (7,6.27321) (8,7.04448) (9,6.21519) (10,6.80554) (11,7.41049) (12,7.9434) (13,8.03783) (14,8.56236) (15,8.90496) (16,9.31621) };
      \addlegendentry{Bip. ShockH-RS \cite{lehmann2023bipartite}}
      \addplot[mark=star,color=colorMorphisHash] coordinates { (1,1.0) (2,1.99648) (3,2.99012) (4,3.98851) (5,4.98907) (6,5.96566) (7,6.95795) (8,7.93707) (9,7.01035) (10,7.77682) (11,8.56753) (12,9.2963) (13,9.52298) (14,10.2569) (15,10.7942) (16,11.4211) };
      \addlegendentry{MorphisHash-RS \cite{hermann2025morphishash}}
      \addplot[no marks,draw=none,opacity=0] coordinates { (1,1.0) };
      \addlegendentry{\phantom{Dummy}}
      \addplot[no marks,draw=none,opacity=0] coordinates { (1,1.0) };
      \addlegendentry{\phantom{Dummy}}
      \addplot[no marks,draw=none,opacity=0] coordinates { (1,1.0) };
      \addlegendentry{\phantom{Dummy}}
      \addplot[no marks,draw=none,opacity=0] coordinates { (1,1.0) };
      \addlegendentry{\hspace{-7mm}\textbf{Recursive Splitting}}
      \addplot[mark=+,color=colorSimdRecSplit,solid] coordinates { (1,1.0) (2,1.96871) (3,2.90405) (4,3.81513) (5,4.68989) (6,5.52338) (7,6.34704) (8,7.09158) (9,6.25277) (10,6.83009) (11,7.38094) (12,7.897) (13,7.95124) (14,8.34378) (15,8.64923) (16,8.96747) };
      \addlegendentry{SIMDRecSplit \cite{bez2023high}}
      \addplot[no marks,draw=none,opacity=0] coordinates { (1,1.0) };
      \addlegendentry{\phantom{Dummy}}
      \addplot[no marks,draw=none,opacity=0] coordinates { (1,1.0) };
      \addlegendentry{\phantom{Dummy}}
      \addplot[no marks,draw=none,opacity=0] coordinates { (1,1.0) };
      \addlegendentry{\phantom{Dummy}}
      \addplot[no marks,draw=none,opacity=0] coordinates { (1,1.0) };
      \addlegendentry{\phantom{Dummy}}
      \addplot[no marks,draw=none,opacity=0] coordinates { (1,1.0) };
      \addlegendentry{\phantom{Dummy}}

      \addplot[color=gray,dashed] coordinates { (8,1) (8,14.5) };
      \node[color=gray] at (axis cs: 9,14) {\tiny HT};
    \end{axis}
\end{tikzpicture}
        \caption{Strong scaling on the 8-core Intel machine.}
        \label{fig:scalingThreadsStrong16}
    \end{subfigure}%
    \hfill
    \begin{subfigure}[t]{0.49\textwidth}
        \centering
\begin{tikzpicture}
    \begin{axis}[
        plotEvalScaling,
        ylabel={Speedup},
        xlabel={Threads},
        xtick distance=4,
        ymax=15,
      ]
      \addplot[mark=diamond,color=colorBbhash,solid] coordinates { (1,1.0) (2,1.44348) (3,1.54746) (4,1.72671) (5,1.91405) (6,2.22412) (7,2.26792) (8,2.34704) (9,2.44991) (10,2.57837) (11,2.97042) (12,3.10456) (13,2.75) (14,2.81977) (15,3.26694) (16,2.92394) };
      \addlegendentry{BBHash \cite{limasset2017fast}}
      \addplot[mark=triangle,color=colorBipartiteShockHash,solid] coordinates { (1,1.0) (2,1.98811) (3,2.90403) (4,3.78442) (5,4.67632) (6,5.48881) (7,6.30164) (8,7.08894) (9,6.23721) (10,6.85321) (11,7.37673) (12,7.98094) (13,8.0878) (14,8.56942) (15,8.95298) (16,9.37817) };
      \addlegendentry{Bip. ShockH-RS \cite{lehmann2023bipartite}}
      \addplot[mark=square,color=colorRustFmph,solid] coordinates { (1,1.0) (2,1.53078) (3,1.86486) (4,2.31447) (5,2.69006) (6,3.03965) (7,3.32645) (8,3.50476) (9,3.60627) (10,3.68) (11,3.73156) (12,3.77307) (13,3.79924) (14,2.8853) (15,2.97286) (16,3.00286) };
      \addlegendentry{FMPH \cite{beling2023fingerprinting}}
      \addplot[mark=diamond,color=colorRustFmphGo,solid] coordinates { (1,1.0) (2,0.702871) (3,0.776314) (4,0.924973) (5,1.06307) (6,1.17974) (7,1.33608) (8,1.45981) (9,1.49837) (10,1.53017) (11,1.56573) (12,1.59565) (13,1.62188) (14,1.08973) (15,1.12926) (16,1.15839) };
      \addlegendentry{FMPHGO \cite{beling2023fingerprinting}}
      \addplot[mark=star,color=colorMorphisHash] coordinates { (1,1.0) (2,2.01719) (3,3.01246) (4,3.99584) (5,5.0101) (6,5.99258) (7,6.99124) (8,7.99703) (9,7.04089) (10,7.82586) (11,8.54757) (12,9.35747) (13,9.59093) (14,10.3072) (15,10.8772) (16,11.5074) };
      \addlegendentry{MorphisHash-RS \cite{hermann2025morphishash}}
      \addplot[mark=phobic,color=colorDensePtHash,densely dotted] coordinates { (1,1.0) (2,2.00312) (3,2.98097) (4,3.96909) (5,4.99378) (6,6.00374) (7,7.00327) (8,7.78421) (9,7.50799) (10,8.31197) (11,9.0125) (12,9.13263) (13,10.4094) (14,11.0043) (15,11.6816) (16,12.3525) };
      \addlegendentry{PHOBIC \cite{hermann2024phobic}}
      \addplot[mark=otimes,color=colorPhast] coordinates { (1,1.0) (2,1.94625) (3,2.86812) (4,3.75769) (5,4.61333) (6,5.41998) (7,6.23159) (8,7.01914) (9,6.73326) (10,7.34408) (11,7.98376) (12,8.59415) (13,8.52386) (14,9.27103) (15,9.86649) (16,10.2906) };
      \addlegendentry{PHast \cite{beling2025phast}}
      \addplot[mark=oplus,color=colorPhastPlus] coordinates { (1,1.0) (2,1.85065) (3,2.67453) (4,3.30441) (5,3.97353) (6,4.6499) (7,5.27987) (8,5.78629) (9,5.5178) (10,5.92312) (11,6.31576) (12,6.6692) (13,6.39488) (14,6.6359) (15,6.29909) (16,7.21865) };
      \addlegendentry{PHast${}^+$ \cite{beling2025phast}}
      \addplot[mark=pentagon,color=colorPthash,solid] coordinates { (1,1.0) (2,1.47356) (3,1.8624) (4,2.12289) (5,2.39814) (6,2.60329) (7,2.99347) (8,2.98018) (9,3.14163) (10,3.05468) (11,3.04862) (12,3.3593) (13,3.238) (14,3.10293) (15,3.04363) (16,3.0061) };
      \addlegendentry{PTHash \cite{pibiri2021pthash}}
      \addplot[mark=rightTriangle,color=colorPthash,solid] coordinates { (1,1.0) (2,1.97131) (3,2.92683) (4,3.83299) (5,4.7619) (6,5.66658) (7,6.58381) (8,7.47913) (9,7.45893) (10,8.11771) (11,8.78995) (12,8.86836) (13,9.07957) (14,9.03798) (15,8.91578) (16,9.1522) };
      \addlegendentry{PTHash-HEM \cite{pibiri2021parallel}}
      \addplot[mark=Mercedes star,color=colorPtrHash] coordinates { (1,1.0) (2,1.95385) (3,2.67454) (4,3.63565) (5,4.21291) (6,4.8367) (7,5.32625) (8,6.35722) (9,6.51451) (10,6.62085) (11,6.75473) (12,6.82018) (13,6.93168) (14,7.03778) (15,7.15432) (16,7.76381) };
      \addlegendentry{PtrHash \cite{grootkoerkamp2025ptrhash}}
      \addplot[mark=+,color=colorSimdRecSplit,solid] coordinates { (1,1.0) (2,2.01084) (3,3.0) (4,3.95248) (5,4.85117) (6,5.67307) (7,6.48387) (8,7.36489) (9,6.49357) (10,7.0962) (11,7.65493) (12,8.23208) (13,8.26027) (14,8.71207) (15,8.97099) (16,9.33979) };
      \addlegendentry{SIMDRecSplit \cite{bez2023high}}
      \addplot[mark=o,color=colorSicHash,solid] coordinates { (1,1.0) (2,1.8875) (3,2.77792) (4,3.53008) (5,4.40566) (6,5.08222) (7,5.76543) (8,6.41007) (9,6.73738) (10,7.02256) (11,7.47744) (12,7.96588) (13,8.33356) (14,8.72121) (15,8.98077) (16,9.219) };
      \addlegendentry{SicHash \cite{lehmann2023sichash}}

      \addplot[color=gray,dashed] coordinates { (8,1) (8,14.5) };
      \node[color=gray] at (axis cs: 9,14) {\tiny HT};

        \legend{}
    \end{axis}
\end{tikzpicture}
        \caption{Weak scaling on the 8-core Intel machine.}
        \label{fig:scalingThreadsWeak16}
    \end{subfigure}

    \caption{
        Multi-threaded construction by number of threads.
        Weak scaling with 10~million keys per thread, strong scaling with 100~million keys.
        We give self-speedups because each approach has a different focus.
    }
    \label{fig:scalingThreads}
\end{figure*}

As a reminder, we run our experiments on an 8 core (16 hardware threads (HT)) processor with 2 memory channels.
In \cref{fig:scalingThreads}, we give parallel measurements using both \emph{weak scaling} and \emph{strong scaling}.
With weak scaling, the number of keys per thread are constant at 10 million.
With strong scaling, the total number of input keys is 100 million and stays the same.
For perfect hashing, strong scaling might be the more important approach, trying to construct a certain perfect hash function as quickly as possible.
However, with strong scaling there might be more threads than it is actually useful.
Weak scaling ensures that each thread always has enough work to do.
Like before, we use the configurations listed in \ifAppendix \cref{appendix:s:table}\else the extended version \cite{lehmann2025modern}\fi.
As such, the absolute construction times are very different, so we only plot self-speedups.
In the following, we first describe approaches with a direct parallelization before looking at the approaches using partitioning.
In addition, \ifAppendix \cref{appendix:multithreaded128}\else the extended version \cite{lehmann2025modern}~\fi gives measurements on a 64 core machine with up to 1.28 billion keys.

\myparagraph{Direct Parallelization\@.}
Some approaches use a parallel implementation of their internal data structures.
An advantage of this technique compared to an external layer of partitioning is that it is transparent to the queries.
Except for PHast \cite{beling2025phast}, this generally does not seem to work well for the approaches that do it.
We see that BBHash \cite{limasset2017fast}, PTHash \cite{pibiri2021parallel}, FMPH \cite{beling2023fingerprinting}, and FMPHGO \cite{beling2023fingerprinting} only achieve a speedup of around 5 when running on 16 threads.
We will later see that PTHash works better with partitioning (i.e., with the approach called PTHash-HEM).
FMPHGO \cite{beling2023fingerprinting} uses a direct parallelization as well, even though it could use internal partitioning \cite{beling2023fingerprinting}.
PHast, in contrast, places all keys of a bucket in a small window of output positions.
By leaving small gaps between different threads, they can work completely independently.
The gaps can be filled afterwards transparently to the queries.

\myparagraph{Internal Partitioning\@.}
Some of the approaches internally partition the input anyway, so in essence they get their parallelization for free.
Note that the scaling behavior of SIMDRecSplit varies strongly depending on the configuration parameters \cite{bez2023high}.
The less space-efficient configurations that it is not actually designed for scale less well because more time is spent partitioning keys to a large number of buckets.
PHOBIC \cite{hermann2024phobic} needs internal partitioning for its interleaved coding.
It scales well with strong scaling and weak scaling, which we attribute to a well implemented partitioning step.
PtrHash \cite{grootkoerkamp2025ptrhash} scales well but does not profit as much from hyperthreading.

\myparagraph{External Partitioning\@.}
Most other approaches parallelize by adding an additional layer of partitioning, which introduces small query and construction time overheads.
PTHash-HEM \cite{pibiri2021parallel} scales best among those.
SicHash \cite{lehmann2023sichash} could theoretically use internal partitioning.
However, the majority of its construction time is spent constructing the BuRR retrieval data structures covering all internal partitions.
While a parallel construction if BuRR is available now \cite{becht2024parallel}, it was not available when SicHash was presented.
It therefore uses an external partitioning step.

\myparagraph{GPU Parallelization\@.}
Even though this is not the focus of this evaluation, we mention two GPU parallel constructions.
PHOBIC-GPU \cite{hermann2024phobic} and GPURecSplit \cite{bez2023high} are, to our knowledge, the only perfect hash functions with a GPU construction.
Both approaches achieve a similar peak construction throughput of about 70~million keys per second on an Nvidia RTX 3090 GPU.
In a reasonable construction time, PHOBIC-GPU can achieve a space consumption of about 1.7 bits per key, while GPURecSplit can achieve about 1.5 bits per key.
The respective query implementations are identical to the CPU versions and can only be used on the CPU, so our measurements from \cref{s:query} show that PHOBIC is much faster to query.
Refer to the PHOBIC paper \cite{hermann2024phobic} for details.

\section{Conclusion}\label{s:conclusion}
In this paper, we surveyed state-of-the-art perfect hash function constructions.
We categorized them by their working principles into approaches based on retrieval, brute-force, and fingerprinting.
Historically, retrieval based methods performed the jump from logarithmic to constant number of bits per key.
For this, they tuned the number of choices per key, the retrieval data structure, load factor, and search techniques.
However, right now, purely retrieval-based methods cannot get below 2 bits per key.
With ShockHash and its variants, the lower bound can be reached by combining it with brute-force.
Several authors implement similar variants of perfect hashing through fingerprinting.
For approaches based on brute-force, PTHash combines the ideas from FCH and CHD to achieve fast queries with good space consumption.
PHOBIC then optimizes this approach by determining an optimal bucket assignment function.
PtrHash and PHast are variants of bucket placement radically optimized for query speed, because of which they dominate a wide range of the trade-off.
Finally, we looked at the more space-efficient approaches.
The simple brute-force construction is made practical through the development of RecSplit, which is the first approach that goes a significant step towards the space lower bound.
SIMDRecSplit, ShockHash, and MorphisHash then tune this to search in a much more structured way.
They finally get beaten again by less structured brute-force again using \consensus, which achieves extremely small space consumption.

\myparagraph{Evaluation\@.}
After explaining the approaches, we performed and discussed a wide range of benchmarks.
We gave plots illustrating the Pareto front of the entire trade-off between construction throughput, query throughput, and space consumption.
In essence, the brute-force approaches focus on showing what is possible in terms of storage space, but sacrifice on construction and query performance.
The most space-efficient approach is \consensus.
Hash function construction through fingerprinting appealed through its simplicity and fast queries but is outperformed by more recent techniques now.
Regarding query time, PHast offers the best performance, followed by PtrHash, both while also achieving rather small space.
In general, over the past few years, perfect hashing has seen a large number of algorithmic improvements, leading to massive speedups and enabling larger and larger input sets.

\myparagraph{Future Work\@.}
It is reasonable to assume that approaches that are more space-efficient are also slower to construct.
Future work will likely give additional trade-offs and achieve even smaller space consumption.
Part of this might also be enabled by additional GPU accelerated constructions.
Current approaches have the tendency that the most space-efficient approaches are slower to query.
There is no reason to think that this is a fundamental limitation, especially since smaller data structures should be more cache-efficient.
We believe that in the future, it will be possible to construct MPHFs that are very close to the space lower bound, while still achieving fast queries.
Given the large progress over the 28 years since the last survey, we are excited for the breakthroughs that the next decades might bring.

\begin{acks}
    This paper is based on and has text overlaps with the doctoral thesis of Hans-Peter Lehmann \cite{lehmann2024fast}.
    We would like to thank the anonymous reviewers for their valuable feedback.

    \paragraph{Funding}
    H.-P.L., P.S., and S.W.: This project has received funding from the European Research Council (ERC) under the European Union’s Horizon 2020 research and innovation programme (grant agreement No. 882500).
    This work was also supported by funding from the pilot program Core Informatics at KIT (KiKIT) of the Helmholtz Association (HGF).
    R.P.: Rasmus Pagh is part of BARC, supported by the VILLUM Foundation grant 54451.
    G.E.P. and S.V.: This work is also partially supported by the project ``SEcurity and RIghts In the CyberSpace - SERICS'' (PE00000014 - CUP H73C2200089001) under the National Recovery and Resilience Plan (NRRP) funded by the European Union - NextGenerationEU.
    Views and opinions expressed are
    those of the authors only and do not necessarily reflect those of the
    European Union or the Italian MUR. Neither the European Union nor the
    Italian MUR can be held responsible for them.

    \includegraphics[width=2cm]{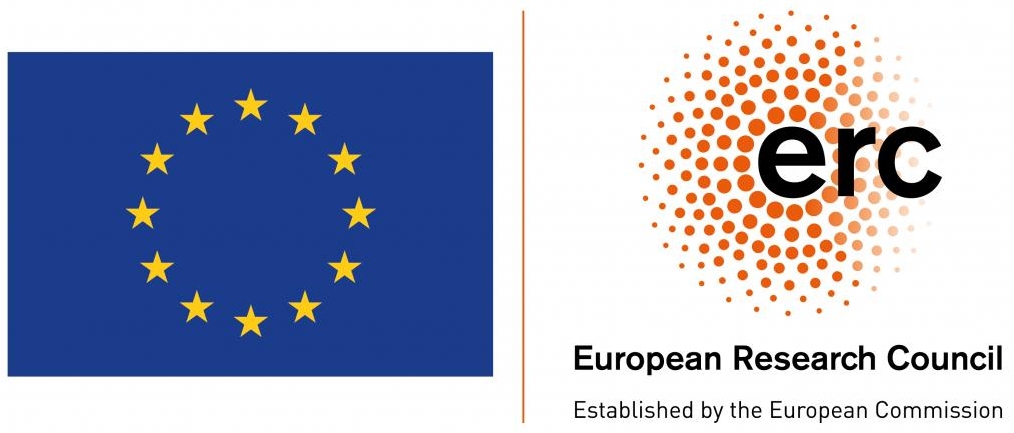}
\end{acks}

\bibliographystyle{ACM-Reference-Format}
\bibliography{paper}

\ifAppendix

\appendix
\newpage
\crefalias{section}{appendix}

\begin{center}
    \large\textbf{Appendix}
\end{center}

\begin{figure}[b]
  \centering
  \includegraphics[scale=0.9]{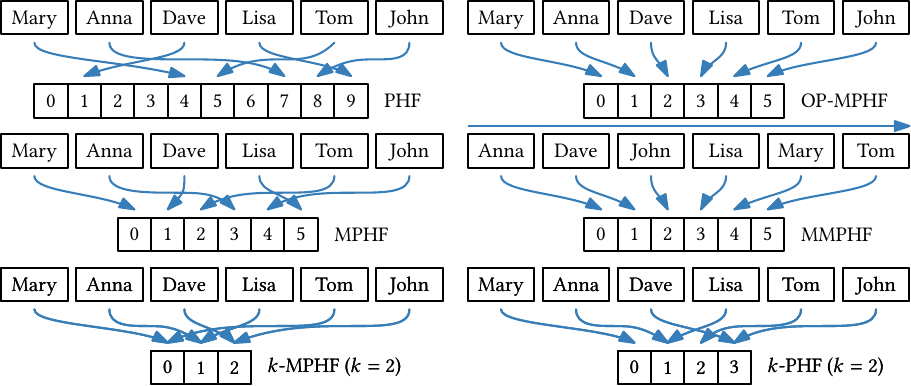}
  \caption{%
      Illustrations of different variants of perfect hashing.
  }
  \label{fig:variants}
\end{figure}

\section{Variants}\label{appendix:s:variants}
There are several variants of perfect hashing, using a different set of techniques to solve, so we only mention them here but do not go into detail.
We give an illustration in \cref{fig:variants}.
A \emph{$k$-perfect hash function} is a generalization of perfect hashing where each output value allows for at most $k$ collisions.
Minimal $k$-perfect hashing has a space lower bound of about $n (\log_2(e) + \log_2(k!/k^k)/k)$ bits \cite{belazzougui2009hash,mairson1983program}.
A similar bound for non-minimal $k$-perfect hashing is not known \cite{belazzougui2009hash}.

An \emph{order-preserving} MPHF maps an arbitrarily ordered list of keys to their position in the
list~\cite{fox1991order}. The name is somewhat misleading as “order” usually
implies an order on the universe (e.g., lexicographical order for strings).
This comes with an expected space consumption of at least $\log_2(n!)/n\approx \log_2 n -
\log_2 e$ bits per key because it needs to differentiate between all $n!$
permutations of the set.
It is simple to get within $n\log_2 e$ of this bound by constructing an MPHF and using it to index a packed array of explicitly stored positions.
We can get almost optimal space consumption by storing the positions in a retrieval data structure, which we explain in \cref{s:retrieval}.

A \emph{monotone} MPHF (MMPHF)~\cite{belazzougui2009monotone} preserves
an \emph{intrinsic} order of the keys implied by an order on the universe. For
example, with a set of strings, the lexicographically first string would have
the hash value 0, and so on. In other words, an MMPHF can answer rank queries on
the input set, but without storing the set, and giving undefined results for
keys not in the set. Relying on the natural order makes MMPHFs much more
space-efficient than order-preserving minimal perfect hash functions. There are
constructions using as few as $\Oh{\log \log \log u}$ bits per key
\cite{belazzougui2009monotone}; recently, the tight bound $\Theta\left(n \min\left\{\log_2\log_2 \frac un, \log_2 n\right\}\right)$
was proven for $u\geq (1+ \varepsilon)n$ with arbitrary $\varepsilon > 0$~\cite{assadi2023tight,kosolobov2024simplified}.
Depending on the input distribution, \emph{learned} MMPHFs can get below these worst-case lower bounds \cite{ferragina2023learned}.
We refer to
\cite{belazzougui2011theoryPractice} for a survey on monotone MPHFs.

\label{appendix:s:dynamicPerfectHashing}
Finally, there has been work on dynamic perfect hashing where the key set can be changed while still avoiding collisions \cite{mortensen2005dynamic,demaine2006dictionariis}. In that case $\Theta(n \log \log u)$ bits are necessary and sufficient \cite{mortensen2005dynamic}.
If the key set is available during updates, the problem is related to augmented open addressing for which tight bounds are known \cite{AugmentedOpenAddressingUpperBound,CellProbeLowerBound}. There is also an approach that avoids collisions as much as possible and rebuilds the perfect hash function if the key set changes too much \cite{lu2006perfect}.
Despite the name ``dynamic perfect hashing'', Dietzfelbinger \etal \cite{dietzfelbinger1994dynamic} actually explain a dynamic set data structure and not a space-efficient perfect hash function.
In this survey, however, we only consider the static case.

\section{The Birth of Perfect Hashing}\label{appendix:s:birth}
In this section, we give a short overview of the origins of perfect hashing.
These approaches already introduce a number of basic principles that are still used in modern constructions, some of which we have explained in \cref{s:commonTools}.
Even if there are faster and more space-efficient approaches today, they illustrate the progress that was made on the topic during the last 50 years of research.

In 1963, for identifying reserved words in an assembler, Grenievski and Turski \cite{greniewski1963external} present a function that can convert tokens to integers without collisions.
Given an input text $T$, their hash function has the form $f_i(T) = a_i f_{i-1}(T) + T[i] + c_i$.
The idea is to determine constants $a_i$ and $c_i$ experimentally such that $f_n$ avoids collisions on the input set.
However, the authors do not describe a way to generalize the idea to arbitrary input sets nor do they call the resulting data structure a \emph{perfect hash function}.
In the first edition of The Art of Computer Programming \cite{knuth1998art} released in 1973, Knuth describes finding a hash function without collisions as an ``amusing puzzle''.
Knuth states that the puzzle can be solved manually in about one day of work if the number of input keys is small enough and gives an example with $n=31$ keys.
He uses this as an introduction to a chapter about hash tables and not as an independent field of research.
The second edition \cite{knuth1998art} of his book released in 1998 then already describes the first practical perfect hash function constructions.

It was Sprugnoli \cite{sprugnoli1977perfect} who first used the terms \emph{perfect hashing} and \emph{minimal perfect hashing} in 1977.
He also describes the first algorithm to systematically construct perfect hash functions.
The idea is to find a linear transformation of the input keys, i.e., $x \mapsto \lfloor (x+c_1)/c_2 \rfloor$, where $c_1$ and $c_2$ are constants determined by the algorithm.
This approach only works well if the keys are uniformly distributed.
To deal with this problem, Sprugnoli then proposes to scramble the input using a modulo operation.
While this does not work for all input sets, it gives a fundamental basis that is still used in many modern perfect hash functions.
Sprugnoli also introduces the ideas of initial hashing, as well as \emph{partitioning} (see \cref{s:commonTools}).
He even already discusses the space usage of his perfect hash functions in terms of machine code that needs to be written for representing the function.
Today, perfect hashing is mainly measured as the amount of space needed in a corresponding data structure, but measuring machine code was common for a long time \cite{schmidt2000gperf}.

Cichelli \cite{cichelli1980minimal} describes a first practical algorithm for determining \emph{minimal} perfect hash functions based on brute-force searching for a simple assignment of letters to numbers.
Similar letter-based approaches are presented later, with the main innovations being to look at letters in different positions in the input string which are less likely to be correlated.
Jaeschke \cite{jaeschke1981reciprocal} gives an algorithm that can already handle input sets of up to 1000 keys.

\begin{table}[pt]
  \caption{%
      Selected configurations for all competitors.
      Measurements are all single-threaded and for $n=100$~million keys.
      We use the configurations marked with $^*$ in \cref{s:scalingN,s:scalingThreads}.
      The \emph{cache} column gives the average number of L1 cache misses per query, excluding loading the queried keys.
  }
  \label{tab:parametersOverview}
  \centering
  \scalebox{0.9}{
    
\addtolength\tabcolsep{-0.5pt}
\begin{centering}
\begin{tabular}[t]{lll rrrr}
    \toprule
    & Approach & Configuration & Space    & Construction & \multicolumn{2}{c}{Query} \\ \cline{6-7}
    &          &               & bits/key & ns/key       & \hspace{5mm}ns & cache \\ \midrule

    \rot{12}{Bucket Placement}
                                   &\mr{FCH}& $c$=$7.0$$^*$ & 7.000 &  1\,100 &  97 &  1.3 \\
                                               && $c$=$3.0$ & 3.000 & 46\,240 &  91 &  1.3 \\ \cline{2-7}
                               &\mr{CHD}& $\lambda$=$4$$^*$ & 2.166 &  1\,522 & 356 & 11.4 \\
                                           && $\lambda$=$6$ & 2.007 & 18\,336 & 346 & 11.5 \\ \cline{2-7}
        &\mr{PTHash}& $\lambda$=$4.0$, $\alpha$=$0.99$, C-C & 3.342 &     516 &  76 &  1.3 \\
                && $\lambda$=$6.0$, $\alpha$=$0.95$, EF$^*$ & 2.150 &  1\,125 & 119 &  4.6 \\ \cline{2-7}
    &\mr{PTHash-HEM}& $\lambda$=$4.0$, $\alpha$=$0.99$, C-C & 3.167 &     528 &  80 &  1.3 \\
                && $\lambda$=$6.0$, $\alpha$=$0.95$, EF$^*$ & 2.150 &  1\,137 & 126 &  4.6 \\ \cline{2-7}
       &\mr{PHOBIC}& $\lambda$=$4.0$, $\alpha$=$1.0$, IC, C & 2.966 &     297 &  87 &  2.3 \\
           && $\lambda$=$6.0$, $\alpha$=$1.0$, IC, Rice$^*$ & 2.076 &     603 & 123 &  7.4 \\ \cline{2-7}
                  &\mr{PHast}& $S$=7, $\lambda$=3.7, EF$^*$ & 1.998 &     750 &  43 &  1.1 \\
                               && $S$=11, $\lambda$=6.3, EF & 1.848 &  5\,944 &  44 &  1.1 \\ \cline{2-7}
     &\mr{PHast$^+$}& $\delta$=2, $S$=8, $\lambda$=4.35, EF & 2.074 &     181 &  38 &  1.1 \\
               && $\delta$=1, $S$=11, $\lambda$=6.6, EF$^*$ & 1.853 &     256 &  41 &  1.1 \\ \cline{2-7}
            &\mr{PtrHash}& $\lambda$=$3.0$, linear, vec$^*$ & 2.990 &     141 &  45 &  1.0 \\
                              && $\lambda$=$4.0$, cubic, EF & 2.118 &     358 &  47 &  1.0 \\
    \midrule

    \rot{8}{Fingerprinting}
                   &\mr{BBHash}& $\gamma$=$5.0$$^*$ & 6.870 & 157 & 127 & 4.0 \\
                                  && $\gamma$=$1.5$ & 3.293 & 273 & 173 & 7.4 \\ \cline{2-7}
                     &\mr{FMPH}& $\gamma$=$5.0$$^*$ & 6.298 &  69 &  88 & 3.3 \\
                                  && $\gamma$=$1.5$ & 3.013 & 103 & 139 & 5.7 \\ \cline{2-7}
    &\mr{FMPHGO}& $\gamma$=$5.0, s$=$4, b$=$16$$^*$ & 6.426 & 490 &  84 & 4.1 \\
                   && $\gamma$=$1.5, s$=$4, b$=$16$ & 2.435 & 457 & 106 & 6.1 \\ \cline{2-7}
    &\mrf{FiPS}& $\gamma$=$5.0$, LS=256, Off=32$^*$ & 6.980 &  72 &  82 & 2.4 \\
                  && $\gamma$=$1.5$, LS=256, Off=16 & 3.119 & 103 & 123 & 5.5 \\
    \midrule

    \rot{14}{Multiple Choice}
                                  &\mr{BPZ}& $c$=$1.25$, $b$=$3$$^*$ & 7.500 &      512 & 235 &  4.2 \\
                                              && $c$=$1.25$, $b$=$6$ & 3.125 &      514 & 219 &  4.8 \\ \cline{2-7}
         &\mrf{SicHash}& $\alpha$=$0.95$, $p_1$=$37$, $p_2$=$44$$^*$ & 2.197 &      285 & 145 &  4.9 \\
                          && $\alpha$=$0.97$, $p_1$=$45$, $p_2$=$31$ & 2.080 &      349 & 140 &  4.4 \\ \cline{2-7}
                    &\mrf{ShockHash-RS}& $\ell$=$40$, $b$=$2000$$^*$ & 1.551 &   4\,128 & 320 & 10.2 \\
                                          && $\ell$=$55$, $b$=$2000$ & 1.526 & 109\,921 & 314 &  9.9 \\ \cline{2-7}
               &\mrf{Bip. ShockHash-RS}& $\ell$=$64$, $b$=$2000$$^*$ & 1.525 &  14\,214 & 352 & 10.0 \\
                                         && $\ell$=$128$, $b$=$2000$ & 1.489 & 371\,678 & 296 &  9.6 \\ \cline{2-7}
                         &\mrf{Bip. ShockHash-Flat}& $\ell$=$64$$^*$ & 1.618 &   2\,202 & 180 &  3.9 \\
                                                     && $\ell$=$100$ & 1.547 &  19\,716 & 174 &  4.1 \\ \cline{2-7}
       &\mrf{MorphisHash}& $\ell$=$40$, $b$=$2000$, $\beta$=$\ell$-4 & 1.522 &  11\,259 & 307 &  9.8 \\
                    && $\ell$=$64$, $b$=$2000$, $\beta$=$\ell$-4$^*$ & 1.489 &  19\,255 & 306 &  8.8 \\ \cline{2-7}
    &\mrf{MorphisHash-Flat}& $\ell$=$60$, $eb$=$3$, $\beta$=$\ell$-2 & 1.611 &   3\,228 & 167 &  3.5 \\
                      && $\ell$=$84$, $eb$=$3$, $\beta$=$\ell$-2$^*$ & 1.547 &  10\,586 & 156 &  3.5 \\
    \midrule

    \rot{6}{Rec. Splitting}
                    &\mr{RecSplit}& $\ell$=$8$, $b$=$100$ & 1.793 &   1\,629 & 212 & 11.5 \\
                           && $\ell$=$14$, $b$=$2000$$^*$ & 1.584 & 290\,374 & 252 & 11.7 \\ \cline{2-7}
               &\mrf{SIMDRecSplit}& $\ell$=$8$, $b$=$100$ & 1.809 &      233 & 242 &  9.7 \\
                           && $\ell$=$16$, $b$=$2000$$^*$ & 1.560 & 277\,118 & 270 & 10.5 \\ \cline{2-7}
    &\mrf{\consensus}& $k$=$512$, $\varepsilon$=$0.1$$^*$ & 1.579 &      575 & 358 & 11.4 \\
                   && $k$=$32768$, $\varepsilon$=$0.0005$ & 1.444 &  71\,301 & 445 & 16.0 \\
    \bottomrule
\end{tabular}
\end{centering}

  }
\end{table}

\section{Selected Configurations}\label{appendix:s:table}
While the Pareto fronts give a good overall picture, they do not provide a guide on how to select the configuration parameters.
In \cref{tab:parametersOverview}, we take the two most promising configurations of each competitor from the Pareto fronts.
We give their construction time, query time, and space consumption.
The table therefore suggests configurations that we recommend for use in an application.
The number of cache misses depends on multiple factors, like the cache line size, memory bandwidth, and whether the fetches can be predicted.
Therefore, the number of cache misses during a query does not necessarily correspond directly to the query performance.
We still include the average number of L1 data cache misses in the table, given as the measured number minus one (due to the use of string input keys).
Remember from \cref{evaluation:s:setup} that the sizes of the L1 and L2 data caches are 48 KiB and 512 KiB per core, and the L3 cache has a size of 16 MiB.
We use the configurations marked with the $^*$ symbol for our evaluations in \cref{s:scalingN,s:scalingThreads}.
In \cref{appendix:smallerInput}, we give the table for just 10\,000 input keys to illustrate the lower order terms.

We start with approaches based on multiple-choice hashing.
SicHash is faster than BPZ in construction and queries, while also achieving better space consumption.
ShockHash-RS is then able to achieve $1.52$ bits per key, reducing the gap to the space lower bound of $\approx 1.443$ bits per key by about $30\%$.
Finally, bipartite ShockHash-RS reduces the space consumption to just $1.489$ bits per key, which is within $3.3$\% of the lower bound with practically feasible construction time.
Using a single CPU thread, bipartite ShockHash-RS achieves a space consumption better than what was previously only achieved using thousands of threads on a GPU \cite{bez2023high}.
Using the same space consumption, MorphisHash improves the ShockHash construction time by 50\%--100\%.
The flat versions of ShockHash and MorphisHash cause significantly fewer cache misses than their counterpart with splitting trees.

We next look at the approaches based on bucket placement.
The partitioned implementation of PTHash, PTHash-HEM, has the same construction time and space consumption when run on a single thread.
However, its queries are about 10\% slower.
Even though PHOBIC uses partitioning as well, with compact coding it loses less query time compared to PTHash.
By using Golomb-Rice coding instead of Elias-Fano coding, PHOBIC improves the query times of PTHash.
FCH and CHD are classical approaches that cannot directly compete with more recent constructions --- FCH is slower to construct and CHD is slower to query.
PtrHash offers, by far, the fastest queries in the table.
At the same time, it has very fast construction and decent space consumption.
This makes PtrHash a good choice optimizing the overall trade-off.
The approaches based on bucket placement achieve the smallest number of cache misses, with FCH, PHast, PtrHash, and the fixed-length coding of PTHash being very close to a single cache miss per query.
PtrHash achieves this even for its Elias-Fano coded seeds through its cache-efficient variant.

We now give a selection of configurations using tree-based search.
Comparing different approaches where each is given about half an hour of construction time, RecSplit is able to produce a perfect hash function with $1.58$ bits per key.
RecSplit starts a chain of approaches that further reduce the space consumption.
SIMDRecSplit improves the space consumption to $1.56$ bits per key.
Then ShockHash goes down to $1.489$ and finally \consensus achieves $1.444$ bits per key --- within 0.1\% of the lower bound.
The most space-efficient variant of \consensus has the largest number of cache misses due to it storing each layer of the tree in an independent data structure.

For perfect hashing through fingerprinting, we give $\gamma=1.5$ and $\gamma=5.0$ for all competitors.
FMPH achieves the best space consumption among those.
FiPS, with its interleaved rank data structure, needs about 0.1 bits per key more space for the smaller configuration.
BBHash needs an additional 0.17 bits per key.
Looking at the construction time and query time, the same configuration with FMPH is consistently faster to construct and query.
This is partially due to its updated implementation compared to the original paper.
Through a small number of retries, FMPHGO achieves much lower space consumption but also much slower construction.
Especially for large values of $\gamma$, the interleaved select data structure in FiPS leads to a significant reduction in cache misses.
However, note that FMPHGO achieves a similar query time even though it has more cache misses.

\begin{table}[pt]
  \caption{%
      Selected configurations for all competitors.
      Measurements are all single-threaded and for 10\,000 input keys to illustrate lower order terms.
      The \emph{cache} column gives the average number of L1 cache misses per query, excluding loading the queried keys.
      Refer to \cref{tab:parametersOverview} for the main comparison with a larger input set.
  }
  \label{tab:parametersOverview10k}
  \centering
  \scalebox{0.9}{
    
\addtolength\tabcolsep{-0.5pt}
\begin{centering}
\begin{tabular}[t]{lll rrrr}
    \toprule
    & Approach & Configuration & Space    & Construction & \multicolumn{2}{c}{Query} \\ \cline{6-7}
    &          &               & bits/key & ns/key       & \hspace{5mm}ns & cache \\ \midrule

    \rot{12}{Bucket Placement}
                                       &\mr{FCH}& $c$=$7.0$ &  7.072 &     233 &  49 & 0.2 \\
                                               && $c$=$3.0$ &  3.072 & 45\,000 &  49 & 0.2 \\ \cline{2-7}
                                   &\mr{CHD}& $\lambda$=$4$ &  2.240 &     548 & 180 & 0.2 \\
                                           && $\lambda$=$6$ &  2.096 &  5\,900 & 187 & 0.2 \\ \cline{2-7}
        &\mr{PTHash}& $\lambda$=$4.0$, $\alpha$=$0.99$, C-C &  2.989 &     368 &  42 & 0.2 \\
                    && $\lambda$=$6.0$, $\alpha$=$0.95$, EF &  2.397 &     743 &  60 & 0.2 \\ \cline{2-7}
    &\mr{PTHash-HEM}& $\lambda$=$4.0$, $\alpha$=$0.99$, C-C &  3.277 &     393 &  42 & 0.2 \\
                    && $\lambda$=$6.0$, $\alpha$=$0.95$, EF &  2.448 &     831 &  63 & 0.2 \\ \cline{2-7}
    &\mr{PHOBIC}& $\lambda$=$6.0$, $\alpha$=$1.0$, IC, Rice & 62.093 &     326 &  62 & 4.3 \\
                  && $\lambda$=$4.0$, $\alpha$=$1.0$, IC, C & 39.654 &     250 &  48 & 1.3 \\ \cline{2-7}
                      &\mr{PHast}& $S$=7, $\lambda$=3.7, EF &  2.582 & 60\,700 &  17 & 0.0 \\
                               && $S$=11, $\lambda$=6.3, EF &  2.371 &  5\,900 &  17 & 0.0 \\ \cline{2-7}
     &\mr{PHast$^+$}& $\delta$=1, $S$=11, $\lambda$=6.6, EF &  2.989 & 44\,500 &  23 & 0.0 \\
                   && $\delta$=2, $S$=8, $\lambda$=4.35, EF &  2.811 & 44\,300 &  19 & 0.0 \\ \cline{2-7}
                &\mr{PtrHash}& $\lambda$=$3.0$, linear, vec &  3.159 &     147 &  17 & 0.0 \\
                              && $\lambda$=$4.0$, cubic, EF &  2.329 & 44\,300 &  20 & 0.0 \\
    \midrule

    \rot{8}{Fingerprinting}
                   &\mr{BBHash}& $\gamma$=$5.0$$^*$ & 7.475 &  2\,100 & 63 & 0.2 \\
                                  && $\gamma$=$1.5$ & 3.878 &  2\,800 & 77 & 0.2 \\ \cline{2-7}
                     &\mr{FMPH}& $\gamma$=$5.0$$^*$ & 6.381 & 60\,000 & 46 & 0.0 \\
                                  && $\gamma$=$1.5$ & 3.174 &      60 & 63 & 0.0 \\ \cline{2-7}
    &\mr{FMPHGO}& $\gamma$=$5.0, s$=$4, b$=$16$$^*$ & 6.528 & 44\,100 & 42 & 0.0 \\
                   && $\gamma$=$1.5, s$=$4, b$=$16$ & 2.560 & 44\,100 & 52 & 0.0 \\ \cline{2-7}
    &\mrf{FiPS}& $\gamma$=$5.0$, LS=256, Off=32$^*$ & 7.194 &      68 & 35 & 0.2 \\
                  && $\gamma$=$1.5$, LS=256, Off=16 & 3.360 &      96 & 56 & 0.2 \\
    \midrule

    \rot{14}{Multiple Choice}
                                 &\mr{BPZ}& $c$=$1.25$, $b$=$3$$^*$ & 7.519 &      199 &  92 & 0.2 \\
                                             && $c$=$1.25$, $b$=$6$ & 3.145 &      200 & 100 & 0.2 \\ \cline{2-7}
        &\mrf{SicHash}& $\alpha$=$0.95$, $p_1$=$37$, $p_2$=$44$$^*$ & 2.441 &      652 &  82 & 0.2 \\
                         && $\alpha$=$0.97$, $p_1$=$45$, $p_2$=$31$ & 2.330 &  21\,400 &  82 & 0.2 \\ \cline{2-7}
                   &\mrf{ShockHash-RS}& $\ell$=$40$, $b$=$2000$$^*$ & 1.803 &   4\,300 & 206 & 0.2 \\
                                         && $\ell$=$55$, $b$=$2000$ & 1.776 & 108\,900 & 200 & 0.2 \\ \cline{2-7}
              &\mrf{Bip. ShockHash-RS}& $\ell$=$64$, $b$=$2000$$^*$ & 1.770 &  15\,300 & 231 & 0.2 \\
                                        && $\ell$=$128$, $b$=$2000$ & 1.738 & 405\,800 & 180 & 0.2 \\ \cline{2-7}
                        &\mrf{Bip. ShockHash-Flat}& $\ell$=$64$$^*$ & 2.367 &   2\,300 & 105 & 0.2 \\
                                                    && $\ell$=$100$ & 2.283 &  18\,500 &  98 & 0.2 \\ \cline{2-7}
      &\mrf{MorphisHash}& $\ell$=$40$, $b$=$2000$, $\beta$=$\ell$-4 & 1.525 &  12\,100 & 190 & 0.2 \\
                   && $\ell$=$64$, $b$=$2000$, $\beta$=$\ell$-4$^*$ & 1.499 &  21\,500 & 182 & 0.2 \\ \cline{2-7}
    &\mr{MorphisHash-Flat}& $\ell$=$60$, $eb$=$3$, $\beta$=$\ell$-2 & 2.127 &   5\,500 &  91 & 0.2 \\
                     && $\ell$=$84$, $eb$=$3$, $\beta$=$\ell$-2$^*$ & 2.043 &   9\,600 &  87 & 0.2 \\
    \midrule

    \rot{6}{Rec. Splitting}
                    &\mr{RecSplit}& $\ell$=$8$, $b$=$100$ & 1.978 &   1\,500 & 122 & 0.2 \\
                           && $\ell$=$14$, $b$=$2000$$^*$ & 1.781 & 295\,600 & 153 & 0.2 \\ \cline{2-7}
               &\mrf{SIMDRecSplit}& $\ell$=$8$, $b$=$100$ & 2.004 &      264 & 141 & 0.2 \\
                           && $\ell$=$16$, $b$=$2000$$^*$ & 1.788 & 282\,200 & 156 & 0.2 \\ \cline{2-7}
    &\mrf{\consensus}& $k$=$512$, $\varepsilon$=$0.1$$^*$ & 2.150 &      561 & 123 & 0.2 \\
                   && $k$=$32768$, $\varepsilon$=$0.0005$ & 3.640 &      240 & 117 & 0.2 \\
    \bottomrule
\end{tabular}
\end{centering}

  }
\end{table}

\section{Experiments with Smaller Input Sets}\label{appendix:smallerInput}
\Cref{tab:parametersOverview10k} gives the overview table from \cref{tab:parametersOverview} with a smaller input set of just 10\,000 keys.
This can help to get a feeling for the lower order terms involved in the space consumption and construction time.
Additionally, it gives details on the query performance when the entire data structure fits into the cache.
This is also illustrated by all approaches causing only 0--0.2 L1 cache misses on average, except for PHOBIC, which has higher constant space overhead.
For most competitors, the space consumption stays almost the same.
The only exception is PHOBIC \cite{hermann2024phobic}, which needs significantly more space.
This is caused by its interleaved coding of seeds, which stores a dedicated encoder for each bucket in all partitions.
Using just about 4 partitions with the small input set then has considerable space overhead.
This can be avoided by selecting a different encoding.
Even if the entire data structure fits into the cache, PtrHash \cite{grootkoerkamp2025ptrhash} keeps by far the fastest queries.

\section{Multi-Threaded Construction With 128 Threads}\label{appendix:multithreaded128}
In \cref{fig:scalingThreads128}, we give the multi-threaded construction performance of \cref{fig:scalingThreads} on a larger machine.
The machine is equipped with an AMD EPYC 7702P processor with 64 cores (128 hardware threads), pinned to its base clock speed of 2.0 GHz.
It runs Rocky Linux 9.5 with Linux 5.14.0 and supports AVX2, having 8 memory channels.
For comparison, the 8-core Intel machine we use in most other experiments has 2 memory channels.
With weak scaling, the number of keys per thread are constant at 10 million,
so the total input size increases gradually up to a total of 1.28 billion using 128 threads.

\begin{figure*}[t]
    \centering
    \begin{tikzpicture}
        \pgfplotslegendfromname{legendEvalScaling}
    \end{tikzpicture}
    \vspace{1mm}

    \begin{subfigure}[t]{0.49\textwidth}
        \centering
\begin{tikzpicture}
    \begin{axis}[
        plotEvalScaling,
        ylabel={Speedup},
        xlabel={Threads},
        xtick distance=16,
      ]
      \addplot[mark=diamond,color=colorBbhash,solid] coordinates { (1,1.0) (8,3.93881) (16,5.50292) (24,6.35606) (32,6.93125) (40,7.33791) (48,7.49659) (56,4.18311) (64,7.37351) (72,5.26258) (80,5.83264) (88,4.08623) (96,5.97189) (104,5.39918) (112,4.52197) (120,6.91912) (128,5.01421) };
      \addlegendentry{BBHash \cite{limasset2017fast}}
      \addplot[mark=triangle,color=colorBipartiteShockHash,solid] coordinates { (1,1.0) (8,7.03837) (16,12.1863) (24,15.9501) (32,19.0763) (40,21.3479) (48,24.0592) (56,26.1507) (64,27.7516) (72,25.7765) (80,27.0637) (88,28.4175) (96,28.9553) (104,29.39) (112,29.7673) (120,30.388) (128,30.8172) };
      \addlegendentry{Bip. ShockH-RS \cite{lehmann2023bipartite}}
      \addplot[mark=square,color=colorRustFmph,solid] coordinates { (1,1.0) (8,6.67239) (16,11.0374) (24,13.6864) (32,14.7737) (40,15.3297) (48,15.4126) (56,15.275) (64,15.4964) (72,16.3556) (80,16.707) (88,17.1767) (96,16.6745) (104,17.6736) (112,16.9722) (120,16.6745) (128,15.5245) };
      \addlegendentry{FMPH \cite{beling2023fingerprinting}}
      \addplot[mark=diamond,color=colorRustFmphGo,solid] coordinates { (1,1.0) (8,5.35206) (16,10.2756) (24,13.984) (32,16.9301) (40,18.7718) (48,19.9635) (56,20.7813) (64,21.2632) (72,22.7371) (80,23.0191) (88,23.5607) (96,24.0142) (104,24.4227) (112,24.7644) (120,24.6601) (128,24.4383) };
      \addlegendentry{FMPHGO \cite{beling2023fingerprinting}}
      \addplot[mark=star,color=colorMorphisHash] coordinates { (1,1.0) (8,7.91072) (16,15.3316) (24,21.9297) (32,28.3689) (40,33.8556) (48,41.9296) (56,47.7608) (64,53.3324) (72,48.9402) (80,52.8595) (88,57.4394) (96,61.2054) (104,63.3646) (112,66.5653) (120,69.3328) (128,72.8207) };
      \addlegendentry{MorphisHash-RS \cite{hermann2025morphishash}}
      \addplot[mark=phobic,color=colorDensePtHash,densely dotted] coordinates { (1,1.0) (8,7.1912) (16,12.9875) (24,16.8776) (32,20.4921) (40,27.1398) (48,27.8037) (56,33.2375) (64,29.2157) (72,35.9283) (80,38.7679) (88,40.2757) (96,42.0569) (104,43.6729) (112,44.8931) (120,45.8201) (128,47.5986) };
      \addlegendentry{PHOBIC \cite{hermann2024phobic}}
      \addplot[mark=otimes,color=colorPhast] coordinates { (1,1.0) (8,7.1866) (16,13.1422) (24,17.8182) (32,21.4757) (40,24.8383) (48,27.3755) (56,28.5056) (64,29.1425) (72,30.5693) (80,32.1429) (88,33.0479) (96,34.1636) (104,34.9988) (112,35.7231) (120,36.729) (128,36.4327) };
      \addlegendentry{PHast \cite{beling2025phast}}
      \addplot[mark=oplus,color=colorPhastPlus] coordinates { (1,1.0) (8,6.64568) (16,11.5105) (24,14.3268) (32,16.7418) (40,18.7114) (48,19.5669) (56,19.6814) (64,19.5507) (72,20.1705) (80,20.5401) (88,20.9236) (96,20.9422) (104,21.3991) (112,21.3216) (120,21.5954) (128,20.0162) };
      \addlegendentry{PHast${}^+$ \cite{beling2025phast}}
      \addplot[mark=pentagon,color=colorPthash,solid] coordinates { (1,1.0) (8,2.31102) (16,2.30225) (24,2.60315) (32,2.79796) (40,2.90664) (48,3.01059) (56,3.01528) (64,3.02038) (72,3.04545) (80,3.04199) (88,3.51899) (96,3.44036) (104,3.07821) (112,3.43102) (120,3.07042) (128,3.02913) };
      \addlegendentry{PTHash \cite{pibiri2021pthash}}
      \addplot[mark=rightTriangle,color=colorPthash,solid] coordinates { (1,1.0) (8,8.05694) (16,15.9903) (24,23.3855) (32,29.8439) (40,37.1249) (48,41.213) (56,44.526) (64,46.4906) (72,47.0549) (80,52.1987) (88,52.9399) (96,53.9716) (104,54.3956) (112,56.919) (120,57.4929) (128,56.7856) };
      \addlegendentry{PTHash-HEM \cite{pibiri2021parallel}}
      \addplot[mark=Mercedes star,color=colorPtrHash] coordinates { (1,1.0) (8,7.274) (16,12.6564) (24,15.4093) (32,18.7078) (40,20.6076) (48,20.7504) (56,22.8929) (64,22.9368) (72,22.5056) (80,23.1139) (88,23.3848) (96,22.9808) (104,23.1139) (112,23.3848) (120,20.8953) (128,20.5722) };
      \addlegendentry{PtrHash \cite{grootkoerkamp2025ptrhash}}
      \addplot[mark=+,color=colorSimdRecSplit,solid] coordinates { (1,1.0) (8,7.4503) (16,13.9335) (24,19.1932) (32,23.4828) (40,27.4708) (48,30.8863) (56,33.1851) (64,36.0394) (72,31.8188) (80,33.765) (88,34.8826) (96,36.2271) (104,35.7798) (112,36.6084) (120,36.4549) (128,35.3795) };
      \addlegendentry{SIMDRecSplit \cite{bez2023high}}
      \addplot[mark=o,color=colorSicHash,solid] coordinates { (1,1.0) (8,7.13803) (16,13.4628) (24,18.2887) (32,21.6787) (40,24.7354) (48,26.3653) (56,27.9143) (64,28.2602) (72,28.6148) (80,29.734) (88,28.7955) (96,27.7445) (104,26.9574) (112,26.9574) (120,15.5566) (128,14.0952) };
      \addlegendentry{SicHash \cite{lehmann2023sichash}}

      \addplot[color=gray,dashed] coordinates { (64,1) (64,85) };
      \node[color=gray] at (axis cs: 72,80) {\tiny HT};

      \legend{}
    \end{axis}
\end{tikzpicture}
        \caption{Strong scaling on the 64-core AMD machine.}
        \label{fig:scalingThreadsStrong128}
    \end{subfigure}%
    \hfill
    \begin{subfigure}[t]{0.49\textwidth}
        \centering
\begin{tikzpicture}
    \begin{axis}[
        plotEvalScaling,
        ylabel={Speedup},
        xlabel={Threads},
        xtick distance=16,
      ]
      \addplot[mark=diamond,color=colorBbhash,solid] coordinates { (1,1.0) (8,1.68486) (16,2.941) (24,2.94166) (32,3.58191) (40,3.9829) (48,4.57048) (56,5.07011) (64,5.79022) (72,5.69632) (80,5.72321) (88,5.82428) (96,5.76101) (104,5.72385) (112,5.85125) (120,5.73396) (128,5.63133) };
      \addlegendentry{BBHash \cite{limasset2017fast}}
      \addplot[mark=triangle,color=colorBipartiteShockHash,solid] coordinates { (1,1.0) (8,7.0876) (16,12.1843) (24,15.5047) (32,17.9929) (40,19.7502) (48,21.1574) (56,22.9368) (64,23.727) (72,21.9575) (80,22.6687) (88,23.28) (96,23.9075) (104,23.769) (112,24.2087) (120,24.4021) (128,24.6644) };
      \addlegendentry{Bip. ShockH-RS \cite{lehmann2023bipartite}}
      \addplot[mark=square,color=colorRustFmph,solid] coordinates { (1,1.0) (8,3.67525) (16,2.99114) (24,4.40332) (32,5.25035) (40,5.78915) (48,6.27911) (56,6.80922) (64,7.13675) (72,7.25788) (80,7.12886) (88,7.32019) (96,7.87832) (104,8.00398) (112,7.92437) (120,7.76569) (128,7.17035) };
      \addlegendentry{FMPH \cite{beling2023fingerprinting}}
      \addplot[mark=diamond,color=colorRustFmphGo,solid] coordinates { (1,1.0) (8,0.980608) (16,0.710974) (24,1.13036) (32,1.42562) (40,1.68331) (48,1.92413) (56,2.12356) (64,2.30748) (72,2.50059) (80,2.65664) (88,2.79668) (96,2.91891) (104,3.09737) (112,3.20545) (120,3.26521) (128,3.30952) };
      \addlegendentry{FMPHGO \cite{beling2023fingerprinting}}
      \addplot[mark=star,color=colorMorphisHash] coordinates { (1,1.0) (8,7.98975) (16,15.5056) (24,22.2437) (32,28.7395) (40,34.5452) (48,42.8298) (56,48.704) (64,54.2901) (72,49.7513) (80,53.8388) (88,58.816) (96,63.0307) (104,64.5164) (112,68.4678) (120,71.4446) (128,74.7485) };
      \addlegendentry{MorphisHash-RS \cite{hermann2025morphishash}}
      \addplot[mark=phobic,color=colorDensePtHash,densely dotted] coordinates { (1,1.0) (8,7.39887) (16,13.6086) (24,18.6258) (32,21.7511) (40,26.8801) (48,27.5448) (56,32.639) (64,31.6233) (72,35.7959) (80,34.6822) (88,39.594) (96,40.7411) (104,42.3292) };
      \addlegendentry{PHOBIC \cite{hermann2024phobic}}
      \addplot[mark=otimes,color=colorPhast] coordinates { (1,1.0) (8,6.7311) (16,11.2493) (24,16.3144) (32,19.8656) (40,23.0732) (48,25.3733) (56,27.6991) (64,29.4343) (72,29.3776) (80,30.8614) (88,32.0322) (96,33.2021) (104,33.9459) (112,35.2176) (120,35.3033) (128,35.9196) };
      \addlegendentry{PHast \cite{beling2025phast}}
      \addplot[mark=oplus,color=colorPhastPlus] coordinates { (1,1.0) (8,5.5582) (16,8.85419) (24,11.6848) (32,13.6829) (40,15.1186) (48,16.3275) (56,17.1623) (64,17.8055) (72,17.8358) (80,18.4564) (88,18.6937) (96,18.9756) (104,19.0927) (112,19.3891) (120,19.359) (128,19.0126) };
      \addlegendentry{PHast${}^+$ \cite{beling2025phast}}
      \addplot[mark=pentagon,color=colorPthash,solid] coordinates { (1,1.0) (8,1.94353) (16,2.22722) (24,1.85657) (32,1.95606) (40,2.5128) (48,2.64686) (56,2.7825) (64,2.29044) (72,2.98614) (80,3.07082) (88,3.11464) (96,3.13798) (104,3.18577) };
      \addlegendentry{PTHash \cite{pibiri2021pthash}}
      \addplot[mark=rightTriangle,color=colorPthash,solid] coordinates { (1,1.0) (8,7.10028) (16,13.5198) (24,18.999) (32,23.5818) (40,28.4119) (48,31.1095) (56,35.6298) (64,37.3196) (72,40.0409) (80,42.0252) (88,44.4915) (96,46.8365) (104,48.5175) };
      \addlegendentry{PTHash-HEM \cite{pibiri2021parallel}}
      \addplot[mark=Mercedes star,color=colorPtrHash] coordinates { (1,1.0) (8,7.12061) (16,12.5969) (24,15.7923) (32,18.0251) (40,20.6098) (48,21.5676) (56,22.2549) (64,22.9966) (72,24.4047) (80,24.438) (88,24.8025) (96,23.9179) (104,23.9179) (112,16.2691) (120,16.0348) (128,15.7269) };
      \addlegendentry{PtrHash \cite{grootkoerkamp2025ptrhash}}
      \addplot[mark=+,color=colorSimdRecSplit,solid] coordinates { (1,1.0) (8,7.53404) (16,14.2031) (24,19.5951) (32,24.2379) (40,28.1702) (48,31.1639) (56,33.8794) (64,36.7181) (72,33.042) (80,34.163) (88,36.5866) (96,37.4255) (104,37.1234) (112,36.9817) (120,36.4801) (128,40.2172) };
      \addlegendentry{SIMDRecSplit \cite{bez2023high}}
      \addplot[mark=o,color=colorSicHash,solid] coordinates { (1,1.0) (8,9.61499) (16,18.166) (24,24.8226) (32,30.2228) (40,34.7502) (48,37.2526) (56,39.5339) (64,41.8954) (72,43.2253) (80,44.129) (88,45.7773) (96,46.0118) (104,46.5279) (112,46.4416) (120,46.0376) (128,46.1502) };
      \addlegendentry{SicHash \cite{lehmann2023sichash}}

      \addplot[color=gray,dashed] coordinates { (64,1) (64,85) };
      \node[color=gray] at (axis cs: 72,80) {\tiny HT};

      \legend{}
    \end{axis}
\end{tikzpicture}
        \caption{Weak scaling on the 64-core AMD machine.}
        \label{fig:scalingThreadsWeak128}
    \end{subfigure}

    \caption{
        Multi-threaded construction by number of threads, on a 128 core machine.
        Weak scaling with 10~million keys per thread, strong scaling with 100~million keys.
        We give self-speedups because each approach has a different focus.
    }
    \label{fig:scalingThreads128}
\end{figure*}

\section{Experiments on ARM Machine}\label{appendix:arm}
In addition to the x86 machine that we perform our main experiments on, we run the comparison on an ARM machine as well.
In this section, we give an overview over the results.
We use an Ampere Altra Q80-3 processor with 80 cores, pinned to its base clock speed of 3.00GHz.
The sizes of the L1 and L2 data caches are 64 KiB and 1 MiB per core, and the L3 cache has a size of 32 MiB.
The machine runs Rocky Linux 9.4 with Linux 5.14.0.
We use the GNU C++ compiler version 14.2.0 with optimization flags \texttt{-O3 -march=native}.
For the competitors written in Rust, we compile in release mode with \texttt{target-cpu=native}.
Because the machine does not support AVX instructions, we do not include SIMDRecSplit in our evaluation.
There are no competitors that specifically support the vector instructions on ARM machines, but the compiler might still automatically vectorize parts of all competitors.
\Cref{fig:paretoArm} gives our main comparison on the ARM machine.
Overall, the measurements are very similar to the Intel machine.
We outline differences below.
Note that an earlier version of this paper compared the machines without pinning the clock speed.
This led to different results where the ARM machine was generally slower despite its higher base clock speed.
This indicates that perfect hashing profits from Intel's TurboBoost, even though it does not effect the relative performance of the approaches.

\begin{figure}[t]
    \center
    \begin{tikzpicture}
        \pgfplotslegendfromname{legendEvalParetoConstruction}
    \end{tikzpicture}
    \vspace{1mm}

    \begin{subfigure}[t]{0.48\textwidth}
        \input{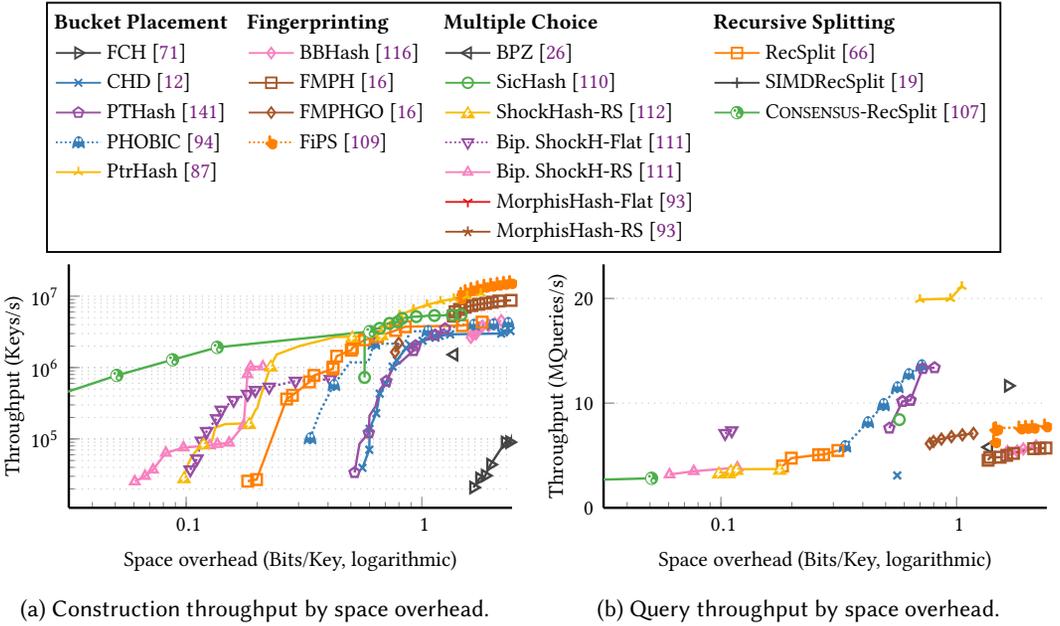}
        \caption{Construction throughput by space overhead.}
        \label{fig:paretoConstructionLogArm}
    \end{subfigure}%
    \hfill
    \begin{subfigure}[t]{0.48\textwidth}
            \begin{tikzpicture}
        \begin{axis}[
            plotEvalPareto,
            xlabel={Space overhead (Bits/Key, logarithmic)},
            ylabel={Throughput (Queries/s)},
            legend columns=2,
            ymin=0,
            xmode=log,
            ymode=log,
            xmin=0.01,
            xmax=4.5,
            ymin=8e5,
            ymax=1e8,
            log x ticks with fixed point,
            ytick distance=10,
          ]
          \addplot[mark=diamond,color=colorBbhash,solid] coordinates { (1.61538,5736247) (1.80848,5738880) (1.92019,6061708) (2.0436,6109108) };
          \addlegendentry{BBHash \cite{limasset2017fast}}
          \addplot[mark=leftTriangle,color=colorBdz,solid] coordinates { (1.3698,5581603) };
          \addlegendentry{BPZ \cite{botelho2013practical}}
          \addplot[mark=flippedTriangle,color=colorBipartiteShockHashFlat,densely dotted] coordinates { (0.10421,7293946) (0.11116,7301401) };
          \addlegendentry{Bip. ShockH-Flat \cite{lehmann2023bipartite}}
          \addplot[mark=triangle,color=colorBipartiteShockHash,solid] coordinates { (0.06027,2925174) (0.06684,2984896) (0.07648,3229139) (0.11643,3550253) };
          \addlegendentry{Bip. ShockH-RS \cite{lehmann2023bipartite}}
          \addplot[mark=x,color=colorChd,solid] coordinates { (0.59856,2868781) (0.6006,2945941) (0.8577,2974419) };
          \addlegendentry{CHD \cite{belazzougui2009hash}}
          \addplot[mark=rightTriangle,color=colorFch,solid] coordinates { (1.95731,11254924) };
          \addlegendentry{FCH \cite{fox1992faster}}
          \addplot[mark=square,color=colorRustFmph,solid] coordinates { (1.3602,5719187) (1.38346,6327111) (2.2566,6342360) };
          \addlegendentry{FMPH \cite{beling2023fingerprinting}}
          \addplot[mark=diamond,color=colorRustFmphGo,solid] coordinates { (0.76993,8366100) (0.79839,8951750) (0.86238,9103322) (0.95126,10058338) (1.05833,10338054) (1.17988,10614584) };
          \addlegendentry{FMPHGO \cite{beling2023fingerprinting}}
          \addplot[mark=fingerprint,color=colorFiPS,densely dotted] coordinates { (1.45692,6432936) (1.45983,8190679) (1.55242,8634087) (1.64282,8738203) (1.86575,8778860) (1.99088,9264406) (2.25759,9323140) (2.39705,9352787) (2.53922,9414422) (2.68343,9423294) (2.82988,9469696) (2.97752,9550186) (3.1817,9786651) (3.356,9967108) (3.53396,10158472) (3.71446,10309278) (3.89678,10437323) (4.08203,10539629) (4.26838,10631511) (4.45618,10708931) (4.64503,10796804) (4.83567,10840108) (5.02642,10892059) (5.21806,10956502) (5.41126,11001100) (5.60305,11042402) (5.7975,11080332) (5.99115,11120996) (6.18529,11134617) (6.38061,11155734) (6.57615,11195700) };
          \addlegendentry{FiPS \cite{lehmann2024fast}}
          \addplot[mark=Mercedes star flipped,color=colorMorphisHashFlat] coordinates { (0.09409,8873901) (0.10453,8939746) (0.10783,8949346) (0.11245,9174311) };
          \addlegendentry{MorphisHash-Flat \cite{hermann2025morphishash}}
          \addplot[mark=star,color=colorMorphisHash] coordinates { (0.03597,3836709) (0.03666,3981684) (0.04936,4569339) (0.05185,4653760) (0.05314,4672460) (0.05698,4693733) (0.05738,4733279) (0.06467,4737764) (0.11947,5229305) };
          \addlegendentry{MorphisHash-RS \cite{hermann2025morphishash}}
          \addplot[mark=phobic,color=colorDensePtHash,densely dotted] coordinates { (0.49045,5220569) (0.50377,5365382) (0.52978,5636978) (0.53646,5954153) (0.826,10155377) (0.91005,10266940) (0.93766,11393414) };
          \addlegendentry{PHOBIC \cite{hermann2024phobic}}
          \addplot[mark=otimes,color=colorPhast,mark repeat*=4] coordinates { (0.3605,32393909) (0.36063,33090668) (0.36324,33478406) (0.36845,33602150) (0.37675,33875338) (0.38644,34305317) (0.39898,34340659) (0.43107,34686090) (0.47198,36643459) (0.47212,37091988) (0.47333,37551633) (0.59433,40600893) (0.61187,41000410) };
          \addlegendentry{PHast \cite{beling2025phast}}
          \addplot[mark=oplus,color=colorPhastPlus,mark repeat*=4] coordinates { (0.67561,31308703) (0.67729,31505986) (0.74447,34083162) (0.75317,34164673) };
          \addlegendentry{PHast${}^+$ \cite{beling2025phast}}
          \addplot[mark=pentagon,color=colorPthash,solid] coordinates { (0.49412,7945967) (0.49919,7994883) (0.57313,9911785) (0.61338,10115314) (0.70361,11910433) (0.71841,12363996) (0.82723,12450199) (0.86296,13651877) (0.88456,13724951) };
          \addlegendentry{PTHash \cite{pibiri2021pthash}}
          \addplot[mark=Mercedes star,color=colorPtrHash] coordinates { (0.7262,16989466) (0.93185,17027073) (1.09916,17491691) (1.30481,18470631) };
          \addlegendentry{PtrHash \cite{grootkoerkamp2025ptrhash}}
          \addplot[mark=square,color=colorRecSplit,solid] coordinates { (0.18271,3945862) (0.19914,4700352) (0.25843,4964996) (0.27357,4972403) (0.3147,5411548) };
          \addlegendentry{RecSplit \cite{esposito2020recsplit}}
          \addplot[mark=shockhash,color=colorShockHash,solid] coordinates { (0.09778,3287527) (0.11005,3611933) (0.11279,3615721) (0.17757,3641925) };
          \addlegendentry{ShockHash-RS \cite{lehmann2023shockhash}}
          \addplot[mark=o,color=colorSicHash,solid] coordinates { (0.57182,7927070) (0.57184,8158603) };
          \addlegendentry{SicHash \cite{lehmann2023sichash}}
          \addplot[mark=consensus,color=colorConsensus] coordinates { (0.00212,2201382) (0.05077,2657666) (0.13558,2820078) };
          \addlegendentry{\consensus-RecSplit \cite{lehmann2025consensus}}

          \legend{}
        \end{axis}
    \end{tikzpicture}%
        \caption{Query throughput by space overhead.}
        \label{fig:paretoQueryLogArm}
    \end{subfigure}

    \caption{%
        Trade-off of construction time, space consumption, and query time on the ARM machine.
        Single-threaded measurements with $n=100$~million keys.
        For some approaches, we only show markers for every fourth point to increase readability.
    }
    \label{fig:paretoArm}
\end{figure}

\myparagraph{Construction\@.}
On the ARM machine, ShockHash gets slower.
The reason for this is that the splittings profit most from SIMD parallelism, and the approaches do not have an ARM implementation.
This is also the reason why SIMDRecSplit is missing completely.
This makes Bipartite ShockHash-Flat better relatively, given that it does not need splittings.
The construction of FiPS gets slightly faster than on the Intel machine.

\myparagraph{Queries\@.}
PHast and MorphisHash have slightly faster queries on the ARM machine, while PtrHash and PHOBIC are slower.
Overall, the performance is very similar.

\fi %

\end{document}